\documentclass[a4paper,11pt]{article}
\usepackage{graphicx,epsf,amssymb,amsbsy,amsfonts,amssymb,amsmath,jheppub}
\usepackage{hyperref}
\pdfoutput=1
\usepackage{tikz-feynman}
\tikzfeynmanset{compat=1.0.0}
\usetikzlibrary{shapes.misc}
\usetikzlibrary{decorations.pathmorphing}
\tikzset{snake it/.style={decorate, decoration=snake}}
\tikzset{cross/.style={cross out, draw=black, minimum size=2*(#1-\pgflinewidth), inner sep=0pt, outer sep=0pt},
cross/.default={10pt}}

\title{Thermal Correlators and Bosonization Dualities in Large $N$ Chern Simons Matter Theories}
\author[1]{Sudip Ghosh,} 
\author[2]{Subhajit Mazumdar}
\affiliation[1]{Okinawa Institute of Science and Technology, 1919-1 Tancha, Onna-son, Okinawa 904-0495, Japan}
 \affiliation[2]{Racah Institute of Physics, Hebrew University, Jerusalem 91904, Israel}
\emailAdd{sudip112phys@gmail.com, mazumdar.subhajit@mail.huji.ac.il
 }

\date{}

\abstract{We consider $3$-dimensional conformal field theories with $U(N)_{\kappa}$ Chern Simons gauge fields coupled to bosonic and fermionic matter fields transforming in the fundamental representation of the gauge group. In these CFTs, we compute in the t'Hooft large $N$ limit and to all orders in the t'Hooft coupling $\lambda= N/ \kappa$, the thermal two-point correlation functions of the spin $s=0$, $s=1$ and $s=2$ gauge invariant conformal primary operators. These are the lowest dimension single trace scalar, the $U(1)$ current and the stress tensor operators respectively. Our results furnish additional tests of the conjectured bosonization dualities in these theories at finite temperature.}

\begin{document}
\maketitle
\flushbottom

\section{Introduction} \label{intro}

The consideration of finite temperature effects is of central importance in the application of quantum field theories (QFTs) for studying a wide variety of physical problems. In condensed matter settings, these include for example the study of critical phenomena using the tools of conformal field theory (CFT), transport properties and thermalization. In the context of high energy physics and cosmology, some of these applications involve for instance the study of finite temperature dynamics in gauge theories, quark-gluon plasmas in heavy-ion collision experiments, and thermal phase transitions in the early universe. The presence of finite temperature effects are also of great interest in the context of the AdS/CFT correspondence \citep{Maldacena:1997re}, where thermal aspects of a class of strongly coupled quantum field theories living on the boundary of asymptotically anti de-Sitter, $AdS$ spacetimes, are holographically related to the physics of black holes in the bulk dual gravitational theory \citep{Witten:1998zw}.  

As in the case zero temperature, some of the basic observables in finite temperature QFTs are correlation functions of local operators. Generically in interacting non-supersymmetric QFTs, these are difficult to compute outside the regime of perturbation theory in the associated coupling parameters and therefore exact solutions are often hard to come by. In this paper we consider a class of interacting CFTs, without supersymmetry, in $3$-dimensions, where remarkably thermal correlation functions can be computed to all orders in the coupling constant. More specifically here, these CFTs are Chern Simons theories based on the gauge group $U(N)$ and involve coupling to either bosonic or fermionic matter fields that transform under the fundamental representation of the gauge group \citep{Giombi:2011kc},\citep{Aharony:2011jz}. There has been a plethora of recent activity regarding the study of these theories. Part of the interest in these theories stems from the fact that they are effectively solvable in the t'Hooft large $N$ limit, thereby allowing the calculation of many observables to any order in the t'Hooft coupling parameter, $\lambda$. Moreover, these theories have been conjectured to exhibit a fascinating web of strong-weak coupling bosonization dualities which can be regarded as generalisations of level-rank dualities in pure Chern Simons theory. Several computations, mainly at large $N$ but exact in the t'Hooft coupling, have been performed in these theories which furnish explicit checks of these dualities. For example at zero temperature these include the calculation of planar correlation functions of spin-$s$ current operators in \citep{Aharony:2011jz},\citep{Aharony:2012nh},\citep{GurAri:2012is},\citep{Geracie:2015drf},\citep{Turiaci:2018dht},\citep{Yacoby:2018yvy},\citep{Kalloor:2019xjb}   and the $S$-matrix elements of fundamental bosons and/or fermions obtained after deforming the conformal theories by turning on mass terms in \citep{Jain:2014nza},\citep{Dandekar:2014era},\citep{Inbasekar:2015tsa},\citep{Inbasekar:2017ieo},\citep{Inbasekar:2019wdw}. These results have also revealed several novel features, such as the modification of standard crossing symmetry rules for scattering amplitudes. On the other hand at finite temperature, which will be our main area of focus in this paper, the exact in $\lambda$, thermal free energies have also been obtained for these theories in \citep{Giombi:2011kc},\citep{Jain:2012qi},\citep{Aharony:2012ns},\cite{Jain:2013py},\citep{Yokoyama:2012fa},\citep{Jain:2013gza},\citep{Choudhury:2018iwf},\citep{Dey:2019ihe}. There are also a rich variety of thermodynamic phases that have been studied including Gross-Witten-Wadia type phase transitions \citep{Gross:1980he},\citep{Wadia:1980cp}. At finite temperature, many of the intriguing new aspects of these theories have their origin in the presence of a non-trivial holonomy for the gauge fields around the thermal circle. Chern Simons matter theories are also interesting in the context of gauge/gravity dualities, since in the large $N$ limit, these have been conjectured in \citep{Giombi:2011kc}, \citep{Aharony:2011jz} to be holographically dual to parity violating versions of Vasiliev's classical higher spin theories of gravity in $4$-dimensional AdS spacetimes \citep{Fradkin:1987ks},\citep{Vasiliev:1992av},\citep{Vasiliev:1995dn},\citep{Vasiliev:1999ba}. 

In this paper we will focus on the computation of finite temperature $2$-point functions in momentum space of gauge invariant single trace operators in the Chern Simons matter theories alluded to above. In particular we will consider the lightest gauge invariant scalar primary operator, the $U(1)$ current and the stress tensor operator in these theories. The thermal $2$-point function of the spin $s=1$ current operator has been previously obtained in the so called Chern Simons regular fermion theory, exactly in the t'Hooft coupling and perturbatively to leading order in t'Hooft coupling in the Chern Simons critical boson theory in \cite{GurAri:2016xff}. This provided a new test of the conjectured $3$-$d$ bosonisation dualities at finite temperature. In this paper we will generalise their result for Critical Bosons to all orders in t'Hooft coupling and also present a much simplified and compact representation of their $2$-point function in the Regular Fermion theory. Moreover, by considering the spin $s=0$ and $s=2$ cases in this paper, one of our main goals here is to furnish additional non-trivial checks of bosonization dualities at finite temperature in these theories. Let us also note that in the absence of Chern Simons gauge fields, the CFTs that we consider are simply vector models at finite temperature. We refer the reader to \citep{Katz:2014rla}, \citep{Romatschke:2019ybu} for calculations of thermal correlation functions in for example $O(N)$ vector models. Thermal correlation functions in Chern Simons theories involving massive matter fields in the fundamental representations have also been studied recently in \cite{Mishra:2020wos}. 

This paper is structured as follows. We begin with a summary of our main results in Section \ref{summary} where we list out the particular Chern Simons matter theories under study and then present the thermal $2$-point functions of the spin $s=0,s=1$ and $s=2$ gauge invariant operators in these theories. In Section \ref{holonomy}  we review some aspects of the thermal holonomy in these theories and how its effects are  taken into account in our computations. Section \ref{s0}, contains the detailed calculation of spin $s=0$ thermal $2$-point functions at large $N$ and exactly in the t'Hooft coupling for all the theories that we consider in this work. Here we show explicitly that our results for the thermal correlators are indeed consistent with the conjectured bosonization dualities. In Section \ref{u1} we compute the thermal correlator for the $U(1)$ current operator and also check its duality transformation properties. The thermal $2$-function of the stress tensor is studied in Section \ref{s2} of the Appendix. We end in Section \ref{futuredirec} by outlining several interesting directions for future work. The appendices in Sections \ref{notconv} and \ref{usefulstuff} specify some of the notations and conventions used throughout the paper and some useful formulae along with further details of some of the calculations in the main text.


\section{Summary of Results}\label{summary}

We have computed the thermal $2$-point functions of gauge invariant operators with spin $s=0,1,2$, i.e., the scalar singlet, the $U(1)$ current and the stress tensor operators in $U(N)$ Chern Simons gauge theories coupled to matter fields transforming under the fundamental representation of the gauge group. All our results are obtained in the large $N$ limit but incorporate the all orders contribution with respect to the t'Hooft coupling $\lambda=N/\kappa$ in these theories. For the convenience of the reader we provide a brief synopsis of our main results in this section. We first begin with a listing of the particular theories that are the subject of our study in this paper. 

\subsection{List of theories} 
\label{theorylist}

The class of theories under consideration here are $3$-$d$ conformal field theories involving non-Abelian Chern Simons gauge fields coupled to either fundamental bosonic or fundamental fermionic matter fields. We will take the gauge group to be $U(N)$.  Since we shall be primarily working at leading order in the large $N$ limit throughout this paper, the results to be reported here will also continue to hold if the gauge group is instead $SU(N)$. Now the actions for the theories we deal with are given below in equations \ref{rb} to \ref{cf}. %

\begin{itemize}

\item Regular Bosons (RB)
\begin{equation}
\label{rb}
\begin{split}
S_{RB}(\phi)= S_{CS,\kappa_{B}}+ \int d^{3}x \hspace{0.1cm} \left[D_{\mu}\phi^{\dagger}D^{\mu}\phi +\frac{x^{B}_{6}}{3! N^{2}} (\phi^{\dagger}\phi )^{3} \right]
\end{split}
\end{equation}

\item Critical Bosons (CB)  
\begin{equation}
\label{cb}
\begin{split}
S_{CB}(\phi,\sigma_{B})= S_{CS,\kappa_{B}}+ \int d^{3}x \hspace{0.1cm} \left[D_{\mu}\phi^{\dagger}D^{\mu}\phi + \sigma_{B}\phi^{\dagger}\phi\right]
\end{split}
\end{equation}
\item Regular Fermions (RF) 
\begin{equation}
\label{rf}
\begin{split}
S_{RF}(\psi)= S_{CS,\kappa_{F}}+ \int d^{3}x \hspace{0.1cm} \bar{\psi}\gamma_{\mu}D^{\mu}\psi
\end{split}
\end{equation}

\item Critical Fermions (CF) 
\begin{equation}
\label{cf}
\begin{split}
S_{CF}(\psi,\sigma_{F})= S_{CS,\kappa_{F}}+ \int d^{3}x \hspace{0.1cm} \left[\bar{\psi}\gamma_{\mu}D^{\mu}\psi +\sigma_{F}\bar{\psi}\psi +\frac{N}{3! }x^{F}_{6}\sigma_{F}^{3}\right]
\end{split}
\end{equation}
\end{itemize}

where $D_{\mu}= \partial_{\mu} +A_{\mu}$, and $A_{\mu}=A_{\mu}^{a}T^{a}$ is the Chern Simons gauge field. $T^{a}$, $a \in (1,\ldots N^{2})$ are $U(N)$ generators. We will use here the convention of \citep{Aharony:2012ns}, where $A_{\mu}$ is an anti-hermitian matrix. The Chern Simons action denoted by $S_{CS,\kappa}$ in the above is given by
\begin{equation}
\label{cs}
\begin{split}
 S_{CS,\kappa} = -\frac{i \kappa}{8\pi} \int d^{3}x \hspace{0.1cm} \epsilon^{\mu\nu\rho}  \left(A^{a}_{\mu} \partial_{\nu} A^{a}_{\rho} + \frac{1}{3} f_{abc}A^{a}_{\mu}A^{a}_{\nu}A^{a}_{\rho} \right)
 \end{split}
\end{equation}

$\kappa \in \mathbf{Z}$ is the Chern Simons level and $f_{abc}$ are the structure constants of the Lie algebra satisfied by the $U(N)$ generators. We will subsequently study the above theories in the t'Hooft large $N$-limit with $\lambda= N/ \kappa$ kept fixed. We will also employ lightcone gauge in all of our calculations where $A_{-}= \frac{A_{1}-i A_{2}}{\sqrt{2}}=0$.  As first noted in \citep{Giombi:2011kc}, the advantage of this gauge choice is that the cubic interaction terms in the Chern Simons action in equation \ref{cs} vanish. Moreover, the gluon propagators are tree level exact and do not receive corrections from matter loops in this gauge, thereby leading to significant computational simplifications. 

 Now as noted in the introduction the Chern Simons matter theories listed above have been shown to be related to each other via a remarkable web of bosonization dualities. In particular the Regular Boson and Critical Boson theories have been conjectured to be dual to the Critical Fermion and Regular Fermion theories respectively under the following mapping of parameters \citep{Aharony:2012nh},\citep{Aharony:2012ns},\citep{Choudhury:2018iwf} 
\begin{equation}
\label{dualmap}
\begin{split}
 N_{F}= |\kappa_{B}|-N_{B}, \quad \kappa_{F}=-\kappa_{B}, \quad |\lambda_{F}|=1-|\lambda_{B}|, \quad x_{6}^{B}=8\pi^{2}\left(1-|\lambda_{F}|\right)^{2} \left(3- 8 \pi \lambda_{F}x_{6}^{F} \right)
\end{split}
\end{equation}

At finite temperature we also need to take into account the mapping between the thermal mass of the fundamental bosons and that of the fundamental fermions. In \citep{Aharony:2012ns} it was shown that this map is given by $\mu_{F}(\lambda_{F})= \mu_{B}(\lambda_{B}-\mathrm{sgn}(\lambda_{B}))$, where $m_{F}=\beta^{-1} \mu_{F}$ is the thermal mass in the Chern Simons theories coupled to fundamental fermions and $m_{B}=\beta^{-1} \mu_{B}$ is the thermal mass for the elementary vector bosons, with $\beta$ being the inverse temperature. 

\subsection{Momentum space thermal $2$-point functions}

Let us now present our principal results on thermal $2$-point functions in momentum space, starting with the spin $s=0$ case in section \ref{spin0} and followed by the $s=1$ and $s=2$ cases in sections \ref{s1} and \ref{spin2} respectively. Below we denote the lowest dimension gauge invariant scalar, the $U(1)$ current and the stress tensor operators that we have considered in the Regular Boson theory by $J^{(0)}_{B}, J^{(1)}_{B}$ and $J^{(2)}_{B}$ respectively. In the Regular Fermion theory these operators will be denoted as $J^{(0)}_{F}, J^{(1)}_{F}$ and $J^{(2)}_{F}$. For Critical Bosons we use the notation $\tilde{J}^{(0)}_{B}, \tilde{J}^{(1)}_{B}$ and $\tilde{J}^{(2)}_{B}$. Similarly in case of Critical Fermions, we denote the operators of interest via $\tilde{J}^{(0)}_{F}, \tilde{J}^{(1)}_{F}$ and $\tilde{J}^{(2)}_{F}$. 

Here we have also restricted ourselves to a special kinematic regime where the external momentum $q$, assigned to the operator insertions in $2$-point correlation functions satisfies $q^{\pm} =\frac{q^{1} \pm i \hspace{0.03cm}q^{2}}{\sqrt{2}}=0$. This kinematic choice enables solving the relevant Schwinger-Dyson equations, used to derive the thermal $2$-point functions, exactly in the t'Hooft coupling $\lambda$ at large $N$. 

\subsubsection{Spin $s=0$}
\label{spin0}

\subsubsection*{Regular Boson}

\begin{equation}
\label{j0rb}
\begin{split}
\left \langle J^{(0)}_{B}(-q)J^{(0)}_{B}(q)\right \rangle_{\beta}
& = \frac{ N _{B}}{4 \pi \lambda_{B} q_{3}} \tan \left(  \frac{q_{3}\beta}{2} \mathcal{F}_{B}(q_{3}\beta,\mu_{B})\right) 
\end{split}
\end{equation}

where
\begin{equation}
\label{j0rb1}
\begin{split}
\mathcal{F}_{B}(q_{3}\beta,\mu_{B})= \int_{0}^{\infty} dx \hspace{0.1cm} x \hspace{0.05cm}\mathcal{H}_{B}(x, q_{3}\beta,\mu_{B})
\end{split}
\end{equation}

and
\begin{equation}
\label{j0rb2}
\begin{split}
\mathcal{H}_{B} (x, q_{3}\beta,\mu_{B})=  \frac{4   \hspace{0.03cm} i}{\pi} \mathrm{sgn}(\lambda_{B}) \frac{\left[  \log\left(\sinh\left(\frac{1}{2}\sqrt{x^{2}+\mu_{B}^{2}}-\frac{i\pi|\lambda_{B}|}{2}\right)\right)- \mathrm{c.c}  \right]}{\sqrt{x^{2}+\mu_{B}^{2}}\left(4 x^{2}+4 \mu_{B}^{2}+ q_{3}^{2}\beta^{2}\right)}
\end{split}
\end{equation}

In the above $\mathrm{c.c.}$ denotes complex conjugations. $q_{3} = 2\pi n_{q}\hspace{0.03cm} \beta^{-1}$ is the momentum along the thermal circle and $n_{q} \in \mathbf{Z}$  are Matsubara frequency modes. $\beta^{-1}\mu_{B}$ is the thermal mass of the fundamental bosons. The integral in \ref{j0rb1} can be performed as shown in \ref{fintegcalc}.

\subsubsection*{Regular Fermion}

\begin{equation}
\label{j0rf} 
\begin{split}
 \left\langle J^{(0)}_{F}(-q) J^{(0)}_{F}(q)\right\rangle_{\beta} & = \frac{ i \hspace{0.03cm}N_{F} q_{3}}{4 \pi \lambda_{F}} \left( \frac{q_{3}\beta\left(1-e^{i q_{3}\beta\hspace{0.03cm} \mathcal{F}_{F}(q_{3}\beta,\mu_{F})}\right) - 2 i \hspace{0.03cm}\mathrm{sgn}(\lambda_{F})\mu_{F}\left(1+e^{i q_{3}\beta\hspace{0.03cm} \mathcal{F}_{F}(q_{3}\beta,\mu_{F})}\right)}{q_{3}\beta\left(1+e^{i q_{3}\beta\hspace{0.03cm} \mathcal{F}_{F}(q_{3}\beta,\mu_{F})}\right) - 2 i \hspace{0.03cm}\mathrm{sgn}(\lambda_{F})\mu_{F}\left(1-e^{i q_{3}\beta\hspace{0.03cm} \mathcal{F}_{F}(q_{3}\beta,\mu_{F})}\right)}\right)
\end{split}
\end{equation}

where
\begin{equation}
\label{j0rf1} 
\begin{split}
\mathcal{F}_{F}(q_{3}\beta,\mu_{F})= \int_{0}^{\infty} dx \hspace{0.1cm} x \hspace{0.05cm}\mathcal{H}_{F}(x, q_{3}\beta,\mu_{F})
\end{split}
\end{equation}

and
\begin{equation}
\label{j0rf2}
\begin{split}
\mathcal{H}_{F} (x, q_{3}\beta,\mu_{F})=  \frac{4   \hspace{0.03cm} i}{\pi} \mathrm{sgn}(\lambda_{F}) \frac{\left[  \log\left(\cosh\left(\frac{1}{2}\sqrt{x^{2}+\mu_{F}^{2}}-\frac{i\pi|\lambda_{F}|}{2}\right)\right)- \mathrm{c.c}  \right]}{\sqrt{x^{2}+\mu_{F}^{2}}\left(4 x^{2}+4 \mu_{F}^{2}+ q_{3}^{2}\beta^{2}\right)}
\end{split}
\end{equation}

where $\beta^{-1}\mu_{F}$ is the thermal mass of the fundamental fermion. The integral in \ref{j0rf1} can also be performed by following analogous steps as shown in \ref{fintegcalc}.

\subsubsection*{Critical Boson}

\begin{equation}
\label{j0cb}
\begin{split}
\left \langle \tilde{J}_{B}^{(0)}(q) \tilde{J}_{B}^{(0)}(-q)\right \rangle_{\beta} & = - 4 \pi N_{B} \lambda_{B} \hspace{0.04cm} q_{3} \cot\left(\frac{q_{3}\beta }{2} \hspace{0.04cm}\mathcal{F}_{B}(q_{3}\beta, \mu_{B,c})\right)
\end{split}
\end{equation}

Here $\beta^{-1} \mu_{B,c}$ is the thermal mass in the Critical Boson theory. $\mathcal{F}_{B}(q_{3}\beta, \mu_{B,c})$ takes the same form given by \ref{j0rb1}. 

\subsubsection*{Critical Fermion}

\begin{equation}
\label{j0cf} 
\begin{split}
 \left\langle \tilde{J}^{(0)}_{F}(-q) \tilde{J}^{(0)}_{F}(q)\right\rangle_{\beta} & = \frac{ 4 i  \pi N_{F} \lambda_{F}}{q_{3}} \left( \frac{q_{3}\beta\left(1+e^{i q_{3}\beta\hspace{0.03cm} \mathcal{F}_{F}(q_{3}\beta,\mu_{F,c})}\right) - 2 i \hspace{0.03cm}\mathrm{sgn}(\lambda_{F})\mu_{F,c}\left(1-e^{i q_{3}\beta\hspace{0.03cm} \mathcal{F}_{F}(q_{3}\beta,\mu_{F,c})}\right)}{q_{3}\beta\left(1-e^{i q_{3}\beta\hspace{0.03cm} \mathcal{F}_{F}(q_{3}\beta,\mu_{F,c})}\right)  - 2 i \hspace{0.03cm}\mathrm{sgn}(\lambda_{F})\mu_{F,c}\left(1+e^{i q_{3}\beta\hspace{0.03cm} \mathcal{F}_{F}(q_{3}\beta,\mu_{F,c})}\right)}\right)
\end{split}
\end{equation}

Here $\beta^{-1}\mu_{F,c}$ is the thermal mass in the Critical Fermion theory. $\mathcal{F}_{F}(q_{3}\beta, \mu_{F,c})$ takes the same form given by \ref{j0rf1}. 


\subsubsection{Spin $s=1$}
\label{s1}

\subsubsection*{Regular Bosons} 

\begin{equation}
\label{u1rb}
\begin{split}
\left \langle  J_{B,-}^{(1)} (-q) J_{B,+}^{(1)} (q)\right\rangle_{\beta} & = \frac{i N_{B}}{16 \pi \lambda_{B} q_{3}\beta^{2}} \left(e^{i q_{3}\beta \hspace{0.04cm} \mathcal{F}_{B}(q_{3}\beta,\mu_{B})}-1\right) \left(q_{3}^{2}\beta^{2}+4\mu_{B}^{2}\right) - \frac{ N_{B} }{4 \pi  \lambda_{B}} \mathcal{X}(0)
\end{split}
\end{equation} 

where 
\begin{equation} 
\label{cdef}
\begin{split}
\mathcal{X}(0) = \frac{\lambda_{B}}{\beta} \left (\mu_{B}+\frac{i}{|\lambda_{B}|\pi} \left[ \mathrm{Li}_{2}\left(e^{-\mu_{B}+i \pi |\lambda_{B}|}\right)-\mathrm{Li}_{2}\left(e^{-\mu_{B} -i  \pi |\lambda_{B}|}\right)\right] \right)
\end{split}
\end{equation}

\subsubsection*{Regular Fermions}

\begin{equation} 
\label{u1rf}
\begin{split}
& \left\langle J^{(1)}_{F,-}(-q) J^{(1)}_{F,+}(q)\right\rangle_{\beta} \\
& = \frac{i N_{F}}{16 \pi \lambda_{F}q_{3}\beta^{2}} \left[\left( e^{i q_{3}\beta \hspace{0.04cm}\mathcal{F}_{F}(q_{3}\beta, \mu_{F})}-1\right) \left(q_{3}^{2}\beta^{2} -4 \mu_{F}^{2}\right)+4 i  \hspace{0.03cm} \mathrm{sgn}(\lambda_{F})  \hspace{0.03cm}q_{3}\beta \mu_{F}  \hspace{0.03cm} e^{i q_{3}\beta \hspace{0.04cm}\mathcal{F}_{F}(q_{3}\beta, \mu_{F})}\right] \\
\end{split}
\end{equation}

\subsubsection*{Critical Bosons}

\begin{equation}
\label{u1cb}
\begin{split}
\left \langle  \tilde{J}_{B,-}^{(1)} (-q) \tilde{J}_{B,+}^{(1)} (q)\right\rangle_{\beta} & = \frac{i N_{B}}{16 \pi \lambda_{B} q_{3}\beta^{2}} \left(e^{i q_{3}\beta \hspace{0.04cm} \mathcal{F}_{B}(q_{3}\beta,\mu_{B,c})}-1\right) \left(q_{3}^{2}\beta^{2}+4\mu_{B,c}^{2}\right) 
\end{split}
\end{equation} 

\subsubsection*{Critical Fermions}

\begin{equation} 
\label{u1cf}
\begin{split}
& \left\langle \tilde{J}^{(1)}_{F,-}(-q) \tilde{J}^{(1)}_{F,+}(q)\right\rangle_{\beta} \\
& = \frac{i N_{F}}{16 \pi \lambda_{F}q_{3}\beta^{2}} \left[\left(  e^{i q_{3}\beta \hspace{0.04cm}\mathcal{F}_{F}(q_{3}\beta, \mu_{F,c})}-1\right) \left(q_{3}^{2}\beta^{2} -4 \mu_{F,c}^{2}\right)+4 i  \hspace{0.03cm} \mathrm{sgn}(\lambda_{F})  \hspace{0.03cm}q_{3}\beta \mu_{F,c}  \hspace{0.03cm} e^{i q_{3}\beta \hspace{0.04cm}\mathcal{F}_{F}(q_{3}\beta, \mu_{F,c})}\right] \\
\end{split}
\end{equation}

\subsubsection{Spin $s=2$}
\label{spin2}

\subsubsection*{Regular Bosons} 

\begin{equation}
\label{s2rbs}
\begin{split}
&  \left\langle  J^{(2)}_{B,--}(-q) J^{(2)}_{B,++}(q)\right\rangle_{\beta}\\
& = \frac{i N_{B}}{128 \pi \lambda_{B} q_{3}\beta^{4}} \left(1-e^{ i q_{3}\beta \hspace{0.04cm}\mathcal{F}_{B}(q_{3}\beta,\mu_{B})}\right)\left(q_{3}^{2}\beta^{2}+4\mu_{B}^{2}\right)^{2} - \frac{ N_{B}}{48 \pi \lambda_{B}} \mathcal{X}^{2}(0)\left( 3 i q_{3} -4 \mathcal{X}(0)\right)\\
& + \frac{N_{B} }{12 \pi \beta^{3} } \mu_{B}^{3} -\frac{i N_{B} }{4 \pi^{2} |\lambda_{B}| \beta^{3} } \int_{\mu_{B}}^{\infty}  dx \hspace{0.1cm} x \left[ \mathrm{Li}_{2}\left(e^{-x+i \pi |\lambda_{B}|}\right)- \mathrm{Li}_{2}\left(e^{-x-i \pi |\lambda_{B}|}\right)\right]
\end{split}
\end{equation}

\subsubsection*{Regular Fermions} 

\begin{equation}
\label{s2rf}
\begin{split}
 & \left\langle J^{(2)}_{F,--}(-q) J^{(2)}_{F,++}(q)\right\rangle_{\beta}\\
& = \frac{i N_{F} }{128 \pi\lambda_{F} q_{3}\beta^{4}} \left(1- e^{i q_{3}\beta \hspace{0.04cm}\mathcal{F}_{F}\left(q_{3}\beta,\mu_{F}\right)}\right)\left(q_{3}^{2}\beta^{2}+4 \mu_{F}^{2}\right)\left(q^{2}_{3}\beta^{2}-4  \mu_{F}^{2}+4 i q_{3}\beta \hspace{0.04cm}\mathrm{sgn}(\lambda_{F}) \mu_{F}\right)\\
& + \frac{i N_{F} }{96 \pi\lambda_{F} \beta^{3}} \left[ 2 \mu_{F}^{2}\left(3 q_{3}\beta - 2i \hspace{0.04cm} \mathrm{sgn}(\lambda_{F}) \mu_{F} \right) - 3 i \hspace{0.04cm} q_{3}^{2}\beta^{2} \hspace{0.04cm} \mathrm{sgn}(\lambda_{F}) \mu_{F} \right]  \\
&+ \frac{N_{F} }{12 \pi  \beta^{3}} \mu_{F}^{3} - \frac{i N_{F} }{4 \pi^{2} \lambda_{F} \beta^{3}} \int_{\mu_{F}}^{\infty} dx \hspace{0.1cm} x \left[ \mathrm{Li}_{2}\left(-e^{-x+i \pi \lambda_{F}}\right) - \mathrm{Li}_{2}\left(-e^{-x-i \pi \lambda_{F}}\right) \right] 
\end{split}
\end{equation}

\subsubsection*{Critical Bosons} 

\begin{equation}
\label{s2cb}
\begin{split}
&  \left\langle  \tilde{J}^{(2)}_{B,--}(-q) \tilde{J}^{(2)}_{B,++}(q)\right\rangle_{\beta}\\
& = \frac{i N_{B}}{128 \pi \lambda_{B} q_{3}\beta^{4}} \left(1-e^{ i q_{3}\beta \hspace{0.04cm}\mathcal{F}_{B}(q_{3}\beta,\mu_{B,c})}\right)\left(q_{3}^{2}\beta^{2}+4\mu_{B,c}^{2}\right)^{2} \\
& + \frac{N_{B} }{12 \pi \beta^{3} } \mu_{B,c}^{3} -\frac{i N_{B} }{4 \pi^{2} |\lambda_{B}| \beta^{3} } \int_{\mu_{B,c}}^{\infty}  dx \hspace{0.1cm} x \left[ \mathrm{Li}_{2}\left(e^{-x+i \pi |\lambda_{B}|}\right)- \mathrm{Li}_{2}\left(e^{-x-i \pi |\lambda_{B}|}\right)\right]
\end{split}
\end{equation}

\subsubsection*{Critical Fermions} 

\begin{equation}
\label{s2cf}
\begin{split}
& \left\langle  \tilde{J}^{(2)}_{F,--}(-q) \tilde{J}^{(2)}_{F,++}(q)\right\rangle_{\beta}\\
& = \frac{i N_{F} }{128 \pi\lambda_{F} q_{3}\beta^{4}} \left(1- e^{i q_{3}\beta \hspace{0.04cm}\mathcal{F}_{F}\left(q_{3}\beta,\mu_{F,c}\right)}\right)\left(q_{3}^{2}\beta^{2}+4 \mu_{F,c}^{2}\right)\left(q^{2}_{3}\beta^{2}-4  \mu_{F}^{2}+4 i q_{3}\beta \hspace{0.04cm}\mathrm{sgn}(\lambda_{F}) \mu_{F,c}\right)\\
& + \frac{i N_{F} }{96 \pi\lambda_{F} \beta^{3}} \left[ 2 \mu_{F,c}^{2}\left(3 q_{3}\beta - 2i \hspace{0.04cm} \mathrm{sgn}(\lambda_{F}) \mu_{F,c} \right) - 3 i \hspace{0.04cm} q_{3}^{2}\beta^{2} \hspace{0.04cm} \mathrm{sgn}(\lambda_{F}) \mu_{F,c} \right]  \\
&+ \frac{N_{F} }{12 \pi  \beta^{3}} \mu_{F}^{3} - \frac{i N_{F} }{4 \pi^{2} \lambda_{F} \beta^{3}} \int_{\mu_{F,c}}^{\infty} dx \hspace{0.1cm} x \left[ \mathrm{Li}_{2}\left(-e^{-x+i \pi \lambda_{F}}\right) - \mathrm{Li}_{2}\left(-e^{-x-i \pi \lambda_{F}}\right) \right] 
\end{split}
\end{equation}

\section{Thermal Holonomy}
\label{holonomy}

An important aspect of gauge theories at finite temperature is the holonomy of the gauge fields around the Euclidean thermal circle. The presence of a nontrivial holonomy crucially affects the infrared dynamics of the finite temperature theory, since it constitutes the lightest degrees of freedom at energies small compared to inverse radius of the thermal circle. In the case of Chern Simons matter theories, the effects of the thermal holonomy has been studied in details in \cite{Aharony:2012ns, Jain:2013py}. In order to incorporate the contribution of the holonomy in our ensuing computation of thermal correlation functions, we will adopt the prescription of \cite{Aharony:2012ns}, which we shall now briefly review.

Let us denote the gauge field holonomy as
\begin{equation}
\label{hol1}
a= i \hspace{0.03cm} \int_{0}^{\beta} dx^{3} \hspace{0.1cm} \mathcal{A}_{3} 
\end{equation}

where $\beta=T^{-1}$, $T$ is the temperature. $x_{3}$ is the Euclidean time coordinate and  $\mathcal{A}_{3}$ is the spatial zero mode of the gauge field component $A_{3}(x)$.  Thus, 
\begin{equation}
\label{hol2}
\mathcal{A}_{3}= \frac{1}{V_{2}}\int_{\mathbb{R}^{2}}d^{2}\vec{x} \hspace{0.1cm} A_{3}(\vec{x},x^{3})
\end{equation}

Here $V_{2}$ denotes the volume of the spatial manifold $\mathbb{R}^{2}$ in the presence of an infrared cutoff, which can be taken to infinity at the end of the calculation. Let us also chose a gauge where $\partial_{3}\mathcal{A}_{3}=0$ and diagonalise $\mathcal{A}_{3}$. 

Now in the large $N$ limit, physical observables can be computed by  treating the holonomy as a spacetime independent background gauge field. This is because at large $N$, the path integral over all possible configurations of eigenvalues of the holonomy matrix will typically be dominated by a saddle point. In \citep{Aharony:2012ns} it was shown that when $V_{2} T^{2}$ is parametrically larger than $N$, with $V_{2}$ being the volume of the spatial manifold, then to leading order in the large $N$ limit, the saddle point configuration for the eigenvalues of the holonomy $a$, tends to the following smooth distribution 
\begin{equation}
\label{hol3}
a_{ii} \rightarrow a(u)= 2\pi |\lambda| u , \quad u \in \left( -\frac{1}{2} ,\frac{1}{2} \right)
\end{equation}

Note that \ref{hol3} is universal since it does not depend on the details of the matter content and  interactions of the particular Chern Simons matter theory under consideration. We will be working with this distribution throughout the rest of this paper. Now the primary effects of the inclusion of the holonomy in our calculations are twofold. First, all loop momenta will get shifted as, 
\begin{equation}
\label{hol4}
p_{\mu} \rightarrow \tilde{p}_{\mu} = p_{\mu} - \frac{a}{\beta} \delta_{\mu,3}
\end{equation}

Secondly, we will often encounter terms involving functions of the holonomy together with a trace over the fundamental indices carried by the matter fields. From  equation \ref{hol3} it follows that in the large $N$ limit, the trace in such terms needs to be replaced by an integral with respect to $u$ as follows,
\begin{equation}
\label{hol5}
\sum_{i=1}^{N} f(a_{ii }) \rightarrow \int_{-1/2}^{1/2} du \hspace{0.1cm} f(2\pi |\lambda| u) 
\end{equation}

In the later sections we will refer to such integrals as holonomy integrals.


\section{Thermal $2$-point Functions: Spin $s=0$}
\label{s0}

In this section we present the computation of  the finite temperature two point function of the lightest gauge invariant scalar primary operator in the spectrum of the theories listed in section \ref{theorylist}. In the RB and RF theories this operator will be denoted by $J^{(0)}_{B}$ and $J^{(0)}_{F}$ respectively. On the other hand in the CB and CF theories, we shall use the notation $\tilde{J}^{(0)}_{B}$ and $\tilde{J}^{(0)}_{F}$ respectively. 

\subsection{Regular Bosons}

In the regular boson theory, the operator $J^{(0)}_{B}$ is the single trace operator given by $\phi^{\dagger}_{i} \phi^{i}$. The conformal dimension of this operator in the large $N_{B}$ limit is not renormalised and is thereby equal to the scaling dimension in the free theory which is given by $\Delta ({J^{(0)}_{B}}) =1$. Now to obtain the finite temperature $2$-point function for $J^{(0)}_{B}$ we will first of all require the thermal propagator of the fundamental boson $\phi ^{i}$. This result which has already been computed in \citep{Aharony:2012ns} in the large $N_{B}$ limit and to all orders in the t'Hooft coupling $\lambda_{B}$, is given by
\begin{equation}
\label{thermalpropboson}
\left\langle \phi_{i}^{\dagger}(-p)\phi^{j}(q)\right\rangle_{\beta} =G^{j}_{i}((p) \hspace{0.04cm} (2\pi)^{3}\delta_{n_{p},n_{q}}\delta^{2}(\vec{p}-\vec{q})
\end{equation}

where 
\begin{equation}
\label{thermalpropboson1}
\begin{split}
& G^{j}_{i}(p) =\left(\frac{1}{\tilde{p}^{2}+\beta^{-2}\mu_{B}^{2}}\right)_{ji} 
\end{split}
\end{equation}

and
\begin{equation}
\label{thermalpropboson2}
\begin{split}
& \tilde{p}_{\mu}=p_{\mu} - \frac{ a }{\beta}\delta_{\mu 3 }, \quad p_{3}= \frac{2\pi n_{p}}{\beta},\quad  q_{3}= \frac{2\pi n_{q}}{\beta}
\end{split}
\end{equation}

Here $(n_{p},n_{q}) \in \mathbf{Z}$ denote discrete Matsubara frequencies corresponding to the momentum along the thermal circle. $a$ is the holonomy of the gauge field along the Euclidean circle. The thermal mass has been denoted by $\beta^{-1} \mu_{B}$. In \citep{Aharony:2012ns} , $\mu_{B}$ was also calculated exactly in the t'Hooft coupling at large $N_{B}$ and was shown to be determined via the following equation 
\begin{equation}
\label{thermalmassrb}
\pm \mu_{B}=- \frac{|\hat{\lambda}\mu_{B}|}{2}-\frac{1}{2\pi i}\frac{|\hat{\lambda}|}{|\lambda_{B}|}\left[\mathrm{Li}_{2}(e^{-\mu_{B}-\pi i |\lambda_{B}|})-\mathrm{Li}_{2}(e^{-\mu_{B}+\pi i |\lambda_{B}|})\right]
\end{equation}

where
\begin{equation}
\hat{\lambda}^{2}=\lambda^{2}_{B}+\frac{x^{B}_{6}}{8\pi^{2}}
\end{equation}

\subsubsection{Finite temperature vertex factor for $J^{(0)}_{B}$}
 
Following the strategy of \citep{Aharony:2012nh} we will gear up towards the computation of the thermal two point function  by first obtaining the finite temperature vertex factor for  $J^{(0)}_{B}$. The two point function can be subsequently evaluated by joining two such vertex factors by a pair of the fundamental boson propagators. Let us define the vertex factor to be 
\begin{equation}
\label{j0vertex}
\left \langle J^{(0)}_{B}(-q) \phi_{i}^{\dagger}(-k) \phi^{j}(p)\right \rangle_{\beta} = V^{(0)}_{B}(q,k)\delta^{j}_{i} (2\pi)^{3} \delta_{n_{q}+n_{k},n_{p}} \delta^{2}(\vec{q}+\vec{p}-\vec{k}) 
\end{equation}

\begin{figure}[ht]
\begin{center}
\begin{tikzpicture}[scale=.5, transform shape]

 \draw[black, thick,->] (-25,0) -- (-23,0);
 \draw[black] (-19.8,1.6) -- (-18,3);
  \draw[black,->] (-21.6,0.2) -- (-19.8,1.6);

   \draw[black] (-21.6,-0.2) -- (-19.8,-1.6);
      \draw[black,->] (-18,-3)--(-19.8,-1.6) ;

     \Huge{ \node at (-24,-1) {$q$};}
       \node at (-18,0) {$=$};
     \node at (-20,3) {$p$};
     \node at (-20,-3) {$k$};
     \draw (-22,0) circle (14pt);

\draw (-22,0) node[cross] {};

  \draw[black] (-17,0) -- (-15,-1.5);
  \draw[black,->] (-13,-3)--(-15,-1.5) ;
  
  \draw[black,->] (-17,0) -- (-15,+1.5);
  \draw[black] (-15,1.5) -- (-13,3);
  
    \node at (-15,3) {$p$};
     \node at (-15,-3) {$k$};
      
     \draw (-17,0) node[cross] {};
\node at (-12.5,0) {$+$};

 \draw[black] (-10,0) -- (-8,-1.5);
  \draw[black,->]  (-6,-3)--(-8,-1.5);
  
  \draw[black,->] (-10,0) -- (-8,+1.5);
  \draw[black,->] (-8,1.5) -- (-6,3);
  
   \node at (-7.5,1.9)[circle,fill,inner sep=5pt]{};
   \node at (-7.5,-1.9)[circle,fill,inner sep=5pt]{};
     \draw (-10,0) node[cross] {};
     \path [draw=black,snake it]
    (-6.8,2.4) -- (-6.8,-2.4);
   \node at (-5.2,3.2) {$p$};
     \node at (-5.2,-3.2) {$k$};
     \node at (-9.2,2) {$\ell$};
     \node at (-9.2,-2) {$\ell-q$};
     \draw (-10,0) circle (14pt);
\end{tikzpicture}
\caption{Diagrammatic representation of the Schwinger-Dyson equation for the exact vertex factor $V^{(0)}_{B}(q,k)$. }
\label{sdysonrbs0}
  \end{center}
\end{figure} 

In \citep{Aharony:2012nh} $V^{(0)}_{B}(q,k)$ was calculated at zero temperature in the special kinematic configuration where $q^{\pm}=0, q_{3}\ne 0$. There it was argued that in this kinematic regime, the seagull vertices coming from the $\phi A_{\mu}A^{\mu}\phi^{\dagger}$ term in the action in equation \ref{rb} do not contribute to the connected scalar $4$-point function of the $\phi_{i}$ operator at large $N_{B}$. As a consequence the Schwinger Dyson equation for this vertex can be solved exactly to all orders in $\lambda_{B}$. Turning on a finite temperature in our case does not change this conclusion. Henceforth all of our computations will be performed in the regime $q^{\pm}=0$. 

Now the Schwinger Dyson equation for the vertex factor in equation \ref{j0vertex} is given by
\begin{equation}
\label{j0sdyson}
V^{(0)}_{B}(q,k)\delta^{j}_{i}=\delta^{j}_{i}+\int \mathcal{D}^{3}\ell\hspace{0.1cm}\left[\mathcal{V}^{a,\mu}(k,\ell)G(\ell)V^{(0)}_{B}(q,\ell)G(\ell+q)\mathcal{V}^{a,\nu}(\ell+q, k+q)\right]_{i}^{j}\mathcal{G}_{\nu\mu}(k-\ell)
\end{equation}

where the integration measure in the above integral is 
\begin{equation}
\label{intmeas}
\int \mathcal{D}^{3}\ell \equiv \int \frac{d^{2}\ell}{(2\pi)^{2}} \hspace{0.1cm} \frac{1}{\beta} \sum_{n=-\infty}^{\infty}
\end{equation}

The first term in the R.H.S. of the above equation corresponds to the vertex factor in the free theory and
\begin{equation}
\label{rbcubicvertex}
(\mathcal{V}^{a,\mu}(k_{1},k_{2}))_{ij} =i (T^{a}\tilde{k}_{1}^{\mu}+\tilde{k}_{2}^{\mu}T^{a})_{ij} 
\end{equation}

is the vertex factor for the cubic interaction terms $\phi^{\dagger}\partial_{\mu}\phi A^{\mu}, \phi\partial_{\mu}\phi^{\dagger}A^{\mu}$ in the action \ref{rb} and $\mathcal{G}_{\mu\nu}(p)$ is the gauge field propagator. In light cone gauge the only non vanishing components of the propagator involve $(\mu,\nu) \in (+,3)$ and is given by
\begin{equation}
\label{gaugeprop}
\mathcal{G}_{+3}(p)=-\mathcal{G}_{3+}(p)=\frac{4\pi i}{\kappa_{B}}\frac{1}{p^{+}}
\end{equation}

Now using the following relation for generators in the fundamental representation of $U(N)$
\begin{equation}
\label{ungen}
T^{a}_{j\alpha_{1}}T^{a}_{\alpha_{4}i}=-\frac{1}{2}\delta_{ji}\delta_{\alpha_{1}\alpha_{4}}
\end{equation}

we can simplify equation \ref{j0sdyson} to get,
\begin{equation}
\label{j0sdyson1}
\begin{split}
 V^{(0)}_{B}(q,k) & =1+q^{3} \int \mathcal{D}^{3}\ell\hspace{0.1cm} (\ell^{+}+k^{+}) \mathrm{Tr}\left[G(\ell)V^{(0)}_{B}(q,\ell)G(\ell+q)\right]\mathcal{G}_{ 3+}(k-\ell)\\
& =1+ \frac{4\pi i}{\kappa_{B}}q_{3} \int \frac{d^{2}\ell}{(2\pi)^{2}}\hspace{0.1cm} \frac{(\ell^{+}+k^{+})}{(\ell^{+}-k^{+})} \hspace{0.1cm}\frac{1}{\beta}\sum_{n=-\infty}^{\infty}\mathrm{Tr}\left[G(\ell)V^{(0)}_{B}(q,\ell)G(\ell+q)\right]
\end{split}
\end{equation}

In the large $N_{B}$ limit, the trace in the above expression can be replaced by an integral over the holonomy. This yields,
\begin{equation}
\label{j0sdyson2}
\begin{split}
V^{(0)}_{B}(q,k)\delta^{j}_{i} & = 1+ \frac{4 i \pi \lambda_{B}}{\beta}q_{3} \int _{-1/2}^{1/2} du \int \frac{d^{2}\ell}{(2\pi)^{2}}\hspace{0.1cm} \frac{(\ell^{+}+k^{+})}{(\ell^{+}-k^{+})} \sum_{n=-\infty}^{\infty}\left(G(\ell)V^{(0)}_{B}(q,\ell)G(\ell+q)\right)
\end{split}
\end{equation}

Now note that in the RHS of equation \ref{j0sdyson2} there is no dependence on $k_{3}$. This implies that the vertex factor $V(q,\ell)$ appearing inside the Matsubara sum does not depend on $\ell_{3}$ and thus can be taken out of the sum over $n$. Then we get,
\begin{equation}
\label{j0sdyson3}
\begin{split}
V^{(0)}_{B}(q,k) & =  1+4 i \pi \lambda_{B} \hspace{0.04cm} q_{3}  \int \frac{d^{2}\ell}{(2\pi)^{2}}\hspace{0.1cm} \frac{(\ell^{+}+k^{+})}{(\ell^{+}-k^{+})} V^{(0)}_{B}(q,\ell) H_{B} (x, q_{3}\beta,\mu_{B})
\end{split}
\end{equation}

where we have defined
\begin{equation}
\label{j0sdyson4}
\begin{split}
H_{B} (x, q_{3}\beta,\mu_{B})=  \frac{1}{\beta} \int _{-1/2}^{1/2} du \sum_{n=-\infty}^{\infty}\frac{1}{\left((\tilde{\ell}+q)^{2}+\beta^{-2}\mu_{B}^{2}\right)\left(\tilde{\ell}^{2}+\beta^{-2}\mu_{B}^{2}\right)}
\end{split}
\end{equation}

with $x=\beta |\vec{\ell}|\equiv \beta \ell_{s}$. The sum over Matsubara modes and the holonomy integral can be easily done as shown in detail in the Appendix in section \ref{msumholint}. Here we simply write the final result
\begin{equation}
\label{j0sdyson5}
\begin{split}
H_{B} (x, q_{3}\beta,\mu_{B})=  \frac{i \hspace{0.03cm}\beta^{3}}{\pi |\lambda_{B}|}  \frac{\left[  \log\left(\sinh\left(\frac{1}{2}\sqrt{x^{2}+\mu_{B}^{2}}-\frac{i\pi|\lambda_{B}|}{2}\right)\right)- \mathrm{c.c}  \right]}{\sqrt{x^{2}+\mu_{B}^{2}}\left(4 x^{2}+4 \mu_{B}^{2}+ q_{3}^{2}\beta^{2}\right)}
\end{split}
\end{equation}

Now with $q^{\pm}=0$, the dependence of $V^{(0)}_{B}(q,\ell)$ on $\ell$ is solely through the magnitude of the spatial vector $\vec{\ell}$, i.e., $ \ell_{s} = \sqrt{2\ell_{-}\ell_{+}}$. This essentially follows from rotational invariance in the spatial plane. Then the angular integral in equation \ref{j0sdyson3} can also be easily performed. Again we show in section \ref{angintegs} that the relevant angular integral in this case evaluates to
\begin{equation}
\label{j0sdyson6}
\begin{split}
\int _{0}^{2\pi} d\theta \hspace{0.1cm} \frac{(\ell^{+}+k^{+})}{(\ell^{+}-k^{+})}  = 2\pi \left [ \Theta\left(\ell_{s}-k_{s}\right)  -\Theta\left(k_{s}-\ell_{s}\right)  \right]
\end{split}
\end{equation}

Then the Schwinger Dyson equation takes the form
\begin{equation}
\label{j0sdyson7}
\begin{split}
V^{(0)}_{B}(q,k) & = 1+ \frac{2 i \lambda_{B} \hspace{0.03cm}  q_{3}}{\beta ^{2}} \int_{ y }^{\infty} d x \hspace{0.1cm} x  \hspace{0.04cm} V^{(0)}_{B}(q, \ell) H_{B} (x, q_{3}\beta,\mu_{B})- \frac{2 i \lambda_{B} \hspace{0.03cm}  q_{3}}{\beta^{2}} \int_{0}^{y} d x \hspace{0.1cm} x  \hspace{0.04cm} V^{(0)}_{B}(q,\ell) H_{B} (x, q_{3}\beta,\mu_{B})   
\end{split}
\end{equation}

where $x=\beta\ell_{s}, y= \beta k_{s}$. Now it is useful to differentiate both sides of the above equation with respect to $y$. This yields
\begin{equation}
\label{j0sdyson8}
\begin{split}
\partial_{y} V^{(0)}_{B}(q,k)= - \frac{ 4 i \lambda_{B}\hspace{0.03cm}  q_{3}}{\beta^{2}} \hspace{0.05cm} y\hspace{0.03cm} V^{(0)}_{B}(q, k)H_{B}(y, q_{3}\beta, \mu_{B})
\end{split}
\end{equation}

Integrating the above we get,
\begin{equation}
\label{j0sdyson9}
V^{(0)}_{B}(q,k) = V^{(0)}_{B}(q)\hspace{0.03cm} e^{ i q_{3}\beta \hspace{0.03cm} \mathcal{F}_{B}(y,q_{3}\beta,\mu_{B})}
\end{equation}

where
\begin{equation}
\label{j0sdyson10}
\begin{split}
\mathcal{F}_{B}(y,q_{3}\beta,\mu_{B}) &= \int_{y}^{\infty} dx \hspace{0.05cm} x \hspace{0.04cm}\mathcal{H}_{B}(x, q_{3}\beta,\mu_{B})  = \frac{4 \lambda_{B}}{\beta^{3}}\int_{y}^{\infty} dx \hspace{0.05cm} x \hspace{0.04cm} H_{B} (x, q_{3}\beta,\mu_{B})
\end{split}
\end{equation}

The above integral can in fact be done explicitly. We will not show it here, but instead refer the reader to section \ref{fintegcalc} of the Appendix which provides further details regarding this.  Now, the integration constant $V^{(0)}_{B}(q)$ in equation \ref{j0sdyson9} can be also straightforwardly determined as follows. Taking $y \rightarrow \infty$ in \ref{j0sdyson7} we get,
\begin{equation}
\label{j0sdyson11}
\begin{split}
V^{(0)}_{B}(q) &= 1 + \frac{2 i \lambda_{B} \hspace{0.03cm}  q_{3}}{\beta^{2}}\int_{0}^{\infty} dx \hspace{0.1cm} x \hspace{0.04cm} V^{(0)}_{B}(q) \hspace{0.04cm}e^{i q_{3}\beta \hspace{0.05cm}\mathcal{F}_{B}(x,q_{3}\beta,\mu_{B})} H_{B}(x, q_{3}\beta, \mu_{B}) \\
& = 1 + \frac{1}{2}V^{(0)}_{B}(q) \left( e^{i q_{3}\beta \hspace{0.05cm}\mathcal{F}_{B}(y,q_{3}\beta,\mu_{B})} \right)_{0}^{\infty}
\end{split}
\end{equation}

Then noting that $\mathcal{F}_{B}(y,q_{3}\beta,\mu_{B})\big|_{y=\infty} =0$ we get from \ref{j0sdyson11}
\begin{equation}
\label{j0sdyson12}
\begin{split}
 V^{(0)}_{B}(q) = \frac{2}{ 1+ e^{ - i q_{3}\beta  \hspace{0.05cm}\mathcal{F}_{B}(q_{3}\beta,\mu_{B})}}
\end{split}
\end{equation}

where $\mathcal{F}_{B}(q_{3}\beta,\mu_{B}) \equiv \mathcal{F}_{B}(y=0, q_{3}\beta,\mu_{B})$. Thus finally the vertex factor $V^{(0)}_{B}(q,k)$ is given by
\begin{equation}
\label{j0sdyson13}
\begin{split}
&  V^{(0)}_{B}(q, k) =  \frac{2 \hspace{0.04cm}e^{i q_{3}\beta \hspace{0.05cm}\mathcal{F}_{B}(y, q_{3}\beta,\mu_{B})}}{ 1+ e^{i q_{3}\beta \hspace{0.05cm}\mathcal{F}_{B}(q_{3}\beta,\mu_{B})}}
\end{split}
\end{equation}

\subsubsection{Thermal $2$-point function}
\label{rb2pts0}

\begin{figure}[ht]
\begin{center}
\begin{tikzpicture}[scale=.5, transform shape]
 
 \Huge{ \node at (2.5,0) {$J^{(0)}_{B}$};}
  \node at (-2.5,0) {$J^{(0)}_{B}$};
 \node at (0,2)[circle,fill,inner sep=12pt]{};
     \draw[black, thick,->] (-7,0) -- (-4.8,0);
     \node at (0,2)[circle,fill,inner sep=12pt]{};
     \node at (0,-2)[circle,fill,inner sep=12pt]{};
      \node at (-5.2,1.0) {$q$};
     \draw (-4.2,0) circle (12pt);
      \draw [black,thick ,domain=20:152] plot ({4.5*cos(\x)}, {-1+3*sin(\x)});
      \draw [black,thick ,domain=340:208] plot ({4.5*cos(\x)}, {1+3*sin(\x)});
   
\draw (-4.2,0) node[cross] {};
\draw (4.2,0) node[cross] {};
\end{tikzpicture}
\caption{Feynman diagram for thermal $2$-point function of $J^{(0)}_{B}$ in the RB theory. The circled cross on the left denotes the exact vertex factor $V^{(0)}_{B}(q,k)$. The cross on the right is the vertex factor in the free theory. The internal lines with black dots denote exact thermal propagators for the fundamental boson $\phi^{i}$.}
\label{2pointrbs0}
  \end{center}
\end{figure}

The thermal $2$-point function can now be obtained by evaluating the Feynman diagram in Figure \ref{2pointrbs0} where the vertex factor $V^{(0)}_{B}(q,k)$ computed in the above section is connected to the corresponding vertex factor in the free theory with a pair of exact thermal propagators for the matter fields $\phi^{i}$. Note that we would be overcounting diagrams if we inserted exact vertices on both sides in Figure \ref{2pointrbs0}. We thus have, 
\begin{equation}
\label{j0rbthermal1}
\begin{split}
\left \langle J^{(0)}_{B}(-q)J^{(0)}_{B}(q)\right \rangle_{\beta} & =\int \frac{d^{2}\ell}{(2\pi)^{2}}\hspace{0.2cm}\frac{1}{\beta}\sum_{n=-\infty}^{\infty} \mathrm{Tr}\left[V^{(0)}_{B}(q,\ell)G(\ell)V^{(0)}_{B, free}(-q,\ell+q)G(\ell+q)\right]
\end{split}
\end{equation}

Now in the Regular Boson theory \ref{rb}, we have $V^{(0)}_{B, free}(q,k)=1$.  Also the angular integrals in this case is trivial since the integrand in equation \ref{j0rbthermal1} is manifestly independent of the angular variables. Then performing the Matsubara sum and holonomy integral and using \ref{j0sdyson13} we obtain  
\begin{equation}
\label{j0rbthermal2}
\begin{split}
\left \langle J^{(0)}_{B}(-q)J^{(0)}_{B}(q)\right \rangle_{\beta}&=  \frac{i N_{B}}{4 \pi \lambda_{B} q_{3}} \left( \frac{1- e^{i q_{3}\beta \hspace{0.05cm}\mathcal{F}_{B}(q_{3}\beta,\mu_{B})}}{1+ e^{i q_{3}\beta \hspace{0.05cm}\mathcal{F}_{B}(q_{3}\beta,\mu_{B})}}\right) \\
& = \frac{ N _{B}}{4 \pi \lambda_{B} q_{3}} \tan \left(  \frac{q_{3}\beta}{2} \mathcal{F}_{B}(q_{3}\beta,\mu_{B})\right) 
\end{split}
\end{equation}


\subsubsection{ Zero temperature limit}

Let us consider the zero temperature limit of the thermal $2$-point function in equation \ref{j0rbthermal2}. This will serve to check our result against that of \citep{Aharony:2012nh}. Now as $\beta\rightarrow \infty$, the thermal mass $\beta^{-1}\mu_{B}$ goes to zero and we find,
\begin{equation}
\label{t0limrb1}
\begin{split}
& \lim_{\beta\rightarrow \infty} \mathcal{H}_{B} (x,q_{3}\beta, \mu_{B}) =  \frac{4 \hspace{0.04cm} \lambda_{B}}{x \left(4 x^{2}+q_{3}^{2}\beta^{2}\right)}
\end{split}
\end{equation}

Consequently it follows from equation \ref{j0sdyson10} that, 
\begin{equation}
\label{t0limrb2}
\begin{split}
\lim_{\beta\rightarrow \infty} \mathcal{F}_{B}(q_{3}\beta,\mu_{B}) = \frac{ \pi}{q_{3}\beta} \lambda_{B}\mathrm{sgn}(q_{3}) 
\end{split}
\end{equation}

Thus in the zero temperature limit, the $2$-point function in equation \ref{j0rbthermal2} becomes
\begin{equation}
\label{t0limrb3}
\begin{split}
 \lim_{\beta\rightarrow \infty} \left \langle J^{(0)}_{B}(-q)J^{(0)}_{B}(q)\right \rangle_{\beta} = \frac{ N _{B}}{4 \pi \lambda_{B} q_{3}} \tan \left( \frac{\pi \lambda_{B} \hspace{0.03cm} \mathrm{sgn}(q_{3})}{2}\right)
\end{split}
\end{equation}

This precisely agrees with the corresponding result obtained in \citep{Aharony:2012nh}.

\subsection{Critical Bosons}
\label{s0cb}
We will now compute the thermal $2$-point function of  $\tilde{J}^{(0)}_{B}$, the lightest gauge invariant scalar operator in the spectrum of the Critical Boson theory. Referring to the action in \ref{cb}, $\tilde{J}^{(0)}_{B}$  is simply the operator corresponding to the Hubbard-Stratonovich field $\sigma_{B}$.  To leading order in large $N_{B}$, the dimension of this operator in not renormalised and is given by $\Delta = 2+ \mathcal{O}(1/N_{B})$. 

In the Critical Boson theory, the thermal propagator for the fundamental bosons again takes the same form as in the case of the Regular Boson theory given in equations \ref{thermalpropboson} and \ref{thermalpropboson1}. However as shown in \citep{Aharony:2012ns}, the thermal mass in the case is determined by the following equation 
\begin{equation}
\label{thermmasscb}
\begin{split}
\lambda_{B} \mu_{B,c}=\frac{i}{\pi}\left[\mathrm{Li}_{2}\left(e^{-\mu_{B,c}-\pi i \lambda_{B}}\right)-\mathrm{Li}_{2}\left(e^{-\mu_{B,c}+\pi i \lambda_{B}}\right)\right]
\end{split}
\end{equation}

At large $N_{B}$ the correlation functions of gauge invariant operators in the Critical Boson theory can be obtained from the corresponding correlators in the Regular Boson theory using the following argument outlined in \citep{Aharony:2012nh}\footnote{Here we are considering the finite temperature generalisation of the arguments applicable in the zero temperature case of \citep{Aharony:2012nh}}. Consider deforming the regular bosons with $x^{B}_{6}=0$ in \ref{rb} by turning on a quartic interaction of the form $\frac{\lambda_{4}}{2N_{B}} \left(\phi^{\dagger} \phi\right)^{2}$.  In $3$-$d$, this is a relevant coupling and thereby triggers an RG flow. We can then reach the Critical Boson CFT at an IR fixed point of this flow by taking $\lambda_{4}\rightarrow \infty$ and tuning the zero temperature IR mass of the fundamental bosons to zero. 
 
Let us now consider the two-point function of $J^{(0)}_{B}$ in the presence of the $\lambda_{4}$ coupling, which we will keep finite for the moment. First of all, it is easy to see that in the presence of this interaction, the fundamental boson propagator receives corrections from a chain of scalar loops that can end in a tadpole. As $\lambda_{4}\rightarrow \infty$, the only effect of this is to modify the thermal mass which is then determined by equation \ref{thermmasscb} above. Secondly, for computing $2$-point functions we need to include planar diagrams in addition to the type considered in Fig.\ref{2pointrbs0}, that involve a chain of intermediate scalar loops in between the two external operator insertions.  Consequently at large $N_{B}$ we get the following result \citep{Aharony:2012nh}, 
\begin{equation}
\label{j0critcboson}
\begin{split}
\left \langle J_{B}^{(0)}(-q) J_{B}^{(0)}(q)\right \rangle_{\beta, \lambda_{4}} &= \left \langle J_{B}^{(0)}(-q) J_{B}^{(0)}(q)\right\rangle_{\beta,\lambda_{4}=0}\sum_{n=0}^{\infty} \left(-\frac{\lambda_{4}}{N_{B}}\left\langle J_{B}^{(0)}(-q) J_{B}^{(0)}(q)\right\rangle_{\beta,\lambda_{4}=0}\right)^{n}\\
\end{split}
\end{equation}

Now it will be convenient to redefine the scalar singlet operator in the Critical Boson theory as $\tilde{J}_{B}^{(0)}= \lambda_{4} J_{B}^{(0)}$. Then using our result for the thermal $2$-point function in \ref{j0rbthermal2} and taking $\lambda_{4}\rightarrow \infty$ in \ref{j0critcboson} we obtain, 
\begin{equation}
\label{j0critcboson1}
\begin{split}
\left \langle \tilde{J}_{B}^{(0)}(q) \tilde{J}_{B}^{(0)}(-q)\right \rangle_{\beta} & = - 4 \pi N_{B} \lambda_{B}\hspace{0.03cm}  q_{3}\hspace{0.04cm}  \cot\left(\frac{q_{3} \beta}{2} \mathcal{F}_{B}(q_{3}\beta, \mu_{B,c})\right)
\end{split}
\end{equation}

In arriving at the above result, we replaced $\mu_{B}$ in \ref{j0rbthermal2} with $\mu_{B,c}$ to account for the change in the thermal mass due to the $\lambda_{4}$ coupling. It is also straightforward to check that \ref{j0critcboson1} agrees with the corresponding zero temperature result in \citep{Aharony:2012nh}.  Note that we could also have reached the same result above by considering a Legendre transform of the Regular Boson theory with respect to the operator $J^{(0)}_{B}$. 


\subsection{Regular Fermions}

We will now compute thermal $2$-point functions of the lowest dimension spin $s=0$ operator in the Regular Fermion theory in the large $N_{F}$ limit and to all orders in the t'Hooft coupling $\lambda_{F}$. As in the bosonic case we will need the thermal propagator of the elementary fermions $\psi^{i}$. This was obtained in \citep{Aharony:2012ns} and is given by 
\begin{equation}
\label{fermionprop}
\left \langle \psi ^{i}(p)\bar{\psi}_{j}(-q)\right \rangle = S(p )\delta^{i}_{j}(2\pi)^{3}\delta_{n_{p},n_{q}} \delta^{2}(\vec{p}-\vec{q})
\end{equation}

where 
\begin{equation}
\label{fermionprop1}
\begin{split}
& S(p)  = \frac{-i\tilde{p}_{\mu}\gamma^{\mu}+i g(y) p_{+}\gamma^{+}-f(y) p_{s}\mathbf{1}}{\tilde{p}^{2}+\beta^{-2}\mu_{F}^{2}}, \quad y=\beta p_{s}\\
\end{split}
\end{equation}

and
\begin{equation}
\label{fermionprop2}
\begin{split}
& \tilde{p}_{\mu}=p_{\mu}-\frac{a}{\beta}\delta_{\mu 3}, \quad p_{3}=\frac{2\pi n_{p}}{\beta} ,\quad q_{3}=\frac{2\pi n_{q}}{\beta}
\end{split}
\end{equation}

In the above $(n_{p},n_{q}) \in \mathbf{Z}+1/2$, since we impose anti-periodic boundary conditions along the thermal circle for the fermions. $a$ is the gauge field holonomy and the thermal mass has been denoted by $\beta^{-1}\mu_{F}$. The functions $f(y), g(y)$ and the dimensionless thermal mass $\mu_{F}$ are determined by the following equations
\begin{equation}
\label{propterms}
\begin{split}
& \pm\mu_{F}(\lambda_{F})= \lambda_{F}\mu_{F}+\frac{1}{\pi i}\left[\mathrm{Li}_{2}\left(-e^{-\mu_{F}-\pi i \lambda_{F}}\right)-\mathrm{c.c.}\right]\\
& y f(\lambda_{F},y) =-\lambda_{F}\sqrt{y^{2}+\mu_{F}^{2}}+\frac{1}{\pi i} \left[\mathrm{Li}_{2}\left(-e^{-\sqrt{y^{2}+\mu_{F}^{2}}+\pi i \lambda_{F}}\right)-\mathrm{c}.\mathrm{c}\right]\\
& y^{2} g(\lambda_{F},y)=y^{2} f^{2}(\lambda_{F},y)- \mu_{F}^{2}(\lambda_{F})
\end{split}
\end{equation}

Our convention for the gamma matrices is provided in section \ref{notconv} of the Appendix.

\subsubsection{Finite temperature vertex factor for $J^{(0)}_{F}$}

The general methodology for computing thermal correlators of gauge invariant operators in the fermionic theories will be the same as in the bosonic theories considered in the previous section.  The technical steps involved in our analysis will close follow that of \citep{GurAri:2012is} and \citep{GurAri:2016xff}. Thus let us first calculate the relevant vertex factors which we define here as follows 
\begin{equation}
\label{j0fvertex}
\left \langle J^{(0)}_{F}(-q)\psi_{i}(p)\bar{\psi}^{j}(-k)\right\rangle_{\beta} = V^{(0)}_{F}(q,k)\delta^{j}_{i} (2\pi)^{3} \delta_{n_{q}+n_{k},n_{p}} \delta^{2}(\vec{q}+\vec{k}-\vec{p})
\end{equation}

The Schwinger-Dyson equation satisfied by the above vertex factor is given by
\begin{equation}
\label{j0fsdyson}
V^{(0)}_{F}(q,k)\delta^{j}_{i}= \delta^{j}_{i}+\int \mathcal{D}^{3}\ell \hspace{0.04cm} \left[\mathcal{V}^{a,\mu}(k,\ell)S(\ell)V^{(0)}_{F}(q,\ell)S(\ell+q)\mathcal{V}^{a,\nu}(\ell+q,k+q)\right]^{j}_{i} \mathcal{G}_{\nu\mu}(k-\ell)
\end{equation}

where the first term in the R.H.S. above is the vertex factor in the free theory and $\mathcal{V}^{a,\mu}=T^{a}\gamma^{\mu}$, is the vertex factor due to the interaction term $\bar{\psi} \gamma^{\mu}A^{a}_{\mu}T^{a}\psi $ in the action \ref{rf}. Now using the relation for $U(N)$ generators in equation \ref{ungen} and the following identities 
\begin{equation}
\label{gammaidentity}
\begin{split}
 \gamma^{[3|}\gamma^{\mu}\gamma^{|+]}=-2\delta^{\mu}_{-}\mathbf{1}, \quad\quad\quad [\gamma^{3},\gamma^{+}]=2\gamma^{+}
\end{split}
\end{equation}

it is easy to show that for $q^{\pm}=0$, the integrand in equation \ref{j0fsdyson} takes the form
\begin{equation}
\label{j0fsdyson1}
\begin{split}
& \left[\mathcal{V}^{a,\mu}(k,\ell)S(\ell )V^{(0)}_{F}(q,\ell)S(\ell+q)\mathcal{V}^{a,\nu}(\ell+q,k+q)\right]^{j}_{ i} \mathcal{G}_{\nu\mu}(k-\ell) \\
&= -\frac{1}{2}\delta^{j}_{i}\frac{1}{((\tilde{\ell}+q)^{2}+\beta^{-2}\mu_{F}^{2})(\tilde{\ell}^{2}+\beta^{-2}\mu_{F}^{2})}\mathrm{Tr}_{N}\left[\gamma^{[3|}A\gamma^{|+]}\right] \mathcal{G}_{3+}(k-\ell)\\
\end{split}
\end{equation}

where,
\begin{equation}
\label{traceterm}
A=\left(-i\tilde{\ell}_{\mu}\gamma^{\mu}+ig(x)\ell_{+}\gamma^{+}-f(x)\ell_{s}\mathbf{1}\right)V^{(0)}_{F}(q,\ell)\left(-i(\tilde{\ell}+q)_{\mu}\gamma^{\mu}+i g(x) \ell_{+}\gamma^{+}-f(x)\ell_{s}\mathbf{1}\right)
\end{equation}

with $x=\beta \ell_{s}$. $f(x), g(x)$ were given in equation \ref{propterms} and we have defined 
\begin{equation}
\gamma^{[3+}A\gamma^{|+]} \equiv \gamma^{3}A\gamma^{+}-\gamma^{+}A\gamma^{3}
\end{equation}

Now any $2\times 2$ matrix $M$ can be expressed as $M=m_{\mu}\gamma^{\mu}+m_{I}\mathbf{1}$. Then using \ref{gammaidentity} it follows that
\begin{equation}
\label{gammarel}
\gamma^{[3|}M\gamma^{|+]}=2 m_{I}\gamma^{+}-2m_{-}\mathbf{1}
\end{equation}

This suggests that we can decompose the vertex factor $V^{(0)}_{F}(q,k)$ as
\begin{equation}
\label{j0fsdyson2}
V^{(0)}_{F}(q, k)=V^{(0)}_{F,+}(q, k)\gamma^{+}+V^{(0)}_{F,I}(q, k)\mathbf{1}
\end{equation}


Gathering the above results we find that the  Schwinger-Dyson equation \ref{j0fsdyson} becomes, 
\begin{equation}
\label{j0fsdyson3}
\begin{split}
V^{(0)}_{F}(q,k)\delta^{j}_{i}&=\delta^{j}_{i}-\frac{2\pi i \lambda_{F}}{\beta }\delta^{j}_{i}\int _{-1/2}^{1/2}du\int \frac{d^{2}\ell}{(2\pi)^{2}} \hspace{0.04cm}\frac{1}{(\ell-k)^{+}} \sum_{n=-\infty}^{\infty}\frac{\gamma^{[3|}A\gamma^{|+]}}{((\tilde{\ell}+q)^{2}+\beta^{-2}\mu_{F}^{2})(\tilde{\ell}^{2}+\beta^{-2}\mu_{F}^{2})}
\end{split}
\end{equation}

Then comparing the coefficients of $\gamma^{+}$ and $\mathbf{1}$ on both sides we get,
\begin{equation}
\label{j0fsdyson4}
\begin{split}
& V^{(0)}_{F,+}(q,k)  = -\frac{4\pi i \lambda}{\beta }\int _{-1/2}^{1/2}du\int \frac{d^{2}\ell}{(2\pi)^{2}} \hspace{0.04cm}\frac{1}{(\ell-k)^{+}} \sum_{n=-\infty}^{\infty}\frac{c_{1}V^{(0)}_{F,+}(q,\ell)+c_{2}V_{I}(q,\ell)}{((\tilde{\ell}+q)^{2}+\beta^{-2}\mu_{F}^{2})(\tilde{\ell}^{2}+\beta^{-2}\mu_{F}^{2})}\\
& V^{(0)}_{F,I}(q,k) =1-\frac{4\pi i \lambda}{\beta }\int _{-1/2}^{1/2}du\int \frac{d^{2}\ell}{(2\pi)^{2}} \hspace{0.04cm}\frac{1}{(\ell-k)^{+}} \sum_{n=-\infty}^{\infty}\frac{c_{3}V^{(0)}_{F,+}(q,\ell)+c_{4}V^{(0)}_{F,I}(q,\ell)}{((\tilde{\ell}+q)^{2}+\beta^{-2}\mu_{F}^{2})(\tilde{\ell}^{2}+\beta^{-2}\mu_{F}^{2})}
\end{split}
\end{equation}

where
\begin{equation}
\label{cicoeff}
\begin{split}
& c_{1}=-\ell^{+}\left(q_{3}-2 i \ell_{s} f(x)\right),\quad c_{2}=-\tilde{\ell}_{3}(\tilde{\ell}_{3}+q_{3})+\ell_{s}^{2}\left(f^{2}(x)+g(x)-1\right) \\
& c_{3}=2\hspace{0.04cm}(\ell^{+})^{2},\quad c_{4}= -\ell^{+}\left(q_{3}+2 i \ell_{s} f(x)\right)
\end{split}
\end{equation}

From the set of equations in  \ref{j0fsdyson4}  we observe that $V^{(0)}_{F,+}(q,k)$ and $V^{(0)}_{F,I}(q,k)$ are independent of $k_{3}$. Hence these factors do not affect the Matsubara sum in the above integrands. Then carrying out the holonomy integral and Matsubara sum we get,

\begin{equation}
\label{j0fsdyson5}
\begin{split}
V^{(0)}_{F,+}(q,k)  = - \frac{4 i \pi  \lambda_{F} }{\beta^{2}}\int \frac{d^{2}\ell}{(2\pi)^{2}} \hspace{0.04cm}\frac{1}{(\ell-k)^{+}}& \left[2 x^{2}(g(\beta\ell_{s})-1)V^{(0)}_{F,I}(q,\ell)-\beta\ell^{+}\left(q_{3}\beta -2\tilde{f}(\beta\ell_{s})\right) V^{(0)}_{F,+}(q,\ell)\right] \times\\
& \hspace{6.0cm}H_{F}(x,q_{3}\beta,\mu_{F})
\end{split}
\end{equation}

\begin{equation}
\label{j0fsdyson6}
\begin{split}
V^{(0)}_{F,I}(q,k)  =1- \frac{4 i \pi  \lambda_{F}}{\beta^{2}} \int \frac{d^{2}\ell}{(2\pi)^{2}} \hspace{0.04cm}\frac{1}{(\ell-k)^{+}}\hspace{0.04cm} & \left[2 \beta^{2}(\ell^{+})^{2}V^{(0)}_{F,+}(q,\ell)-\beta \ell^{+} (q_{3}\beta+2\tilde{f}(\beta\ell_{s}))V^{(0)}_{F,I}(q,\ell)\right]\times \\
& \hspace{6.0cm} H_{F}(x,q_{3}\beta, \mu_{F})
\end{split}
\end{equation}

where we have defined
\begin{equation}
\label{j0fsdyson7}
\begin{split}
H_{F}(x,q_{3}\beta, \mu_{F}) & = \frac{1}{\beta} \int_{-1/2} ^{1/2} du \sum_{n=-\infty}^{\infty}\frac{1}{((\tilde{\ell}+q)^{2}+\beta^{-2}\mu_{F}^{2})(\tilde{\ell}^{2}+\beta^{-2}\mu_{F}^{2})} \\
& =\frac{i \hspace{0.03cm}\beta^{3}}{\pi |\lambda_{F}|}  \frac{\left[\log\left(\cosh\left(\frac{1}{2}\sqrt{x^{2}+\mu_{F}^{2}}-\frac{i \pi |\lambda_{F}|}{2}\right)\right)-\mathrm{c}.\mathrm{c}\right]}{\sqrt{x^{2}+\mu_{F}^{2}}\left(4 x^{2}+4 \mu_{F}^{2}+ q_{3}^{2}\beta^{2}\right)}
\end{split}
\end{equation}

and we have also used
\begin{equation}
\label{j0fsdyson8}
\begin{split}
 \frac{1}{\beta} \int_{-1/2} ^{1/2} du \sum_{n=-\infty}^{\infty}\frac{\tilde{\ell}_{3}(\tilde{\ell}_{3}+q_{3})}{((\tilde{\ell}+q)^{2}+\beta^{-2}\mu_{F}^{2})(\tilde{\ell}^{2}+\beta^{-2}\mu_{F}^{2})} & = \frac{1}{\beta^{2}}(x^{2}+\mu_{F}^{2})H_{F}(x,q_{3}\beta, \mu_{F})
\end{split}
\end{equation}

Now to solve the equations \ref{j0fsdyson5} and \ref{j0fsdyson6} let us consider the ansatz,
\begin{equation}
\label{newvpvi}
\begin{split}
 V^{(0)}_{F,+} (q,k) = \frac{2k^{-}}{k_{s}}v_{+}(q,y), \quad\quad V^{(0)}_{F,I}(q,k)= v_{I}(q,y)
\end{split}
\end{equation}

where $y=\beta k_{s}$. Then using the results in the appendix \ref{angintegs}, the relevant angular integrals in \ref{j0fsdyson5} and \ref{j0fsdyson6} turn out to be 
\begin{equation}
\label{j0fsdyson9}
\begin{split}
& \int_{0}^{2\pi} d\theta \hspace{0.1cm} \frac{1}{(\ell-k)^{+}} = - \frac{2\pi}{k^{+}} \Theta(k_{s}- \ell_{s})\\
& \int_{0}^{2\pi} d\theta \hspace{0.1cm} \frac{\ell^{+}}{(\ell-k)^{+}} = 2\pi \Theta(\ell_{s}-k_{s})\\
& \int_{0}^{2\pi} d\theta \hspace{0.1cm} \frac{(\ell^{+})^{2}}{(\ell-k)^{+}} = 2\pi k^{+}\Theta(k_{s}-\ell_{s})
\end{split}
\end{equation}

Thereby, the Schwinger-Dyson equations take the form
\begin{equation}
\label{j0fsdyson10}
\begin{split}
y\hspace{0.04cm} v_{+}(q,y)  =  \frac{2 i \lambda_{F}}{\beta^{3}} \int_{0}^{y} dx \hspace{0.1cm} x^{2} \hspace{0.04cm}  \left[2 x (g(x)-1)v_{I}(q,x)- \left(q_{3}\beta -2\tilde{f}(x)\right) v_{+}(q,x)\right] H_{F}(x,q_{3}\beta,\mu_{F})
\end{split}
\end{equation}

\begin{equation}
\label{j0fsdyson11}
\begin{split}
v_{I}(q,y)  =1-  \frac{2 i \lambda_{F}}{\beta^{3}} \int_{y}^{\infty} dx \hspace{0.1cm} x \hspace{0.04cm}  \left[2\hspace{0.03cm} y \hspace{0.04cm} v_{+}(q,x)-  \left(q_{3}\beta +2\tilde{f}(x)\right) v_{I}(q,x)\right] H_{F}(x,q_{3}\beta,\mu_{F})
\end{split}
\end{equation}

where $\tilde{f}(x)= i x f(x)$. Next we differentiate equations \ref{j0fsdyson10} and \ref{j0fsdyson11} with respect to $y$ to get
\begin{equation}
\label{j0fsdyson12}
\begin{split}
\partial_{y} \left(y\hspace{0.04cm} v_{+}(q,y)  \right)=  \frac{2 i \lambda_{F}}{\beta^{3}}\hspace{0.04cm}   y^{2} \left[2 y (g(y)-1)v_{I}(q,y)- \left(q_{3}\beta -2\tilde{f}(y)\right) v_{+}(q,y)\right] H_{F}(y,q_{3}\beta,\mu_{F})
\end{split}
\end{equation}

\begin{equation}
\label{j0fsdyson13}
\begin{split}
\partial_{y} v_{I}(q,y) = \frac{2 i \lambda_{F}}{\beta^{3}} \hspace{0.05cm} y \hspace{0.04cm}  \left[2\hspace{0.03cm} y \hspace{0.04cm} v_{+}(q, y)-  \left(q_{3}\beta +2\tilde{f}( y )\right) v_{I}(q, y)\right] H_{F}(y,q_{3}\beta,\mu_{F})
\end{split}
\end{equation}

Then multiplying equation \ref{j0fsdyson13} by $(q_{3}\beta-2 \tilde{f}(y))/2$ and adding it to equation \ref{j0fsdyson12} we arrive at the following
\begin{equation}
\label{j0fsdyson14}
\begin{split}
\partial_{y}\left[ y\hspace{0.04cm} v_{+}(q,y) + \frac{(q_{3}\beta-2 \tilde{f}(y))}{2}v_{I}(q,y)\right] = & \partial_{y} \left(\frac{q_{3}\beta-2 \tilde{f}(y)}{2} \right) v_{I}(q,y)\\
&  - \frac{ i \lambda_{F}}{\beta^{3}} y \left( 4 y^{2}+4 \mu_{F}^{2}+q_{3}^{2}\beta^{2}\right) v_{I}(q,y) H_{F}(y,q_{3}\beta,\mu_{F})
\end{split}
\end{equation}

Now note that due the following relation which can be derived from the definitions in \ref{propterms} and \ref{j0fsdyson7}, the R.H.S. of the above equation equals zero. 
\begin{equation}
\label{j0fsdyson15}
\begin{split}
\partial_{y}  \tilde{f}(y) = -\frac{ i \lambda_{F}}{\beta^{3}} y \left( 4 y^{2}+4 \mu_{F}^{2}+q_{3}^{2}\beta^{2}\right) H_{F}(y,q_{3}\beta,\mu_{F})
\end{split}
\end{equation}

Thus we get, 
\begin{equation}
\label{j0fsdyson16}
\begin{split}
& \partial_{y}\left[ y\hspace{0.04cm} v_{+}(q,y) + \frac{(q_{3}\beta-2 \tilde{f}(y))}{2}v_{I}(q,y)\right]=0\\
& \implies  y\hspace{0.04cm} v_{+}(q,y) = \chi(q_{3})- \frac{(q_{3}\beta-2 \tilde{f}(y))}{2}v_{I}(q,y)
\end{split}
\end{equation}

where $\chi(q_{3})$ is an integration constant. It can be determined from equations \ref{j0fsdyson10} and \ref{j0fsdyson11} using either $y=\infty$ or $y=0$. Since it is not very important at this moment, let us postpone the explicit evaluation of this constant to the next subsection when we compute the $2$-point function.  Now using \ref{j0fsdyson16} we can easily integrate \ref{j0fsdyson13} to obtain
\begin{equation}
\label{j0fsdyson17}
\begin{split}
v_{I}(q,y)=\frac{1}{q_{3}\beta}\Big(\chi(q_{3})\big(1-\exp\left[ i q_{3}\beta\hspace{0.03cm} \mathcal{F}_{F}(y,q_{3}\beta,\mu_{F})\right]\big)+q_{3}\beta \hspace{0.04cm} \exp\left[i q_{3}\beta\hspace{0.03cm} \mathcal{F}_{F}(y,q_{3}\beta,\mu_{F})\right]\Big)
\end{split}
\end{equation}

where 
\begin{equation}
\label{j0fsdyson18}
\begin{split}
\mathcal{F}_{F}(y,q_{3}\beta,\mu_{F})=\int_{y}^{\infty}dx \hspace{0.04cm}x\hspace{0.04cm} \mathcal{H}_{F}(x,q_{3}\beta,\mu_{F}) =  \frac{4\lambda_{F}}{\beta^{3}} \int_{y}^{\infty}dx \hspace{0.04cm}x\hspace{0.04cm} H_{F}(x,q_{3}\beta,\mu_{F})
\end{split}
\end{equation}

Thus the final result of this section is that the finite temperature vertex factor for $J^{(0)}_{F}$ at large $N_{F}$ and to all orders in $\lambda_{F}$ is 
\begin{equation}
\label{j0fsdyson19}
\begin{split}
 V^{(0)}_{F} (q,k) = \frac{2k^{-}}{k_{s}}v_{+}(q,y) \gamma^{+} + v_{I}(q,y)\mathbf{1}
\end{split}
\end{equation}

where $v_{I}(q,y)$ is now given by equation \ref{j0fsdyson17} and $v_{+}(q,y) $ is determined via \ref{j0fsdyson16}. 


\subsubsection{Thermal $2$-point function}
\label{rf2pts0}

The thermal $2$-point function in the Regular Fermion theory can now be obtained following the same procedure employed for the Regular Boson theory in section \ref{rb2pts0}. Considering the same kind of Feynman diagram as in Figure \ref{2pointrbs0} we take the left and right vertex insertions there to now correspond to the exact vertex factor $V^{(0)}_{F}(q,k)$ and the free theory vertex factor $V^{(0)}_{F,free}(q,k)$ respectively. Then joining these with a pair of exact thermal propgators of the elementary fermions $\psi^{i}$ we get, 
\begin{equation}
\label{j0thermalrf1}
\begin{split}
& \left\langle J^{(0)}_{F}(-q) J^{(0)}_{F}(q)\right\rangle_{\beta}  =-\sum_{i=1}^{N_{F}}\int \frac{d^{2}\ell}{(2\pi)^{2}}\hspace{0.04cm} \frac{1}{\beta} \sum_{n=-\infty}^{\infty}\mathrm{Tr}\left[V^{(0)}_{F}(q,\ell)S_{i}(\ell)V^{(0)}_{F,free}(-q,\ell+q)S_{i}(\ell+q)\right]
\end{split}
\end{equation}

Here the trace inside the integral above is with respect to the spinor indices carried by the gamma matrices contained in the vertex factors and the propagators. The overall minus sign is due to the fermion loop. Then noting that $V^{(0)}_{F,free}(q,k)=1$ and evaluating  the trace explicitly gives
\begin{equation}
\label{j0thermalrf2}
\begin{split}
& \left\langle J^{(0)}_{F}(-q) J^{(0)}_{F}(q)\right\rangle_{\beta}  =-\frac{N_{F}}{\beta}\int_{-1/2}^{1/2} du \int \frac{d^{2}\ell}{(2\pi)^{2}} \sum_{n=-\infty}^{\infty}\frac{2c_{1} V^{(0)}_{F,+}(q,\ell)+2c_{2} V^{(0)}_{F,I}(q,\ell)}{\left((\tilde{\ell}+q)^{2}+\beta^{-2}\mu_{F}^{2}\right)\left(\tilde{\ell}^{2}+\beta^{-2}\mu_{F}^{2}\right)}
\end{split}
\end{equation}  

where $c_{1},c_{2}$ were defined in equation \ref{cicoeff}. Now the angular part of the above integral is again trivial. Then carrying out the Matsubara sum and holonomy integral we find
\begin{equation}
\label{j0thermalrf3}
\begin{split}
& \left\langle J^{(0)}_{F}(-q) J^{(0)}_{F}(q)\right\rangle_{\beta} \\
& =\frac{N_{F}}{4 \pi \lambda_{F} \beta }\int_{0}^{\infty} d x \hspace{0.1cm} x^{2} \left[ (q_{3}\beta-2\tilde{f}(x))v_{+}(q,x)- 2 x (g(x)-1)v_{I}(q,x)\right] \mathcal{H}_{F}(x,q_{3}\beta,\mu_{F})\\
\end{split}
\end{equation}

The above integral turns out to be ultraviolet divergent. We will introduce a UV cutoff $\Lambda$ to regulate this divergence\footnote{The presence of a hard UV cutoff breaks gauge invariance as well as conformal invariance. We can alternatively use dimensional regularisation which will preserve gauge invariance. See \citep{Choudhury:2018iwf} and  \citep{Jain:2012qi} for details regarding implementing dimensional regularisation scheme in Chern Simons matter theories.}. Then using equation \ref{j0fsdyson12} in \ref{j0thermalrf3} we get
\begin{equation}
\label{j0thermalrf3simp}
\begin{split}
& \left\langle J^{(0)}_{F}(-q) J^{(0)}_{F}(q)\right\rangle_{\beta}  =-\frac{i N_{F}}{2 \pi\lambda_{F}\beta}\int_{0}^{\beta\Lambda} d x \hspace{0.1cm} \partial_{x} \left(x \hspace{0.04cm} v_{+}(q,x)\right) = -\frac{i N_{F}}{2 \pi\lambda_{F}\beta}\left( \beta\Lambda \hspace{0.03cm} v_{+}(q,\beta\Lambda)\right)
\end{split}
\end{equation}

Now putting $y=\beta\Lambda$ in equation \ref{j0fsdyson10} and using \ref{j0fsdyson16} and  \ref{j0fsdyson17} we get
\begin{equation}
\label{j0thermalrf4}
\begin{split}
\beta\Lambda \hspace{0.03cm}v_{+}(q,\beta\Lambda)&= \mathcal{I}_{1}+ \chi(q_{3}) (\mathcal{I}_{2}+\mathcal{I}_{3})
\end{split}
\end{equation}

where $\mathcal{I}_{1}, \mathcal{I}_{2}, \mathcal{I}_{3}$ are given by the following integrals
\begin{equation}
\label{j0thermalrf5}
\begin{split}
& \mathcal{I}_{1}= \frac{i}{4 } \int_{0}^{\beta\Lambda} dx \hspace{0.1cm} x \left(q_{3}^{2}\beta^{2}-4 q_{3}\beta \tilde{f}(x)-4\mu_{F}^{2}-4 x^{2}\right)e^{i q_{3}\beta\hspace{0.03cm} \mathcal{F}_{F}(x,q_{3}\beta,\mu_{F})}\mathcal{H}_{F}(x,q_{3}\beta,\mu_{F})\\
& \mathcal{I}_{2}=  -\frac{1}{q_{3}\beta} \hspace{0.03cm} \mathcal{I}_{1}, \quad  \mathcal{I}_{3}=   -\frac{ i }{4 q_{3}\beta} \int_{0}^{\beta\Lambda} dx \hspace{0.1cm}x \left(4x^{2}+4\mu_{F}^{2}+q^{2}_{3}\beta^{2}\right) \mathcal{H}_{F}(x,q_{3}\beta,\mu_{F})
\end{split}
\end{equation}

The above integrals can be evaluated using the following useful relations which can be derived using equations \ref{j0fsdyson15} and \ref{j0fsdyson18},
\begin{equation}
\label{j0thermalrf6}
\begin{split}
& \partial_{x} \left( e^{i q_{3}\beta\hspace{0.03cm} \mathcal{F}_{F}(x,q_{3}\beta,\mu_{F})} \right) = - i q_{3}\beta \hspace{0.03cm} x \hspace{0.03cm} \mathcal{H}_{F}(x,q_{3}\beta,\mu_{F}) e^{i q_{3}\beta\hspace{0.03cm} \mathcal{F}_{F}(x,q_{3}\beta,\mu_{F})}\\
& e^{i q_{3}\beta\hspace{0.03cm} \mathcal{F}_{F}(x,q_{3}\beta,\mu_{F})}\partial_{x} \tilde{f}(x)= \frac{1}{4 q_{3}\beta}\left(4x^{2}+4\mu_{F}^{2}+q_{3}^{2}\beta^{2}\right)\partial_{x}\left( e^{i q_{3}\beta\hspace{0.03cm} \mathcal{F}_{F}(x,q_{3}\beta,\mu_{F})}\right)
\end{split}
\end{equation}

Now using the above relations in the integrands appearing in equation \ref{j0thermalrf5} and integrating by parts we find that $\mathcal{I}_{1}$ simply reduces to boundary terms,
\begin{equation}
\label{j0thermalrf7}
\begin{split}
 \mathcal{I}_{1}&=-\frac{1}{2 } \left(\left(q_{3}\beta-2  \tilde{f}(x)\right) e^{i q_{3}\beta\hspace{0.03cm} \mathcal{F}_{F}(x,q_{3}\beta,\mu_{F})}\right)_{0}^{\beta\Lambda}\\
&=  -\frac{q_{3}\beta}{2 }\left(1-e^{i q_{3}\beta\hspace{0.03cm} \mathcal{F}_{F}(q_{3}\beta,\mu_{F})}\right) -\tilde{f}(0)e^{i q_{3}\beta\hspace{0.03cm} \mathcal{F}_{F}(q_{3}\beta,\mu_{F})}+  \tilde{f}(\beta\Lambda)
\end{split}
\end{equation}

where have used $ \mathcal{F}_{F}(x,q_{3}\beta,\mu_{F})\big|_{x\rightarrow \infty} =0$ and defined $ \mathcal{F}_{F}(x=0,q_{3}\beta,\mu_{F})\equiv \mathcal{F}_{F}(q_{3}\beta,\mu_{F})$. Similarly, using \ref{j0fsdyson14}, the integral $\mathcal{I}_{3}$ can be simplified to yield
\begin{equation}
\label{j0thermalrf8}
\begin{split}
\mathcal{I}_{3}&= \frac{1}{q_{3}\beta}\left(\tilde{f}(\beta\Lambda)-\tilde{f}(0)\right)
\end{split}
\end{equation}

Now to obtain the required $2$-point function we need to determine the constant $\chi(q_{3})$.  Form the Schwinger-Dyson equation \ref{j0fsdyson11} we see that $v_{I}(q, \beta\Lambda)\big|_{\Lambda\rightarrow \infty} =0$. Consequently it follows from \ref{j0fsdyson14} that, 
\begin{equation}
\label{j0thermalrf9}
\begin{split}
\beta\Lambda \hspace{0.03cm} v_{+}(q,\beta\Lambda) = \chi(q_{3})
\end{split}
\end{equation}

Then using \ref{j0thermalrf4} and the results in equations \ref{j0thermalrf7}, \ref{j0thermalrf8} we can solve for $\chi(q_{3})$ to get,
\begin{equation}
\label{j0thermalrf10}
\begin{split}
\chi(q_{3})&= \frac{q_{3}\beta \left( q_{3}\beta-2\tilde{f}(0)\right)e^{i q_{3}\beta\hspace{0.03cm} \mathcal{F}_{F}(q_{3}\beta,\mu_{F})}}{q_{3}\beta \left(1+e^{i q_{3}\beta\hspace{0.03cm} \mathcal{F}_{F}(q_{3}\beta,\mu_{F})}\right)+2 \tilde{f}(0)\left(1-e^{i q_{3}\beta\hspace{0.03cm} \mathcal{F}_{F}(q_{3}\beta,\mu_{F})}\right)}
\end{split}
\end{equation}

We still have a divergent contribution coming from $\tilde{f}(\beta\Lambda)$. This is a linear divergence since from equation \ref{propterms}, we have $\tilde{f}(\beta\Lambda) \sim - i \lambda_{F} \beta\Lambda$ for large $\Lambda$. This divergence can be removed by adding a mass counterterm for the background source for the operator $J^{(0)}_{F}$. The same prescription was used in the zero temperature calculations in \citep{GurAri:2012is}.

Gathering the above results we finally end up with the renormalised finite temperature $2$-point function which is given by, 
\begin{equation}
\label{j0thermalrf11} 
\begin{split}
 \left\langle J^{(0)}_{F}(-q) J^{(0)}_{F}(q)\right\rangle_{\beta} & = \frac{ i \hspace{0.03cm}N_{F} q_{3}}{4 \pi \lambda_{F}} \left( \frac{q_{3}\beta\left(1-e^{i q_{3}\beta\hspace{0.03cm} \mathcal{F}_{F}(q_{3}\beta,\mu_{F})}\right)+2\tilde{f}(0)\left(1+e^{i q_{3}\beta\hspace{0.03cm} \mathcal{F}_{F}(q_{3}\beta,\mu_{F})}\right)}{q_{3}\beta\left(1+e^{i q_{3}\beta\hspace{0.03cm} \mathcal{F}_{F}(q_{3}\beta,\mu_{F})}\right)+2\tilde{f}(0)\left(1-e^{i q_{3}\beta\hspace{0.03cm} \mathcal{F}_{F}(q_{3}\beta,\mu_{F})}\right)}\right)
\end{split}
\end{equation}

where
\begin{equation}
\label{j0thermalrf12} 
\begin{split}
\tilde{f}(0)=-i \lambda_{F} \mu_{F}+\frac{1}{\pi } \left[\mathrm{Li}_{2}\left(-e^{-\mu_{F}+\pi i \lambda_{F}}\right)-\mathrm{c}.\mathrm{c}\right]
\end{split}
\end{equation}

Using the first equation in \ref{propterms}, we get from the above, $\tilde{f}(0)=- i \hspace{0.03cm} \mathrm{sgn}(\lambda_{F})\mu_{F}$.

\subsubsection{Zero temperature limit}

In the low temperature regime, i.e., for $\beta\rightarrow \infty$ we find that 
\begin{equation}
\label{t0limrf1}
\begin{split}
\lim_{\beta\rightarrow \infty} \mathcal{F}_{F}(q_{3}\beta,\mu_{F}) = \frac{ \pi}{q_{3}\beta} \lambda_{F}\mathrm{sgn}(q_{3}) 
\end{split}
\end{equation}

Using this in the expression for the $2$-point function in equation \ref{j0thermalrf11} and neglecting the contribution of the thermal mass we get
\begin{equation}
\label{t0limrf2}
\begin{split}
 \lim_{\beta\rightarrow \infty} \left \langle J^{(0)}_{F}(-q)J^{(0)}_{F}(q)\right \rangle_{\beta} =- \frac{ N _{F}}{4 \pi \lambda_{F} q_{3}} \tan \left( \frac{\pi \lambda_{F} \hspace{0.03cm} \mathrm{sgn}(q_{3})}{2}\right)
\end{split}
\end{equation}

This perfectly matches with the zero temperature $2$-point function of $J^{(0)}_{F}$ computed in \citep{GurAri:2012is}.

\subsection{Critical Fermions}
\label{s0cf}

We will now deal with the operator $\tilde{J}^{(0)}_{F}$ which is the lowest dimension gauge invariant scalar operator in the Critical Fermion theory. This is essentially the operator corresponding to the Hubbard-Stratonovich field $\sigma_{F}$ in the action \ref{cf}. The scaling dimension here is $\Delta=1+\mathcal{O}(1/N_{F})$ and to leading order in large $N_{F}$ there is no anomalous dimension. 

The thermal $2$-point function of $\tilde{J}^{(0)}_{F}$ in the Critical Fermion theory can be obtained by following a strategy similar to the one implemented in the previous section for correlators in the Critical Boson theory. In this case, consider first deforming the Regular Fermion theory by turning on the quartic interaction $\frac{\lambda_{4}}{2 N_{F}} \left( \psi \bar{\psi} \right)^{2}$. The Critical Fermion theory lies at a UV fixed point of the RG flow triggered by this coupling, and can be reached by taking the limit $\lambda_{4}\rightarrow \infty$. At large $N_{F}$, we are also allowed to include the marginal interaction $x^{F}_{6}\left( \psi \bar{\psi} \right)^{3}$ at this fixed point. 

Now one of the effects of turning on the $\lambda_{4}$ and $x^{F}_{6}$ couplings is to modify the thermal mass of the fundamental fermions at the UV fixed point. This was shown in \citep{Aharony:2012ns} to be given by 
\begin{equation}
\label{thermmasscf} 
\begin{split}
\left(1-|\lambda_{F}|+ |\hat{g}_{F}|\right) \mu_{F,c} = \frac{1}{\pi i} \left[\mathrm{Li}_{2}\left(-e^{-\mu_{F,c}- i \pi |\lambda_{F}|} \right)-\mathrm{c}.\mathrm{c}\right]
\end{split}
\end{equation}

where
\begin{equation}
\label{thermmasscf1} 
\begin{split}
\hat{g}_{F} =  \left( 1- 2\pi \lambda_{F} x^{F}_{6} \right)^{-1/2}
\end{split}
\end{equation}

The thermal propagators for the fundamental bosons then take the same form as in equations \ref{fermionprop}, \ref{fermionprop1} with $\mu_{F}$ there appropriately replaced by $\mu_{F,c}$ which is determined by equation \ref{thermmasscf} above.  Also note that $\tilde{f}(x)$ in \ref{propterms} now becomes
\begin{equation}
\label{thermmasscf2} 
\begin{split}
y f(\lambda_{F},y) = \mathrm{sgn}(\lambda_{F}) \mu_{F,c} \hspace{0.03cm} \hat{g}_{F} - \lambda_{F}\sqrt{y^{2}+\mu_{F,c}^{2}}+\frac{1}{\pi i} \left[\mathrm{Li}_{2}\left(-e^{-\sqrt{y^{2}+\mu_{F,c}^{2}}+\pi i \lambda_{F}}\right)-\mathrm{c}.\mathrm{c}\right]\\
\end{split}
\end{equation}

Then adopting the arguments used in the case of the Critical Boson theory, we can easily derive the thermal $2$-point function which turns out to be given by, 
\begin{equation}
\label{j0thermalcf1} 
\begin{split}
 \left\langle \tilde{J}^{(0)}_{F}(-q) \tilde{J}^{(0)}_{F}(q)\right\rangle_{\beta} & = \frac{ 4 i  \pi N_{F} \lambda_{F}}{q_{3}} \left( \frac{q_{3}\beta\left(1+e^{i q_{3}\beta\hspace{0.03cm} \mathcal{F}_{F}(q_{3}\beta,\mu_{F,c})}\right)+2\tilde{f}(0)\left(1-e^{i q_{3}\beta\hspace{0.03cm} \mathcal{F}_{F}(q_{3}\beta,\mu_{F,c})}\right)}{q_{3}\beta\left(1-e^{i q_{3}\beta\hspace{0.03cm} \mathcal{F}_{F}(q_{3}\beta,\mu_{F,c})}\right)+2\tilde{f}(0)\left(1+e^{i q_{3}\beta\hspace{0.03cm} \mathcal{F}_{F}(q_{3}\beta,\mu_{F,c})}\right)}\right)
\end{split}
\end{equation}

where now $\tilde{f}(0)= - i \hspace{0.04cm} \mathrm{sgn}(\lambda_{F}) \mu_{F,c} $ The above result could also have been arrived at by Legendre transforming the Regular Fermion theory with respect to the operator $J^{(0)}_{F}$. We have also checked that the zero temperature limit of the above result agrees with \citep{GurAri:2012is}. 
\subsection{Duality Check}

Let us now check if the results derived above are consistent with the bosonization dualities which the theories considered here are conjectured to exhibit. The relevant mapping of parameters was presented in \ref{dualmap}. Let us note it here again for the sake of completeness. 
\begin{equation}
\label{dualitymap1}
\begin{split}
& N_{F}= |\kappa_{B}|-N_{B}, \quad \kappa_{F}=-\kappa_{B}, \quad |\lambda_{F}|=1- |\lambda_{B}, \quad x_{6}^{B}=8\pi^{2}\left(1-|\lambda_{F}|\right)^{2} \left(3- 8 \pi \lambda_{F}x_{6}^{F} \right); \\
&  \mu_{F}(\lambda_{F}) = \mu_{B,c}(\lambda_{B}), \quad \mu_{F,c}(\lambda_{F}) = \mu_{B}(\lambda_{B})
\end{split}
\end{equation}

\subsubsection*{Regular Fermions and Critical Bosons}

Let us first consider the Regular Fermion and Critical Boson theories. Now the functions $ \mathcal{F}_{F}(q_{3}\beta,\mu_{F})$ and $ \mathcal{F}_{B}(q_{3}\beta,\mu_{B,c})$ appearing in the expressions for the correlators in \ref{j0thermalrf11} and \ref{j0rbthermal2} turn out to be related under the duality map in \ref{dualitymap1} as follows
\begin{equation}
\label{dualtransf1}
\begin{split}
&  \exp \big[ i q_{3}\beta \mathcal{F}_{F}\left(q_{3}\beta,\mu_{F} \right)\big] =  -  \exp\big[ i  q_{3}\beta \mathcal{F}_{B}\left(q_{3}\beta,\mu_{B,c} \right)\big] \left(\frac{q_{3}\beta + 2 \hspace{0.03cm} i \hspace{0.04cm}\mathrm{sgn}(\lambda_{B})\mu_{B,c}}{q_{3}\beta - 2\hspace{0.03cm} i \hspace{0.04cm}\mathrm{sgn}(\lambda_{B})\mu_{B,c}}\right)\\
\end{split}
\end{equation}

Under the duality map above, we also have
\begin{equation}
\label{dualtransf2}
\begin{split}
&  \tilde{f}(0) = -i (\lambda_{B} - \mathrm{sgn}(\lambda_{B}))\mu_{B,c}+\frac{1}{\pi } \left[\mathrm{Li}_{2}\left(e^{-\mu_{B,c}+ i \pi \lambda_{B,c}}\right)-\mathrm{c}.\mathrm{c}\right] = i \hspace{0.03cm} \mathrm{sgn}(\lambda_{B}) \mu_{B,c}
\end{split}
\end{equation}

where we have used the defining equation for the thermal mass in the Critical Boson theory given by \ref{thermmasscb}. Then using \ref{dualitymap1}, \ref{dualtransf1} and \ref{dualtransf2} we find that the thermal $2$-point function of $J^{(0)}_{F}$ in equation \ref{j0thermalrf11} becomes
\begin{equation}
\label{dualcheckrf} 
\begin{split}
 \left\langle J^{(0)}_{F}(-q) J^{(0)}_{F}(q)\right\rangle_{\beta} 
 & = -\frac{N_{B}}{4\pi \lambda_{B}}q_{3} \cot\left (\frac{q_{3} \beta}{2}\mathcal{F}_{B}\left(q_{3}\beta, \mu_{B,c}\right)\right) 
\end{split}
\end{equation}

Comparing \ref{dualcheckrf} and \ref{j0critcboson1} we see that they match provided we redefine the scalar operator $\tilde{J}^{(0)}_{B}$ in \ref{j0critcboson1} as $\tilde{J}^{(0)}_{B} \rightarrow (-4\pi\lambda_{B}) \tilde{J}^{(0)}_{B}$. 

\subsubsection*{Regular Bosons and Critical Fermions}

For the Regular Boson and Critical Fermion theories, the relation between $\mathcal{F}_{B}(q_{3}\beta,\mu_{B})$ and $ \mathcal{F}_{F}(q_{3}\beta,\mu_{F,c})$ is exactly of the same form as in \ref{dualtransf1} under the duality map given in \ref{dualitymap1}. Then implementing the duality map on the thermal $2$-point function of $J^{(0)}_{B}$ in the Regular Boson theory we find
\begin{equation}
\label{dualcheckrf1} 
\begin{split}
 \left\langle J^{(0)}_{B}(-q) J^{(0)}_{B}(q)\right\rangle_{\beta}& = - \frac{  i N_{F} }{4 \pi \lambda_{F} q_{3}} \left( \frac{q_{3}\beta\left(1+e^{i q_{3}\beta\hspace{0.03cm} \mathcal{F}_{F}(q_{3}\beta,\mu_{F,c})}\right)- 2i \hspace{0.03cm} \mathrm{sgn}(\lambda_{F})\mu_{F,c}\left(1-e^{i q_{3}\beta\hspace{0.03cm} \mathcal{F}_{F}(q_{3}\beta,\mu_{F,c})}\right)}{q_{3}\beta\left(1-e^{i q_{3}\beta\hspace{0.03cm} \mathcal{F}_{F}(q_{3}\beta,\mu_{F,c})}\right)-2i \hspace{0.03cm} \mathrm{sgn}(\lambda_{B})\mu_{F,c}\left(1+e^{i q_{3}\beta\hspace{0.03cm} \mathcal{F}_{F}(q_{3}\beta,\mu_{F,c})}\right)}\right)
\end{split}
\end{equation}

Now comparing the above result with \ref{j0thermalcf1} we observe that they match if we redefine the operator $\tilde{J}^{(0)}_{F}$ in the Critical Fermion theory as $\tilde{J}^{(0)}_{F} \rightarrow (-4\pi\lambda_{F}) \tilde{J}^{(0)}_{B}$. 

Accounting for the required changes in the overall normalisations of the operators, we have thus succeeded in establishing a new explicit check of $3$-$d$ bosonization dualities at finite temperature. 


\section{Thermal $2$-point Functions: Spin $s=1$} 
\label{u1}

In this section we will consider the thermal $2$-point function of the $U(1)$ current operator in the Regular Boson and Critical Boson theories. For critical bosons, this $2$-point function has been calculated before in \citep{GurAri:2016xff} perturbatively in the t'Hooft coupling. Our calculation here will however be exact in the t'Hooft coupling. Subsequently in section \ref{u1rff} we will consider the case of Regular and Critical Fermion theories. 

\subsection{Regular Bosons}
\label{u1boson}

The $U(1)$ current operator in the Regular Boson theory is given by
\begin{equation}
\label{u1current}
\begin{split}
J_{B,\mu}^{(1)}(x)= i \phi^{\dagger}(x) \left(\overset{\leftarrow}{D_\mu}-\overset{\rightarrow}{D_\mu}\right) \phi(x)
\end{split}
\end{equation}

where $\overset{\leftarrow}{D_\mu} =\overset{\leftarrow}\partial_{\mu}-A_{\mu}$, $\overset{\rightarrow}{D_\mu} =\overset{\rightarrow}\partial_{\mu}+A_{\mu}$ and $\mu \in (-,+,3)$. Now in computing the momentum space $2$-point functions $\left\langle J_{B,\mu}^{(1)} \hspace{0.05cm} J_{B,\nu}^{(1)}\right\rangle_{\beta}$, we will only consider the case where $(\mu,\nu)=(-,+)$. This is because the correlators involving the components $(\mu,\nu)=(-,-), (-,3), (+,3)$ and $(+,+)$ all vanish due to rotational symmetry in the spatial directions. Further as in the previous sections, we will be working in the kinematic regime where the external momentum $q$ through the current is such that $q^{\pm}=0$. Then its easy to see from the Ward identities for current conservation that the $2$-point function with $(\mu,\nu)=(3,3)$ also vanishes in momentum space upto contact terms. 

In light cone gauge where $A_{-}=0$, the components of the current of interest here are then given by 

\begin{equation}
\label{u1minus}
J_{B,-}^{(1)}(x)= i \left (\partial_{-} \phi (x)\right)^{\dagger} \phi(x)- i \hspace{0.03cm}\phi^{\dagger}(x) \left(\partial_{-}\phi(x) \right)
\end{equation}
\begin{equation}
\label{u1plus}
 J_{B,+}^{(1)}(x)= i \left(\partial_{+} \phi(x)\right)^{\dagger}\phi(x)- i \phi^{\dagger}(x)\left( \partial_{+}\phi(x) \right) -2 i \phi^{\dagger} (x)A_{+} (x)\phi (x)
\end{equation}

Let us also note down the momentum space expressions of the currents in equations \ref{u1minus} and \ref{u1plus}. Using our convention for finite temperature Fourier transforms in Appendix \ref{notconv} and restricting to the kinematic configuration $q^{\pm}=0$ we have
\begin{equation}
\label{u1minusmom}
\begin{split}
J_{B, -}^{(1)}(q_{3}) &= \frac{1}{\beta} \sum_{n} \int \frac{d^{2}\vec{k}}{(2\pi)^{2}} \hspace{0.1cm} \phi^{\dagger}(k_{3}, \vec{k})\left(- 2 k_{-}\right)\phi(q_{3}-k_{3},-\vec{k})     
\end{split}
\end{equation}
\begin{equation}
\label{u1plusmom}
\begin{split}
J_{B,+}^{(1)}(q_{3}) &= \frac{1}{\beta} \sum_{n} \int \frac{d^{2}\vec{k}}{(2\pi)^{2}} \hspace{0.1cm} \phi^{\dagger}(k_{3}, \vec{k})\left(- 2 k_{+}\right)\phi(q_{3}-k_{3},-\vec{k}) \\
&  +  \frac{1}{\beta^{2}} \sum_{n_{1},n_{2}} \int \frac{d^{2}\vec{k}d^{2}\vec{p}}{(2\pi)^{4}} \hspace{0.1cm} \phi^{\dagger}(k_{3}, \vec{k} )\left( -2 i \right) A^{a}_{+}(\vec{p},p_{3})T^{a}\phi(q_{3}-k_{3}-p_{3},-\vec{k}-\vec{p}) 
\end{split}
\end{equation}

where $(n, n_{1},n_{2})\in \mathbf{Z}$ are Matsubara frequency modes. 

\subsubsection{Finite temperature vertex factor for $J_{B,-}^{(1)}$}

We will now compute the exact  finite temperature vertex factor for $J_{B,-}^{(1)}$  in the large $N_{B}$ limit. For this let us define, 
\begin{equation}
\label{j1mvertex}
\begin{split}
\left\langle J_{B,-}^{(1)}(-q) (\phi^{\dagger})^{j}(-k) \phi_{i}(p) \right\rangle_{\beta} = V^{(1)}_{B}(q,k) \delta^{j}_{i} (2\pi)^{3} \delta_{n_{q}+n_{k}, n_{p}} \delta^{(2)}(\vec{q}+\vec{k}-\vec{p}) \\
\end{split}
\end{equation}

As done in the previous section, we can obtain $V^{(1)}_{B}(q,k)$ by solving the relevant Schwinger-Dyson equation which in this case is given by the following, 
\begin{equation}
\label{j1mvertex1}
\begin{split}
 V^{(1)}_{B}(q,k) \delta_{i}^{j} &= V^{(1)}_{B, free}(q,k)\delta_{i}^{j}+ \int \mathcal{D}^{3}\ell\hspace{0.1cm}\left[\mathcal{V}^{a,\mu}(k,\ell)G(\ell) V^{(1)}_{B}(q,\ell) G(\ell+q)\mathcal{V}^{a,\nu}(\ell+q, k+q)\right]_{i}^{j}\mathcal{G}_{\nu\mu}(k-\ell) 
\end{split}
\end{equation}

where $\mathcal{V}^{a,\mu}(k,\ell)$ is the $3$-point interaction vertex given in equation \ref{rbcubicvertex}, $G(\ell)$ is the exact thermal propagator for the fundamental bosons and  $\mathcal{G}_{\nu\mu}(k-\ell) $ is the gauge field propagator. $V^{(1)}_{B, free}(q,k)$ denotes the vertex factor in the free theory. This can be easily read off from equation \ref{u1minusmom} which yields 
\begin{equation}
\label{j1mvertex2}
\begin{split}
V^{(1)}_{B, tree}(q,k) = 2k_{-}
\end{split}
\end{equation}

Now in order to solve equation the Schwinger-Dyson equation \ref{j1mvertex1} we consider the ansatz
\begin{equation}
\label{j1mvertex3}
\begin{split}
V^{(1)}_{B}(q, k) =  2 k_{-} \hspace{0.03cm}\mathcal{U}_{B}(q, y)
\end{split}
\end{equation} 

where $y=\beta k_{s}$. Note that in the kinematic regime where $q^{\pm}=0$, $\mathcal{U}_{B}(q, y)$ only depends on $q_{3}$ and $|\vec{k}|=k_{s}$. This simply follows from the consideration of symmetry under spatial rotations. Now inserting the above ansatz in equation \ref{j1mvertex1} and performing the holonomy integral in the large $N_{B}$ limit we arrive at
\begin{equation}
\label{j1mvertex4}
\begin{split}
2 k_{-} v(q, y) &= 2 k_{-}  + 8 i \pi \lambda_{B} \hspace{0.04cm} q_{3}  \int \frac{d^{2}\ell}{(2\pi)^{2}}\hspace{0.1cm} \frac{(\ell^{+}+k^{+})\ell^{+}}{(\ell^{+}-k^{+})} \hspace{0.03cm} \mathcal{U}_{B} (q,x) \hspace{0.03cm}H_{B} (x, q_{3}\beta,\mu_{B})
\end{split}
\end{equation}

where $H_{B} (x, q_{3}\beta,\mu_{B})$ was defined in equation \ref{j0sdyson4}. Using the results in Appendix \ref{angintegs},  the angular part of the integral in \ref{j1mvertex4} can be done to give
\begin{equation}
\label{j1mvertex5}
\begin{split}
\int_{0}^{2\pi} d\theta \hspace{0.1cm} \frac{\ell^{+}(\ell^{+}+k^{+})}{(\ell^{+}-k^{+})} = 4\pi k^{+} \Theta(\ell_{s}-k_{s})
\end{split}
\end{equation}

Thus we finally have
\begin{equation}
\label{j1mvertex6}
\begin{split}
\mathcal{U}_{B} (q, y) &= 1 + \frac{4 i \lambda_{B} q_{3}}{\beta^{2}}\int_{y}^{\infty} dx \hspace{0.1cm} x  \hspace{0.05cm} \mathcal{U}_{B} (q, x) H_{B} (x, q_{3}\beta,\mu_{B})
\end{split}
\end{equation}

The solution of the above integral equation is given by
\begin{equation}
\label{j1mvertex7}
\begin{split}
\mathcal{U}_{B} (q, y)&= \exp \left[\frac{4 i \lambda_{B} q_{3}}{\beta^{2}}\int_{y}^{\beta \Lambda} dx \hspace{0.1cm} x  \hspace{0.04cm} H_{B} (x, q_{3}\beta,\mu_{B})\right]= e^{ i q_{3}\beta \mathcal{F}_{B}(y, q_{3}\beta,\mu_{B})}
\end{split}
\end{equation}

where $\mathcal{F}_{B}(y, q_{3}\beta,\mu_{B})$ is given by equation \ref{j0sdyson10}. Therefore the exact finite temperature vertex factor $V^{(1)}_{B} (q, y)$ is given by
\begin{equation}
\label{j1mvertex8}
\begin{split}
V^{(1)}_{B}(q, k) = 2 k_{-} \hspace{0.03cm} \mathcal{U}_{B} (q, y) & = 2 k_{-}\hspace{0.03cm} e^{ i q_{3}\beta \mathcal{F}_{B}(y, q_{3}\beta,\mu_{B})}
\end{split}
\end{equation}

\subsubsection{Finite temperature vertex factor for $J_{B, +}^{(1)}$} 
\label{u1pbose}

Let us define the exact thermal vertex factor for $J_{B, +}^{(1)}$ to be 
\begin{equation}
\label{j1pvertex}
\begin{split}
\left\langle J_{B, +}^{(1)}(-q) (\phi^{\dagger})^{j}(-k) \phi_{i}(p) \right\rangle_{\beta} = U^{(1)}_{B}(q,k) \delta^{j}_{i} (2\pi)^{3} \delta_{n_{q}+n_{k}-n_{p}} \delta^{(2)}(\vec{q}+\vec{k}-\vec{p}) \\
\end{split}
\end{equation}

We  define $U^{(1)}_{B}(q,k)$ to consist of all 1PI Feynman graphs where a gluon line can attach at the vertex together with the vertex factor factor in the free theory, i.e. the one without any gluon lines attached. These diagrams shown in Fig.\ref{u1rbD1} and Fig.\ref{u1rbD2} are essentially the same set of diagrams considered in \cite{Aharony:2012nh} for the computation of current $2$-point function at zero temperature. We are simply considering the corresponding finite temperature version here. 

\begin{figure}[htbp]
\begin{center}
\begin{tikzpicture}[scale=.5, transform shape]
 \draw[black,->] (-3.5,0) -- (-2,0);
 \draw[black,->] (-2,0) -- (0,0);
  \node at (0,0)[circle,fill,inner sep=5pt]{};
 \draw[black,->] (0,0) -- (1,0);
  \draw[black,->] (1,0) -- (3,0);
   \draw[black] (2,0) -- (3.5,0);
 \node at (1.5,0)[circle,fill,red,inner sep=4pt]{};
 \draw[draw=black, snake it] (1.5,0) arc (0:180:1.5cm);
 \draw [black,thick ,domain=120:60,->] plot ({2.5*cos(\x)}, {2.5*sin(\x)});
 \Huge{\node at (0,3.2) {$k+q-\ell$};}
 \node at (-2.5,-0.6) {$k$};
 \node at (0.7,-0.6) {$\ell $};
 \node at (2.6,-0.6) {$p$};
 \draw[black,->] (-1.5,-1.8) -- (-1.5,-0.6);
 \node at (-1.5,-2.4) {$q$};
  \Huge{\node at (-6.5,0) {$D_{1}(k,q,p)=$};}
  \Huge{\node at (6.5,0) {$+ ~~reflection$};}
  \draw (-1.5,0) node[cross] {};
\end{tikzpicture}
\caption{Feynman diagram contributing to $U^{(1)}_{B}$. The cross denotes the current vertex and the red dot is the exact vertex shown below in Fig.\ref{phiphiAvertex}. The internal propgator with the black dot is the exact thermal propagator for the fundamental bosons. The reflection diagram, not shown here explicitly, is the one where the current vertex lies to the right of the red vertex.  }
\label{u1rbD1}
  \end{center}
\end{figure} 

\begin{figure}[ht]
\begin{center}
\begin{tikzpicture}[scale=.5, transform shape]
 \draw[black,->] (-3.5,0) -- (-2,0);
 \draw[black,->] (-2,0) -- (0,0);
  \node at (0,0)[circle,fill,inner sep=5pt]{};
 \draw[black,->] (0,0) -- (1,0);
 \draw[black] (1,0) -- (2,0);
  \draw[black,->] (2,0) -- (3,0);
   \draw[black] (3,0) -- (4.5,0);
   \draw[black,->] (4.5,0) -- (6,0);
   \draw[black] (6,0) -- (6.9,0);
 
 \draw[draw=black, snake it] (1.6,0) arc (180:0:1.71cm);
 \draw [black,thick ,domain=120:60,->] plot ({1.5+4.0*cos(\x)}, {4.0*sin(\x)});
 \Huge{\node at (10,0) {$+ ~~reflection$};}
 \draw [black,thick ,domain=160:100,->] plot ({3.5+2.5*cos(\x)}, {2.5*sin(\x)});
   \Huge{\node at (-6.5,0) {$D_{2}(k,q,p)=$};}
   \node at (-2.5,-0.6) {$k$};
    \node at (3,0)[circle,fill,inner sep=5pt]{};
    \draw[black,->] (3.0,0) -- (4,0);
   \node at (4,-0.7) {$\ell_{2}$};
   
 \node at (0.7,-0.6) {$\ell_{1}$};
 \node at (5.9,-0.6) {$p$};
 \node at (1.5,-2.0) {$q$};
 \Huge{\node at (1.5,2.5) {$\ell_{3}$};}
  \draw (1.5,0) node[cross] {};
  \Huge{\node at (1.5,4.6) {$k-\ell_{1}$};}
  \draw[black,->] (1.5,-1.5) -- (1.5,-0.6);
  \draw[draw=black, snake it] (-1.6,0) arc (180:0:3.3cm);
\end{tikzpicture}
\caption{Feynman diagram contributing to $U^{(1)}_{B}$. The internal matter propagators are again exact in $\lambda_{B}$. In the reflected diagram the current vertex lies to the right of the quartic seagull vertex. }
\label{u1rbD2}
  \end{center}
\end{figure}

\begin{figure}[hbt!]
\begin{center}
\begin{tikzpicture}[scale=.5, transform shape]
 \draw[black,->] (-4,0) -- (-3,0);
 \draw[black,->] (-3,0) -- (0,0);
 \draw[black] (0,0) -- (1,0);
 
  \draw[black,->] (-1,1.4) -- (-1,0.6);
  
   \Huge{\node at (2,0) {$=$};}
   \Huge{\node at (-0.5,1) {$r$}; }
   \draw (-1.5,0) node {};
  
   \node at (1,-0.6) {$p$};
   \node at (-4,-0.6) {$k$};
    \path [draw=black,snake it]
    (-1.5,2.4) -- (-1.5,0);
    \node at (-0.5,2.4) {$a,\mu$};
     \node at (-1.5,0)[circle,fill,red,inner sep=4pt]{};
     \node at (6.5,1) {$r$}; 
      \node at (6.5,-0.6) {$p$};
   \node at (4.5,-0.6) {$k$};

    \node at (9.3,1.7) {$r$}; 
      \node at (10,-0.6) {$p$};
   \node at (8,-0.6) {$k$};
       \draw[black,->] (8.8,2.2) -- (8.8,1.4);
   
    \node at (13.5,1.7) {$r$}; 
      \node at (15,-0.6) {$p$};
   \node at (12,-0.6) {$k$};
   
       \draw[black,->] (13.9,2.2) -- (13.9,1.4);
    \draw[black,->] (6,1.4) -- (6,0.6);
    \draw[black,->] (4,0) -- (5,0);
    \draw[black] (5,0) -- (5.5,0);
 \draw[black,->] (5.5,0) -- (6.5,0);
 \draw[black] (6.5,0) -- (7,0);
 
 \Huge{\node at (7.5,0) {$+$};}
  \draw[black,->] (8,0) -- (8.25,0);
  \draw[black] (8.25,0) -- (9,0);
 \draw[black,->] (9,0) -- (10.5,0);
 
 \draw[black] (10.5,0) -- (11,0);
 
 \Huge{\node at (11.5,0) {$+$};}
  \draw[black,->] (12,0) -- (12.5,0);
  \draw[black] (12.5,0) -- (13,0);
 \draw[black] (13,0) -- (14,0);
 \draw[black,->] (14,0) -- (15,0);
 
 \path [draw=black,snake it]
    (5.5,2.4) -- (5.5,0);
    \path [draw=black,snake it]
    (8.5,2.4) -- (8.5,0);
    
    \path [draw=black,snake it]
    (14.5,2.4) -- (14.5,0);

    \draw[draw=black, snake it] (8.5,0) arc (180:0:0.8cm);
    \draw[draw=black, snake it] (14.5,0) arc (0:180:0.8cm);
    
     \Huge{\node at (-7,0) {$\mathcal{K}^{a,\mu}(k, p)$};}
     \Huge{\node at (-4.5,0) {$=$};}
\end{tikzpicture}
\caption{Exact finite temperature vertex for $\left\langle A^{a,\mu}(-r)(\phi^{\dagger})^{j}(-k)  \phi_{i}(p) \right\rangle_{\beta}$ }
\label{phiphiAvertex}
  \end{center}
\end{figure}

Now let us evaluate the diagrams labelled by $D_{1}$ and $D_{2}$ in Fig.\ref{u1rbD1} and Fig.\ref{u1rbD2} respectively. $D_{1}$ is given by the following,
\begin{equation}
\label{j1pvertex1}
\begin{split}
D_{1} (k,q,p)  = & \int  \mathcal{D}^{3}\ell \left[  (-2 i T^{a}) \mathcal{G}_{3+}(k+q-\ell)G(\ell) \mathcal{K}^{a,3}(\ell,k+q)\right]_{i}^{j} + \text{reflection}
\end{split}
\end{equation}

where the reflection term comes from the diagram with the current vertex in Fig.\ref{u1rbD1} appearing to the right of the $3$-point vertex denoted there by the red dot. The factor of $(-2i T^{a})$ in the above integral arises from the Feynman rule for the current vertex which can be determined from the expression of the current in equation \ref{u1plus}. $ \mathcal{K}^{a,\mu}$ is the exact vertex shown in Fig.\ref{phiphiAvertex} and is defined as
\begin{equation}
\label{phiphiavertex}
\begin{split}
\left\langle A^{a,\mu}(-r)(\phi^{\dagger})^{j}(-k)  \phi_{i}(p) \right\rangle_{\beta} = ( \mathcal{K}^{a,\mu}(k,p))^{j}_{ i} (2\pi)^{3}\delta_{n_{r}+n_{k}-n_{p}} \delta^{(2)}(\vec{r}+\vec{k}-\vec{p}) 
\end{split}
\end{equation}

Using the Feynman rules for the cubic and seagull vertices in Fig.\ref{phiphiAvertex} we have,
\begin{equation}
\label{tbpp3}
\begin{split}
( \mathcal{K}^{a,\mu}(k,p))^{j}_{ i}= & i \left(T^{a} \tilde{k}^{\mu} + \tilde{p}^{\mu}T^{a}\right)_{i}^{j}   + i  \int \mathcal{D}^{3}\ell \left[ \{T^{a},T^{b}\}\mathcal{G}_{\sigma 3}(r+k-\ell)G(\ell) \left(T^{a}\tilde{ \ell}^{\sigma} + \tilde{p}^{\sigma}T^{a}\right)\right]_{i}^{j} \delta^{\mu 3} + \\
& i  \int \mathcal{D}^{3}\ell \left[ \left(T^{a} \tilde{k}^{\sigma} + \tilde{\ell}^{\sigma}T^{a}\right) \mathcal{G}_{3\sigma}(k-\ell)G(\ell) \{T^{a},T^{b}\}\right]_{i}^{j}\delta^{\mu 3}
\end{split}
\end{equation}

Simplifying the integrands in the above using \ref{gaugeprop} and \ref{ungen} we obtain
\begin{equation}
\label{aphiphidagvertex}
\begin{split}
( \mathcal{K}^{a,\mu}(k,p))_{j i} & = i (\tilde{k}^{\mu}T^{a} +\tilde{p}^{\mu}T^{a})_{ji}+ \frac{2\pi}{\kappa_{B}}   \int \mathcal{D}^{3}\ell\hspace{0.1cm} \left(\frac{\ell^{+}+k^{+}}{(\ell-k)^{+}}-\frac{\ell^{+}+p^{+}}{(\ell-p)^{+}}\right) \mathrm{Tr}(G(\ell))  T^{a}_{ji}\delta^{\mu 3}\\
& = i (\tilde{k}^{\mu}T^{a} +\tilde{p}^{\mu}T^{a})_{ji}+ \mathcal{C}(k,p) T^{a}_{ji}\delta^{\mu 3}
\end{split}
\end{equation}

Now $\mathcal{C}(k,p)$ is given by
\begin{equation}
\label{cqk}
\begin{split}
 \mathcal{C}(k,q) & = \frac{2\pi}{\kappa_{B}}   \int \mathcal{D}^{3}\ell\hspace{0.1cm} \mathrm{Tr}(G(\ell)) \left(\frac{\ell^{+}+k^{+}}{(\ell-k)^{+}}-\frac{\ell^{+}+q^{+}}{(\ell-q)^{+}}\right) \\
 & = \frac{2\pi N_{B} }{\kappa_{B}} \int_{-1/2}^{1/2} du \int \frac{d^{2}\ell}{(2\pi)^{2}} \hspace{0.1cm}\left(\frac{\ell^{+}+k^{+}}{(\ell-k)^{+}}-\frac{\ell^{+}+q^{+}}{(\ell-q)^{+}}\right) \frac{1}{\beta} \sum_{n=-\infty}^{\infty} \frac{1}{\ell_{s}^{2}+\frac{\mu_{B}^{2}}{\beta^{2}}+\frac{4\pi^{2}}{\beta^{2}}(n-|\lambda_{B}|u)^{2}}\\
& = 2 \pi \lambda_{B}\int \frac{d^{2}\ell}{(2\pi)^{2}} \hspace{0.1cm}\left(\frac{\ell^{+}+k^{+}}{(\ell-k)^{+}}-\frac{\ell^{+}+q^{+}}{(\ell-q)^{+}}\right) \mathcal{M}_{B}(x,\mu_{B})
\end{split}
\end{equation}

where $x=\beta \ell_{s}$ and 
\begin{equation}
\label{cqk1}
\begin{split}
\mathcal{M}_{B}(x,\mu_{B}) & = \frac{1}{\beta} \int_{-1/2}^{1/2} du \sum_{n=-\infty}^{\infty} \frac{1}{\left(\ell_{s}^{2}+\frac{\mu_{B}^{2}}{\beta^{2}}+\frac{4\pi^{2}}{\beta^{2}}(n-|\lambda_{B}|u)^{2}\right)} \\
& = \frac{i \beta}{2 \pi |\lambda_{B}|} \frac{\left(\log\left[\sinh\left(\frac{1}{2}\sqrt{x^{2}+\mu_{B}^{2}}-\frac{i \pi |\lambda_{B}|}{2}\right)\right]-\mathrm{c.c}\right)}{ \sqrt{x^{2}+\mu_{B}^{2}}} 
 \end{split}
\end{equation}

Performing the angular part of the integral in \ref{cqk} using the results in section \ref{angintegs} of the Appendix we get
\begin{equation}
\label{cqk2}
\begin{split}
 \mathcal{C}(k,p) & =\lambda_{B}  \int_{0}^{\infty} d \ell_{s} \hspace{0.1cm} \ell_{s}  \left [ \Theta\left(\ell_{s}-k_{s}\right)  -\Theta\left(k_{s}-\ell_{s}\right) - \Theta\left(\ell_{s}-p_{s}\right)  + \Theta\left(p_{s}-\ell_{s}\right)  \right] \mathcal{M}_{B}(x,\mu_{B}) 
\end{split}
\end{equation}

Now using the following useful relation which can be easily derived from equation \ref{cqk1}, 
\begin{equation}
\label{cqk3}
\begin{split}
\mathcal{M}_{B}(x,\mu_{B}) =  \frac{i \beta}{2 \pi |\lambda_{B}|} \frac{1}{{ \sqrt{x^{2}+\mu_{B}^{2}}} } \left[ - i \pi |\lambda_{B}| +\frac{\sqrt{x^{2}+\mu_{B}^{2}}}{x} \frac{d}{dx} \left( \mathrm{Li}_{2}\left(e^{-\sqrt{x^{2}+\mu_{B}^{2}}+i \pi |\lambda_{B}|}\right)-\text{c.c} \right)\right]
 \end{split}
\end{equation}

we can carry out the radial integral in equation \ref{cqk2}. We then arrive at the result,
\begin{equation}
\label{cqk4}
\begin{split}
 \mathcal{C}(k,p) & = \frac{\lambda_{B}}{\beta} \left[\left(\sqrt{z^{2}+\mu_{B}^{2}}-\sqrt{y^{2}+\mu_{B}^{2}}\right)+\frac{ i}{|\lambda_{B}|\pi} \left(g(z,\mu_{B})-g(y,\mu_{B})\right)\right] \\
 \end{split}
\end{equation}

where 
\begin{equation}
\label{cqk5}
\begin{split}
y=\beta k_{s}, \quad z=\beta q_{s}, \quad g(x,\mu_{B})= \mathrm{Li}_{2}\left(e^{-\sqrt{x^{2}+\mu_{B}^{2}}+i \pi |\lambda_{B}|}\right)-\text{c.c}
 \end{split}
\end{equation}

Gathering the above results and using them in equations \ref{j1pvertex1} and \ref{j1pvertex2} we have
\begin{equation}
\label{j1pvertex3}
\begin{split}
D_{1} (k,q,p) = & - \frac{4\pi }{\kappa_{B}} \int  \mathcal{D}^{3}\ell \hspace{0.1cm} \frac{1}{(\ell-k)^{+}}\left[  i (\tilde{\ell}_{3}+k_{3}+q_{3})+\mathcal{C}(\ell,k)\right] \mathrm{Tr}(G(\ell)) \delta_{i}^{j} \\
&  + \frac{4\pi }{\kappa_{B}}  \int \mathcal{D}^{3}\ell \hspace{0.1cm} \frac{1}{(\ell-k)^{+}} \left[  i(\tilde{\ell}_{3}+k_{3})-\mathcal{C}(\ell,k)\right]  \mathrm{Tr}(G(\ell)) \delta_{i}^{j} \\
& = - \frac{4\pi }{\kappa_{B}}  \int \mathcal{D}^{3}\ell \hspace{0.1cm} \frac{1}{(\ell-k)^{+}} \left[ i q_{3} +2\mathcal{C}(\ell,k)\right] \mathrm{Tr}(G(\ell)) \delta_{i}^{j} 
\end{split}
\end{equation}

where we have used $q^{\pm}=0$ and $\mathcal{C}(\ell,k)=-\mathcal{C}(k,\ell)$. Then carrying out the holonomy and angular integrals we get
\begin{equation}
\label{j1pvertex4}
\begin{split}
D_{1} (k,q,p)  & = \frac{2\lambda_{B} }{k^{+}}  \int_{0}^{k_{s}} d\ell_{s}  \hspace{0.05cm} \ell_{s} \hspace{0.1cm} \left[ i q_{3} +2\mathcal{C}(\ell,k)\right] \mathcal{M}_{B}(x,\mu_{B}) \delta_{i}^{j}
\end{split}
\end{equation} 

The radial integral can also be straightforwardly done and yields
\begin{equation}
\label{j1pvertex5}
\begin{split}
D_{1} (k,q,p)  & =  \frac{1}{ k^{+}}\mathcal{C}(0,k)\left(i q_{3}+ \mathcal{C}(0,k)\right)  \delta_{i}^{j}
\end{split}
\end{equation}

Next let us consider the diagram labelled by $D_{2}$ in Fig.\ref{u1rbD2}. More explicitly this can be expressed as
\begin{equation}
\label{j1pvertex2}
\begin{split}
D_{2} (k,q,p)  = & \int  \mathcal{D}^{3}\ell_{1} \mathcal{D}^{3}\ell_{2} \left[  \mathcal{V}^{a,+} (k,\ell_{1})\mathcal{G}_{3+}(k-\ell)G(\ell_{1})(-2 i T^{b}) \mathcal{G}_{3+}(\ell_{1}+q-\ell_{2})G(\ell_{2}) \{ T^{a},T^{b}\}\right]_{i}^{j} \\
& + \text{reflection}
\end{split}
\end{equation}

where the reflection term is due to the diagram where the current vertex is situated to the right of the seagull vertex in Fig.\ref{u1rbD2}. Upon simplifying the integrand in \ref{j1pvertex2} we get,
\begin{equation}
\label{j1pvertex6}
\begin{split}
D_{2} (k,q,p)   = & - \frac{1 }{2}\left(\frac{4\pi i}{\kappa_{B}}\right)^{2} \int  \mathcal{D}^{3}\ell_{1}  \mathcal{D}^{3}\ell_{2} \hspace{0.1cm}  \frac{(\ell_{1}^{+}+ k^{+})}{(\ell_{1}-k)^{+} (q+\ell_{1}-\ell_{2})^{+}} \left[\delta_{ji} \mathrm{Tr}(G(\ell_{1})) \mathrm{Tr}(G(\ell_{2}))+ (G(\ell_{2})G(\ell_{1}))_{ji}\right]\\
& + \text{reflection} \\
\end{split}
\end{equation}

Now in the large $N_{B}$ limit, the factor of $(G(\ell_{2})G(\ell_{1}))_{ji}$ appearing in the above integral is subleading compared to the disconnected piece $\mathrm{Tr}(G(\ell_{1})) \mathrm{Tr}(G(\ell_{2}))$. We will neglect such contributions from here on. Then taking $q^{\pm}=0$ we have 
\begin{equation}
\label{j1pvertexD2}
\begin{split}
D_{2} (k,q,p)   = -\left(\frac{4\pi i}{\kappa}\right)^{2} \int  \mathcal{D}^{3}\ell_{1}  \mathcal{D}^{3}\ell_{2} \hspace{0.1cm}  \frac{(\ell_{1}^{+}+ k^{+}) }{(\ell_{1}-k)^{+} (\ell_{1}-\ell_{2})^{+}} \delta_{ji} \mathrm{Tr}(G(\ell_{1})) \mathrm{Tr}(G(\ell_{2}))
\end{split}
\end{equation}

In this case, the angular part of the integral in \ref{j1pvertexD2} is
\begin{equation}
\label{j1pvertex7}
\begin{split}
\int_{0}^{2\pi} d\theta_{1}\int_{0}^{2\pi} d\theta_{2} \hspace{0.1cm} \frac{\left(\ell_{1}+k\right)^{+}}{\left(\ell_{1}-k\right)^{+}\left(\ell_{1}-\ell_{2}\right)^{+}}= \frac{(-2)(2\pi)^{2}}{k^{+}} \Theta(k_{s}-\ell_{s,1})\Theta(\ell_{s,1}-\ell_{s,2})
\end{split}
\end{equation}

We are then left with the holonomy and radial integrals. Executing these respectively leads to, 
\begin{equation}
\label{j1pvertex8}
\begin{split}
D_{2} (k,q,p)  & =- \frac{8 \lambda^{2}}{k^{+}} \int_{0}^{k_{s}} d\ell_{s,1} \hspace{0.1cm} \ell_{s,1} \hspace{0.05cm}\mathcal{M}_{B}(x_{1},\mu_{B}) \int_{0}^{\ell_{s,1}} d\ell_{s,2} \hspace{0.1cm} \ell_{s,2} \hspace{0.05cm} \mathcal{M}_{B}(x_{2},\mu_{B}) \\
& = -\frac{1}{k^{+}} \mathcal{C}^{2}(0,k)
\end{split}
\end{equation}

where $y =  \beta k_{s}$. Finally adding the net contributions of $D_{1}$ and $D_{2}$ to the vertex factor in the free theory,  we get the exact finite temperature vertex factor $U^{(1)}_{B}$ to be given by
\begin{equation}
\label{j1pvertex9}
\begin{split}
U^{(1)}_{B} (q,k)  & = 2 k^{-} + D_{1} (p,q,k) +D_{2} (p,q,k) \\
& = \frac{k_{s}^{2} +  i q_{3} \hspace{0.03cm}\mathcal{C}(0,k)}{k^{+}}
\end{split}
\end{equation}

\subsubsection{Thermal $2$-point function}

\begin{figure}[ht]
\begin{center}
\begin{tikzpicture}[scale=.5, transform shape]
 
 \Huge{ \node at (2.5,0) {$J^{(1)}_{B,-}$};}
  \node at (-2.5,0) {$J^{(1)}_{B,+}$};
 \node at (0,2)[circle,fill,inner sep=12pt]{};
     \draw[black, thick,->] (-7,0) -- (-5.0,0);
     \node at (0,2)[circle,fill,inner sep=12pt]{};
     \node at (0,-2)[circle,fill,inner sep=12pt]{};
      \node at (-5.2,1.0) {$q$};
     \draw (-4.2,0) circle (14pt);
      \draw [black,thick ,domain=20:152] plot ({4.5*cos(\x)}, {-1+3*sin(\x)});
      \draw [black,thick ,domain=340:208] plot ({4.5*cos(\x)}, {1+3*sin(\x)});
      \node at (4.2,0) [rectangle,draw]  {};
\draw (-4.2,0) node[cross] {};
\draw (4.2,0) node[cross] {};
\end{tikzpicture}
\caption{Feynman diagram for $U(1)$ current $2$-point function. The internal lines with black dots are exact finite temperature propagators for the fundamental bosons. The circled cross on the left is the vertex for $J^{(1)}_{B,-}$. The boxed cross on the right denotes the vertex factor corresponding to $J^{(1)}_{B,+}$. }
\label{j1rb2pt}
  \end{center}
\end{figure} 

The finite temperature $2$-point function of the $U(1)$ current in the Regular Boson theory can be obtained from the Feynman diagram in Fig.\ref{j1rb2pt} where the vertex factors  $V^{(1)}_{B}(q, k)$ and $U^{(1)}_{B}(q, k)$ are connected via the exact thermal propagators of the fundamental bosons. We thus have,
\begin{equation}
\label{u1rbtherm}
\begin{split}
\left \langle  J_{B,-}^{(1)} (-q) J_{B,+}^{(1)} (q)\right\rangle_{\beta} & =\int \frac{d^{2}\ell}{(2\pi)^{2}}\hspace{0.04cm} \frac{1}{\beta} \sum_{n=-\infty}^{\infty} \mathrm{Tr}\left[V^{(1)}_{B}(q,\ell)G(\ell)U^{(1)}_{B}(-q,\ell+q)G (\ell+q)\right] \\
\end{split}
\end{equation}

Plugging the results for the vertex factors from equations \ref{j1mvertex8} and \ref{j1pvertex9} into the above and subsequently doing the Matsubara sum, angular and holonomy integrals, the integral expression in equation \ref{u1rbtherm} becomes
\begin{equation}
\label{u1rbtherm1}
\begin{split}
\left \langle  J_{B,-}^{(1)} (-q) J_{B,+}^{(1)} (q)\right\rangle_{\beta} & = \frac{N_{B}}{ 4 \pi \lambda_{B}\beta} \int_{0}^{\beta\Lambda} dx \hspace{0.1cm} x \left( x^{2}- i q_{3}\beta^{2}\mathcal{C}(0,\ell) \right) e^{  i q_{3}\beta \mathcal{F}_{B} (x, q_{3}\beta,\mu_{B})} \hspace{0.05cm} \mathcal{H}_{B}(x, q_{3}\beta,\mu_{B})  
\end{split}
\end{equation}

where we have employed a cutoff $\Lambda$ to regulate UV divergences in the above integral. Now let us note the following identities,
\begin{equation}
\label{u1rbtherm2}
\begin{split}
& \partial_{x} \left( e^{i q_{3}\beta\hspace{0.03cm} \mathcal{F}_{B}(x,q_{3}\beta,\mu_{B})} \right) = - i q_{3}\beta \hspace{0.03cm} x \hspace{0.03cm} \mathcal{H}_{B}(x,q_{3}\beta,\mu_{B}) e^{i q_{3}\beta\hspace{0.03cm} \mathcal{F}_{B}(x,q_{3}\beta,\mu_{B})}\\
& e^{i q_{3}\beta\hspace{0.03cm} \mathcal{F}_{B}(x,q_{3}\beta,\mu_{B})}\partial_{x} \mathcal{C}(0,\ell)= \frac{i}{4 q_{3}\beta^{2}}\left(4x^{2}+4\mu_{B}^{2}+q_{3}^{2}\beta^{2}\right)\partial_{x}\left( e^{i q_{3}\beta\hspace{0.03cm} \mathcal{F}_{B}(x,q_{3}\beta,\mu_{B})}\right)
\end{split}
\end{equation}

The utility of the above relations lies in the fact that plugging them in equation \ref{u1rbtherm1}, the integral can be easily performed via integration by parts. It turns out that the final contribution only comes from boundary terms and is given by
\begin{equation}
\label{u1rbtherm3}
\begin{split}
& \left \langle  J_{B,-}^{(1)} (-q) J_{B,+}^{(1)} (q)\right\rangle_{\beta}  \\
&= \frac{  N_{B} }{4 \pi  \lambda_{B} } \left[ \frac{i\Lambda^{2}}{q_{3}\beta} + \left(  \mathcal{C}(0, \ell) e^{i q_{3}\beta\hspace{0.03cm} \mathcal{F}_{B}(x,q_{3}\beta,\mu_{B})}\right)_{0}^{\beta\Lambda} - \frac{i}{4 q_{3}\beta^{2}} \left( \left(4x^{2}+4\mu_{B}^{2}+q_{3}^{2}\beta^{2}\right) e^{i q_{3}\beta\hspace{0.03cm} \mathcal{F}_{B}(x,q_{3}\beta,\mu_{B})}\right)_{0}^{\beta\Lambda}\right]
\end{split}
\end{equation}

Since $\mathcal{F}_{B}(x,q_{3}\beta,\mu_{B})\big|_{x\rightarrow\infty} =0$ and $ \mathcal{C}(0, 0)=0$, we then get
\begin{equation}
\label{u1rbtherm4}
\begin{split}
\left \langle  J_{B,-}^{(1)} (-q) J_{B,+}^{(1)} (q)\right\rangle_{\beta} & = -\frac{i N_{B}}{16 \pi \lambda_{B} q_{3}\beta^{2}} \left(1-e^{i q_{3}\beta \hspace{0.04cm} \mathcal{F}_{B}(q_{3}\beta,\mu_{B})}\right) \left(q_{3}^{2}\beta^{2}+4\mu_{B}^{2}\right) + \frac{ N_{B} }{4 \pi  \lambda_{B}} \mathcal{C}(0,\Lambda)\\
\end{split}
\end{equation} 

The divergent contribution is now contained in $ \mathcal{C}(0, \Lambda)$. Let us isolate the divergent piece by first defining $ \mathcal{C}(0,k)= \mathcal{X}(y)-\mathcal{X}(0)$ where $\mathcal{X}(y)$ is given by 
\begin{equation}
\label{u1rbtherm5}
\begin{split}
\mathcal{X}(y) =\frac{\lambda_{B}}{\beta} \left[\sqrt{y^{2}+\mu_{B}^{2}}+\frac{ i}{\pi |\lambda_{B}|} \left( \mathrm{Li}_{2}\left[e^{-\sqrt{x^{2}+\mu_{B}^{2}}+i \pi |\lambda_{B}|}\right]-\text{c.c}\right) \right] 
\end{split}
\end{equation} 

Then for $\Lambda\rightarrow \infty$ we have $\mathcal{X}(\beta\Lambda) \sim \lambda_{B} \Lambda$. In order to remove this divergence we can introduce a mass counterterm for the background gauge field that couples to the $U(1)$ current. Thus the renormalised thermal $2$-point function is 
\begin{equation}
\label{u1rbtherm6}
\begin{split}
\left \langle  J_{B,-}^{(1)} (-q) J_{B,+}^{(1)} (q)\right\rangle_{\beta} & = -\frac{i N_{B}}{16 \pi \lambda_{B} q_{3}\beta^{2}} \left(1-e^{i q_{3}\beta \hspace{0.04cm} \mathcal{F}_{B}(q_{3}\beta,\mu_{B})}\right) \left(q_{3}^{2}\beta^{2}+4\mu_{B}^{2}\right) - \frac{ N_{B} }{4 \pi  \lambda_{B}} \mathcal{X}(0)\\
\end{split}
\end{equation} 

where
\begin{equation}
\label{u1rbtherm7}
\begin{split}
\mathcal{X}(0) =\frac{\lambda_{B}}{\beta} \left[\mu_{B}+\frac{ i}{\pi |\lambda_{B}|} \left( \mathrm{Li}_{2}\left[e^{-\mu_{B}+i \pi |\lambda_{B}|}\right]-\text{c.c}\right) \right] 
\end{split}
\end{equation} 

\subsection{Critical Bosons}
\label{u1thermcb}

In section \ref{j0cb} we had discussed how the thermal correlators in the Critical Boson theory can be obtained by deforming the Regular Boson theory by a relevant quartic interaction and then flowing to the IR where this coupling becomes large. One of the important effects of turning on this coupling there was the inclusion of additional planar Feynman diagrams where a chain of scalar loops can be attached in between the two external operator insertions. However in case of spinning external operators, these diagrams do not contribute. This is because they involve $2$-point functions of the form $\left\langle J^{(s)} J^{(0)}\right\rangle_{\beta}$ which vanish due to symmetry under spatial rotations unless $s=0$. 

As a result we can straightforwardly get the finite temperature $2$-point function of the $U(1)$ current in the Critical Boson from our previous analysis involving regular bosons. We simply need to account for the change in the thermal mass which was given previously in equation \ref{thermmasscb}. Then we have,
\begin{equation}
\label{u1cbtherm1}
\begin{split}
\left \langle  \tilde{J}_{B,-}^{(1)} (-q) \tilde{J}_{B,+}^{(1)} (q)\right\rangle_{\beta} & = -\frac{i N_{B}}{16 \pi \lambda_{B} q_{3}\beta^{2}} \left(1-e^{i q_{3}\beta \hspace{0.04cm} \mathcal{F}_{B}(q_{3}\beta,\mu_{B,c})}\right) \left(q_{3}^{2}\beta^{2}+4\mu_{B,c}^{2}\right) 
\end{split}
\end{equation} 

Note that this is of the same form as in \ref{u1rbtherm6}. The last term there, i.e., $\mathcal{X}(0)$ vanishes in the Critical Boson theory due to \ref{thermmasscb}. Expanding the above result around $\lambda_{B}=0$, we can verify that it agrees with the perturbative result derived in \citep{GurAri:2016xff} upto an extra term $\frac{i N_{B}\lambda_{B}q_{3}}{8 \pi}$ which was included in \citep{GurAri:2016xff} to cancel an anomaly.  



\subsection{Regular Fermions} 
\label{u1rff}

The thermal $2$-point function for the $U(1)$ current in the Regular Fermion theory was first obtained in \cite{GurAri:2016xff}, to which we refer the reader for details of the computation\footnote{Note that some of our notations in this section will be slightly different from the ones used in \cite{GurAri:2016xff}. This is just to ensure consistency with the rest of our paper.}.  Our purpose in including this section here will be to show that the integral expression for the $2$-point function presented in  \cite{GurAri:2016xff} can be considerably simplified. As a byproduct, it will be easier to check that the final result transforms appropriately under bosonization dualities. 

The $U(1)$ current in the Regular Fermion theory is given by
\begin{equation}
\label{u1regferm}
\begin{split}
J_{F,\mu}^{(1)}(x)= i \bar{\psi}(x) \gamma_{\mu} \psi (x)
\end{split}
\end{equation} 

Then the momentum space expression for the current is 
\begin{equation}
\label{u1regferm1}
\begin{split}
J_{F,\mu}^{(1)}(\vec{q},q_{3}) = \frac{1}{\beta} \sum_{n} \int \frac{d^{2}\vec{k}}{(2\pi)^{2}} \hspace{0.1cm} \bar{\psi} (\vec{k},k_{3}) (i \gamma_{\mu}) \psi (\vec{q}-\vec{k},q_{3}-k_{3})
\end{split}
\end{equation}

\subsubsection{Finite temperature vertex factor for $J_{F,-}^{(1)}$}

Just as in the case of the U(1) current in the bosonic theory in section \ref{u1rb}, we will only consider the two point function $\left \langle J_{F,\mu}^{(1)} J_{F,\nu}^{(1)} \right\rangle_{\beta}$ with $(\mu,\nu) =(-,+)$.  In order to compute this we then require the corresponding finite temperature vertex factors. For $J_{F,-}^{(1)}$, this is defined as
\begin{equation}
\label{j1rfvertex}
\begin{split}
\left\langle J_{F,-}^{(1)}(-q) \psi_{i}(p) \bar{\psi}^{j}(-k)\right\rangle_{\beta} = V^{(1)}_{F}(q,k) \delta_{i}^{j}(2\pi)^{3} \delta^{(3)}(q+k-p)
\end{split}
\end{equation}

In \cite{GurAri:2016xff}, the Schwinger-Dyson equation for $V^{(1)}_{F}(q,k)$ was solved at large $N_{F}$ and exactly in the t'Hooft coupling $\lambda_{F}$. The resulting solution was found to be given by the following expression
\begin{equation}
\label{j1rfvertex1}
\begin{split}
V^{(1)}_{F}(q,k) & = v^{(1)}_{+}(q,y)\gamma^{+}+ \beta k^{+}v^{(1)}_{I}(q,y)\mathbf{1} \\
\end{split}
\end{equation}

where
\begin{equation}
\label{j1rfvertex2}
\begin{split}
& v^{(1)}_{I}(q,y) = \frac{i}{q_{3}\beta} \left(1-e^{i q_{3}\beta \hspace{0.03cm}\mathcal{F}_{F}(y,q_{3}\beta,\mu_{F})}\right) \\
\end{split}
\end{equation}

\begin{equation}
\label{j1rfvertex3}
\begin{split}
& v^{(1)}_{+}(q,y) = i- \frac{i(q_{3}\beta-2\tilde{f}(y))}{2 q_{3}\beta} \left(1-e^{i q_{3}\beta \hspace{0.03cm}\mathcal{F}_{F}(y,q_{3}\beta,\mu_{F})}\right) 
\end{split}
\end{equation}

and $\mathcal{F}_{F}(y,q_{3}\beta,\mu_{F})$ has been defined previously in equation \ref{j0fsdyson16}. 

\subsubsection{Thermal $2$-point function}

We again consider a Feynman diagram of the type shown in Fig.\ref{2pointrbs0} where we now connect the exact vertex factor for $J^{(1)}_{F,-}$ to the vertex factor for  $J^{(1)}_{F,+}$ with exact thermal propagators for the fundamental fermions. Note that to avoid overcounting of diagrams we should take the vertex factor for $J^{(1)}_{F,+}$ to be that of the free theory. As a result, the thermal $2$-point function in the Regular Fermion theory is given by
\begin{equation}
\label{j1fermth}
\begin{split}
& \left\langle J^{(1)}_{F,-}(-q) J^{(1)}_{F,+}(q)\right\rangle_{\beta}  =-\sum_{i=1}^{N_{F}}\int \frac{d^{2}\ell}{(2\pi)^{2}}\hspace{0.04cm} \frac{1}{\beta} \sum_{n=-\infty}^{\infty} \mathrm{Tr}\left[V^{(1)}_{F}(q,\ell)S_{i}(\ell)V^{(1)}_{F, free}(-q,\ell+q)S_{i}(\ell+q)\right] \\
\end{split}
\end{equation} 

Here $V^{(1)}_{F, free}(q,k)=i\gamma^{-}$ is the vertex factor for $J^{(1)}_{F,+}$ in the free theory. The trace is over the gamma matrices and the overall negative sign comes from the fermion loop in the Feynman diagram contributing to this correlator. Then using the result for the vertex factors in equations \ref{j1rfvertex2} and \ref{j1rfvertex3} we have,
\begin{equation}
\label{j1fermth1}
\begin{split}
&  \left\langle J^{(1)}_{F,-}(-q) J^{(1)}_{F,+}(q)\right\rangle_{\beta} \\
& = \frac{ N_{F}}{8 \pi \lambda_{F} q_{3} \beta^{2}} \int_{0}^{\beta\Lambda} dx \hspace{0.1cm} x \left( 4x^{2}+4\mu_{F}^{2}+q_{3}^{2}\beta^{2}\right) \tilde{f}(x) \mathcal{H}_{F}(x, q_{3}\beta,\mu_{F}) \hspace{0.1cm} + \\
& \hspace{0.5cm}\frac{ N_{F}}{4 \pi \lambda_{F} \beta}  \int_{0}^{\beta\Lambda} dx \hspace{0.1cm} x \left(x^{2}+\mu_{F}^{2}\right)e^{i q_{3}\beta \hspace{0.03cm}\mathcal{F}_{F}(y,q_{3}\beta,\mu_{F})} \mathcal{H}_{F}(x, q_{3}\beta,\mu_{F}) \hspace{0.1cm} - \\
& \hspace{0.5cm} \frac{ N_{F}}{8 \pi \lambda_{F} q_{3} \beta^{2}}  \int_{0}^{\beta\Lambda} dx \hspace{0.1cm} x \left[ 4x^{2}+4\mu_{F}^{2}-q_{3}\beta \left(q_{3}\beta-2\tilde{f}(x)\right)\right] \tilde{f}(x) \hspace{0.04cm} e^{i q_{3}\beta \hspace{0.03cm}\mathcal{F}_{F}(y,q_{3}\beta,\mu_{F})} \mathcal{H}_{F}(x, q_{3}\beta,\mu_{F})
\end{split}
\end{equation} 

where the cutoff $\Lambda$ has been put in to regulate the UV divergent integrals. This is precisely the same result that was first obtained in \citep{GurAri:2016xff}. Now  following the strategy adopted in section \ref{rf2pts0}, we will show that the above integral can be evaluated to yield a rather simpler expression.  Using the relations given in equations \ref{j0fsdyson15} and \ref{j0thermalrf6}, and appropriately integrating by parts we can show that the ultimate result is once again solely due to boundary terms as follows,
\begin{equation} 
\label{j1fermth2}
\begin{split}
& \left\langle J^{(1)}_{F,-}(-q) J^{(1)}_{F,+}(q)\right\rangle_{\beta} \\
 &= \frac{i N_{F}}{8 \pi \lambda_{F}q_{3}^{2}\beta^{3}} \left[ \left(q_{3}\beta-2\tilde{f}(x)\right)\left(2\left(x^{2}+\mu_{F}^{2}\right)+q_{3}\beta \tilde{f}(x)\right) e^{i q_{3}\beta \hspace{0.04cm}\mathcal{F}_{F}(x,q_{3}\beta, \mu_{F})} \right]_{0}^{\beta\Lambda} - \\
 & \frac{i N_{F}}{16 \pi \lambda_{F}q_{3}^{2}\beta^{3}} \left[\left( 4x^{2}+4\mu_{F}^{2}+q_{3}^{2}\beta^{2}\right) \left(q_{3}\beta-2\tilde{f}(x)\right) e^{i q_{3}\beta \hspace{0.04cm}\mathcal{F}_{F}(x,q_{3}\beta, \mu_{F})} \right]_{0}^{\beta\Lambda} + \frac{i N_{F}}{4 \pi \lambda_{F}q_{3}\beta^{2}} \left ( \tilde{f}^{2}(x)\right)_{0}^{\beta\Lambda} \\
 & = \frac{i N_{F}}{16 \pi \lambda_{F}q_{3}\beta^{2}} \left[\left( - 1 + e^{i q_{3}\beta \hspace{0.04cm}\mathcal{F}_{F}(q_{3}\beta, \mu_{F})}\right) \left(q_{3}^{2}\beta^{2} -4 \mu_{F}^{2}\right)+4 i \mathrm{sgn}(\lambda_{F}) q_{3}\beta \mu_{F}e^{i q_{3}\beta \hspace{0.04cm}\mathcal{F}_{F}(q_{3}\beta, \mu_{F})}\right] \\
& +  \frac{i N_{F}}{4 \pi \lambda_{F}\beta} \tilde{f}(\beta\Lambda)
\end{split}
\end{equation}

where we have used $\tilde{f}(0) = - i \mathrm{sgn}(\lambda_{F}) \mu_{F}$ and $ \mathcal{F}_{F}(x,q_{3}\beta,\mu_{F})\big|_{x\rightarrow \infty} =0$. The divergent piece in the above result is $\tilde{f}(\beta\Lambda) \sim - i \lambda_{F} \beta \Lambda$. Removing this by adding a mass counterterm for the background gauge field that can couple to the $U(1)$ current, we obtain the following renormalised thermal $2$-point function. 
\begin{equation} 
\label{j1fermth3}
\begin{split}
& \left\langle J^{(1)}_{F,-}(-q) J^{(1)}_{F,+}(q)\right\rangle_{\beta} \\
&= \frac{i N_{F}}{16 \pi \lambda_{F}q_{3}\beta^{2}} \left[\left( - 1 + e^{i q_{3}\beta \hspace{0.04cm}\mathcal{F}_{F}(q_{3}\beta, \mu_{F})}\right) \left(q_{3}^{2}\beta^{2} -4 \mu_{F}^{2}\right)+4 i \hspace{0.03cm} \mathrm{sgn}(\lambda_{F}) \hspace{0.03cm} \mu_{F} q_{3}\beta \hspace{0.03cm} e^{i q_{3}\beta \hspace{0.04cm}\mathcal{F}_{F}(q_{3}\beta, \mu_{F})}\right]
\end{split}
\end{equation}

It was argued in \citep{GurAri:2016xff} that the above result should be appended by including an anomaly cancelling term $\left(-\frac{i N_{F}}{4\pi}- \frac{i N_{F}\lambda_{F}}{8}\right) q_{3}$. 

\subsection{Critical Fermions}
\label{s1cf}

The thermal $2$-point function of spin $s \ge 1$ operators in the Critical Fermion should take the same form as the corresponding correlators in Regular Fermion theory after accounting for the required modification of the thermal mass for the fundamental fermions. This again follows from the discussion in section \ref{u1thermcb} where we argued that additional planar diagrams which can arise due to the quartic interaction term $\left(\bar{\psi} \psi\right)^{2}$ vanish on grounds on rotational symmetry. 

Therefore the exact finite temperature $2$-point function of the $U(1)$ current in the Critical Fermion theory is given by
\begin{equation} 
\label{j1cb}
\begin{split}
& \left\langle J^{(1)}_{F,-}(-q) J^{(1)}_{F,+}(q)\right\rangle_{\beta} \\
&= \frac{i N_{F}}{16 \pi \lambda_{F}q_{3}\beta^{2}} \left[\left( - 1 + e^{i q_{3}\beta \hspace{0.04cm}\mathcal{F}_{F}(q_{3}\beta, \mu_{F,c})}\right) \left(q_{3}^{2}\beta^{2} - 4 \mu_{F,c}^{2}\right)+4 i \hspace{0.03cm} \mathrm{sgn}(\lambda_{F}) \hspace{0.03cm} \mu_{F,c} q_{3}\beta \hspace{0.03cm}e^{i q_{3}\beta \hspace{0.04cm}\mathcal{F}_{F}(q_{3}\beta, \mu_{F,c})}\right]
\end{split}
\end{equation}

\subsection{Duality Check}

Since the thermal correlators in this case take the same form in the regular and critical versions of the theories of interest, we will only show here the effects of the duality transformations on the Regular Fermion and Critical Boson theories.  Now applying the duality map in \ref{dualitymap1} and using \ref{dualtransf1} on the thermal $2$-point function in \ref{j1fermth3} we get
\begin{equation} 
\label{dualchecku1rf}
\begin{split}
 \left\langle J^{(1)}_{F,-}(-q) J^{(1)}_{F,+}(q)\right\rangle_{\beta} \rightarrow  -\frac{i N_{B}}{16 \pi \lambda_{B} q_{3}\beta^{2}} \left(1-e^{i q_{3}\beta \hspace{0.04cm} \mathcal{F}_{B}(q_{3}\beta,\mu_{B})}\right) \left(q_{3}^{2}\beta^{2}+4\mu_{B}^{2}\right)  +\frac{i N_{B}}{8 \pi \lambda_{B}} q_{3}
\end{split}
\end{equation}

Comparing the above with the result in \ref{u1thermcb}, we see that the $2$-point functions match across the Regular Fermion and Critical Boson theories upto the addition of the term $ \frac{i N_{B}}{8 \pi \lambda_{B}} q_{3}$. Now this is a contact term, since upon Fourier transforming to position space this yields a derivative of delta function. We can remove this contact term ambiguity by adding a Chern Simons term for the background gauge field that couples to the $U(1)$ current. However, note that in the presence of Chern Simons interactions, some contact terms can indeed be physical. We refer the reader to \citep{Closset2012} for a detailed discussion of issues pertaining to Chern Simons contact terms in $3$-$d$ QFTs. 





\section{Discussions and Future Directions}
\label{futuredirec}

We conclude this paper by highlighting a number of avenues which we regard as being worthy of further exploration. First of all, let us discuss some potential generalisations of the results presented here. An immediate generalisation would be to consider distributions for the thermal holonomy other than the universal form considered in this paper.  It should be fairly straightforward to include a finite chemical potential as well for the conserved $U(1)$ current. We can also consider mass deformations of the Chern Simons matter theories that we have studied here. Although this will drive us away from the conformal fixed point where we have chosen to focus our attention throughout this paper, it certainly opens up the possibility of exploring a much richer structure.  For example, it will be interesting to consider thermal correlators in various phases such as the Higgsed and condensed phases. Another interesting direction would be to generalise our calculation of $2$-point functions to the case of higher spin currents greater than $2$. There are tentative hints of some general structures in our finite temperature results for spin $s=1$ and $s=2$ and it will be good to check how much, if any, of this structure survives for higher spins and gain a physical understanding thereof. It will be fruitful to even attempt to compute higher point thermal correlators. The simplest yet nontrivial case will then involve $3$-point thermal correlators of spin $s$ current operators, which can be obtained by generalising the corresponding zero temperature calculation that has already been performed in \citep{Aharony:2012nh}. 

Let us now mention some interesting applications of the results in our paper. In conformal field theories at finite temperatures local primary operators can have nontrivial one-point functions. In the class of CFTs that we have considered in this paper, we can extract such thermal one-point functions starting from thermal $2$-point functions by considering an OPE limit. For example, it will be very interesting to obtain the thermal one-point functions of the single trace spin $s$ currents $J^{(s)}$ from our expressions for the $2$-point functions of gauge invariant operators. Now in momentum space the OPE regime corresponds to a large spatial momentum expansion. However the correlators that we have computed are all in the special kinematic regime where the external spatial momentum has been set to zero. Therefore it will be worthwhile to try to extend our calculations to the case of generic spatial momentas. Although the relevant Schwinger-Dyson equations can not be exactly solved in the t'Hooft coupling at large $N$ in this case, we can hope to make progress by analysing the Schwinger-Dyson equations in the limit $|\vec{q}|\beta \gg 1$, with $q$ being the external momentum in the correlator.  As an alternative approach towards obtaining thermal $1$-point functions in Chern Simons matter theories, we can employ analytic conformal bootstrap techniques which have been recently developed for finite temperature CFTs, \citep{Iliesiu:2018fao}. Chern Simons matter theories at zero temperature have also been studied recently via conformal bootstrap methods in \citep{Aharony:2018npf}.  It will be interesting to generalise the results obtained in that paper to finite temperature using the formalism developed in \citep{Iliesiu:2018fao}. 

Another area worth exploring will be the finite temperature real time dynamics of these Chern Simons matter theories. In this regard we can consider analytically continuing the Matsubara frequencies in the Euclidean thermal $2$-point functions that we have computed to obtain the corresponding finite temperature Lorentzian correlators. These can be potentially useful for studying transport phenomena,\citep{Geracie:2015drf},\citep{GurAri:2016xff}. Further, as noted in the introduction, $3$-$d$ Chern Simons theories with fundamental matter fields are also supposed to be holographically dual to Vasiliev's higher spin theories of gravity in $AdS_{4}$. We will like to understand what lessons can we learn about the physics of black holes in the bulk higher spin theory \citep{Didenko:2009td} from our finite temperature computations in the boundary field theory side. Can we see for example poles that may correspond to quasinormal modes in the spectral functions obtained from our thermal $2$-point functions ? 

Finally let us mention another direction that we find worth pursuing further. The Chern Simons matter theories of interest to us here have an infinite tower of higher spin currents which are exactly conserved in the strict large $N$ limit. Including $1/N$ corrections leads to a breaking of the higher spin symmetries associated with these currents. Now the Ward identities for these weakly broken higher spin symmetries turn out to impose powerful constraints. For example it completely determines the structure of all zero temperature $3$-point functions involving spin $s$ currents as shown in \citep{Maldacena:2011jn}.\citep{Maldacena:2012sf}. In the context of this paper, it will be fruitful to analyse what constraints do the breaking of higher spin symmetries impose on thermal correlations functions. 

We hope to address the issues discussed above in future work. 



\section*{Acknowledgments}
 
We would like to thank Shiraz Minwalla, Gautam Mandal, Barak Kol, Suvrat Raju, Frank Ferrari, Sachin Jain, Yasha Neiman, Pinaki Banerjee and Alexandre Serantes for many useful discussions and insightful comments. S.G. gratefully acknowledges the hospitality of the International Solvay Institutes, Brussels during Strings 2019; ICTS-TIFR, Bangalore; Korea Institute for Advanced Study, Seoul and Yukawa Institute for Theoretical Physics, Kyoto during various stages of completion of this work. S.M. would like to thank TIFR, Mumbai; MITP, Mainz; International Solvay Institutes, Brussels during Strings 2019 and the organizers of The Fifth Israeli-Indian Conference on String Theory for hospitality while this work was in progress. S.G. is supported by the Quantum Gravity Unit of the Okinawa Institute of Science and Technology Graduate University (OIST). This work of S.M. was partially supported by the ``Quantum Universe'' I-CORE program of the Israeli Planning and Budgeting Committee. 


\appendix

\label{appendix}

\section{Notation and conventions}
\label{notconv}
Here we list our choice of conventions used in the paper.

\subsubsection*{Choice of Gamma matrices}

We have chosen the gamma matrices $\gamma^{\mu}$, with $\mu \in (1,2,3)$ to be given by Pauli sigma matrices. 
\begin{equation}
\gamma^{\mu}=\{ \sigma^1, \sigma^2, \sigma^3 \}
\end{equation}

In this paper we frequently make use of light-cone coordinates for the spatial directions. The gamma matrices corresponding to the light-cone directions are defined as
\begin{equation}
\label{gammapm}
\gamma^{\pm}=\frac{\gamma^{1}\pm i \hspace{0.04cm}\gamma^{2}}{\sqrt{2}}
\end{equation}

\subsubsection*{Convention for Fourier Transformation}

The Fourier transforms that we consider throughout this paper are mostly at finite temperature. Our convention for this is as follows
\begin{equation}
\phi(x^{3}, \vec{x})=\frac{1}{\beta} \sum_{n}\int \frac{d^{2}\vec{k}}{(2\pi)^{2}} \hspace{0.1cm} e^{i k_{3} x^{3}+i \vec{k}.\vec{x}} \hspace{0.04cm} \phi(k_{3},\vec{k})
\end{equation}

where $k_{3}=\frac{2\pi n}{\beta}$. For bosonic functions, $n\in \mathbf{Z}$ , since in that case we impose periodic boundary conditions along the compact Euclidean time direction.  In case of fermions, $n\in \mathbf{Z}+1/2$  due to anti-periodic boundary conditions around the thermal circle.

\section{Useful formulae}
\label{usefulstuff}

\subsection{Gamma Matrix Identities}
 
The following identities involving gamma matrices have been used in our calculations. 
\begin{equation}
\label{gammaidentity1}
\begin{split}
&  a_{\mu}b_{\nu}\gamma^{[3|}\gamma^{\mu}\gamma^{\nu}\gamma^{|+]}=2\hspace{0.04cm}  a.b\hspace{0.04cm}\gamma^{+}+2(a_{3}b_{-}-a_{-}b_{3})\mathbf{1}\\
& a_{\mu}\gamma^{[3|}\gamma^{\mu}\gamma^{+}\gamma^{|+]}=a_{\mu}\gamma^{[3|}\gamma^{+}\gamma^{\mu}\gamma^{|+]}=2\hspace{0.04cm} a_{-}\hspace{0.04cm} \gamma^{+}
 \end{split}
\end{equation}

The above can be easily derived using the definition \ref{gammapm} and the algebra satisfied by the gamma matrices which is given by,
\begin{equation}
\label{gammacomm}
\begin{split}
[ \gamma^{\mu}, \gamma^{\nu}] = 2 i \hspace{0.03cm} \varepsilon^{\mu\nu\rho} \gamma^{\rho}
 \end{split}
\end{equation}

\subsection{Angular Integrals}
\label{angintegs}

We list here the results of the angular integrals that we have used in our computations. These integral can be performed by contour techniques. We show here the calculation of one simple angular integral for clarity and the same applies for the rest of them.\\

Let us consider the integral, 
\begin{equation}
\label{ang1}
\int_{0}^{2\pi} d\theta \hspace{0.1cm} \frac{1}{(\ell-k)^{+}} 
\end{equation}

Here, $\theta$ is the angle between the spatial vectors $\vec{\ell}$ and $\vec{k}$. Now, we can write $\ell^{+}=\frac{\ell_s}{\sqrt{2}}e^{i \theta}$ and $k^{+}=\frac{k_s}{\sqrt{2}}$, where it is understood that  the vector $\vec{k}$ has been oriented along one of the spatial axes, thereby reducing the complication of introducing a phase for $k^{+}$. Then making the following change of variables $z =e^{i \theta}$, \ref{ang1} takes the form of a contour integral
\begin{eqnarray}\nonumber
\int_{0}^{2\pi} d\theta \hspace{0.1cm} \frac{1}{(\ell-k)^{+}}
&=&\oint_{\Gamma}  \frac{dz}{i \hspace{0.03cm} z} \hspace{0.1cm} \frac{\sqrt{2}}{\ell_{s} \left(z-\frac{k_{s}}{\ell_{s}}\right)}
\end{eqnarray}

where the contour $\Gamma$ is the unit circle $|z|=1$ in the complex plane. Performing the integral using Cauchy's residue theorem, we can easily see that the above reduces to 
\begin{eqnarray}
\label{ang1final}
\int_{0}^{2\pi} d\theta \hspace{0.1cm} \frac{1}{(\ell-k)^{+}}=- \frac{2\pi}{k^{+}} \Theta(k_s-l_s)
\end{eqnarray}

Now similar manipulations yield the following general result.
\begin{equation}
\label{anggen}
\begin{split}
 \int _{0}^{2\pi} d\theta \frac{(\ell^{+})^{s}}{(\ell-k)^{+}} = 2\pi (k^{+})^{s-1} \Theta (\ell_{s}-k_{s})~~;~~\text{where}~~ s \in {\mathbf{Z}}^{+}
 \end{split}
\end{equation}

Then using equations \ref{ang1final} and \ref{anggen}, we find the following the results for the relevant angular integrals encountered in the main text. 
\begin{equation}
\label{ang2}
\begin{split}
\int _{0}^{2\pi} d\theta \hspace{0.1cm} \frac{(\ell^{+}+k^{+})}{(\ell^{+}-k^{+})}  = 2\pi \left [ \Theta\left(\ell_{s}-k_{s}\right)  -\Theta\left(k_{s}-\ell_{s}\right)  \right]
\end{split}
\end{equation} 

\begin{equation}
\label{ang3}
\begin{split}
\int_{0}^{2\pi} d\theta \hspace{0.1cm} \frac{(\ell+k)^{-}}{(\ell-k)^{+}} = - 2\pi \frac{k^{-}}{k^{+}}\left(1+\frac{\ell_{s}^{2}}{k_{s}^{2}}\right)\Theta(k_{s}-\ell_{s})
\end{split}
\end{equation}

\begin{equation}
\label{ang4}
\begin{split}
\int_{0}^{2\pi} d\theta_{1} \int_{0}^{2\pi} d\theta_{2} \hspace{0.1cm} \frac{1}{(\ell_{2}-k)^{+}(\ell_{1}-\ell_{2})^{+}} = \frac{(- 2\pi)^{2} }{(k^{+)^{2}}}\Theta(k_{s}-\ell_{s,2})\Theta(\ell_{s,2}-\ell_{s,1})
\end{split}
\end{equation}



\begin{equation}
\label{ang5}
\begin{split}
& \int_{0}^{2\pi} d\theta_{1} \int_{0}^{2\pi} d\theta_{2}\int_{0}^{2\pi} d\theta_{3} \hspace{0.1cm} \frac{(\ell_{1}+k)^{+}}{(\ell_{1}-k)^{+}(\ell_{1}-\ell_{2})^{+}(\ell_{3}-\ell_{1})^{+}}\\
& =\frac{2(2\pi)^{3}}{(k^{+})^{2}}\Theta(k_{s}-\ell_{s,1})\Theta(\ell_{s,1}-\ell_{s,2})\Theta(\ell_{s,1}-\ell_{s,3})
 \end{split}
\end{equation} 

\begin{equation}
\label{ang6}
\begin{split}
& \int_{0}^{2\pi} d\theta_{1} \int_{0}^{2\pi} d\theta_{2}\int_{0}^{2\pi} d\theta_{3} \hspace{0.1cm} \frac{(\ell_{1}+k)^{+}}{(\ell_{1}-k)^{+}(\ell_{2}-\ell_{3})^{+}(\ell_{3}-\ell_{1})^{+}}\\
& =\frac{2(-2\pi)^{3}}{(k^{+})^{2}}\Theta(k_{s}-\ell_{s,1})\Theta(\ell_{s,3}-\ell_{s,2})\Theta(\ell_{s,1}-\ell_{s,3})
 \end{split}
\end{equation} 

\subsection{Matsubara Sums and Holonomy Integrals}
\label{msumholint}

In this section of the Appendix we collect some results for the Matsubara sums and thermal holonomy integrals which feature in our analysis of finite temperature correlators. Note that the following discussion will be for general $q$, where $q$ is external momentum flowing through the current vertices. The results used in the main text of the paper are then obtained by setting $q^{\pm}=0$ in the following results.

\subsubsection*{Fermionic Theories}

In the Chern Simons theories coupled to fundamental fermionic matter fields, the Matsubara sums that we encounter in the calculation of finite temperature vertex factors and $2$-point functions are as follows
\begin{equation}
\label{matsumf1}
\begin{split}
&  M_{F,1}=\sum_{n=-\infty}^{\infty} \frac{1}{\left((\tilde{\ell}-q)^{2}+\beta^{-2}\mu_{F}^{2}\right)\left(\tilde{\ell}^{2}+\beta^{-2}\mu_{F}^{2}\right)} 
\end{split}
\end{equation}

\begin{equation}
\label{matsumf2}
\begin{split}
&  M_{F,2}=\sum_{n=-\infty}^{\infty} \frac{\tilde{\ell}_{3}}{\left((\tilde{\ell}-q)^{2}+\beta^{-2}\mu_{F}^{2}\right)\left(\tilde{\ell}^{2}+\beta^{-2}\mu_{F}^{2}\right)} \\
\end{split}
\end{equation}

and
\begin{equation}
\label{matsumf3a}
\begin{split}
&  M_{F,3}=\sum_{n=-\infty}^{\infty} \frac{\tilde{\ell}_{3}(\tilde{\ell}_{3}-q_{3})}{\left((\tilde{\ell}-q)^{2}+\beta^{-2}\mu_{F}^{2}\right)\left(\tilde{\ell}^{2}+\beta^{-2}\mu_{F}^{2}\right)} \\
\end{split}
\end{equation}

where the sum with respect to $n$ is over half-integers due to anti-periodic boundary conditions for the fermions along the thermal circle. Now doing these sums using Mathematica we find
\begin{equation}
\label{matsumf3}
\begin{split}
 M_{F,1}& = \frac{\beta}{4}\frac{1}{\left[\left(q_{3}^{2}+q_{s}^{2}-2\vec{q}.\vec{\ell}\right)^{2}+4 q_{3}^{2}\left(\ell_{s}^{2}+m_{F}^{2}\right)\right]} \left[\frac{q_{3}^{2}+q_{s}^{2}-2\vec{q}.\vec{\ell}}{\sqrt{\ell_{s}^{2}+m_{F}^{2}}}S_{F,1}+\frac{q_{3}^{2}-q_{s}^{2}+2\vec{q}.\vec{\ell}}{\sqrt{(\vec{\ell}-\vec{q})^{2}+m_{F}^{2}}}S_{F,2}\right] \\
& + \frac{1}{2}\frac{i q_{3}\beta}{\left[\left(q_{3}^{2}+q_{s}^{2}-2\vec{q}.\vec{\ell}\right)^{2}+4 q_{3}^{2}\left(\ell_{s}^{2}+m_{F}^{2}\right)\right]} \left[S_{F,3}+S_{F,4}\right]
\end{split}
\end{equation}

\begin{equation}
\label{matsumf4}
\begin{split}
M_{F,2}& = \frac{i \beta}{4}\frac{1}{\left[\left(q_{3}^{2}+q_{s}^{2}-2\vec{q}.\vec{\ell}\right)^{2}+4 q_{3}^{2}\left(\ell_{s}^{2}+m_{F}^{2}\right)\right]}\left(q_{3}^{2}+q_{s}^{2}-2\vec{q}.\vec{\ell}\right)\left(S_{F,1}+S_{F,2}\right)\\
& + \frac{ q_{3}\beta}{2}\frac{\sqrt{\ell_{s}^{2}+m^{2}_{F}}\hspace{0.05cm}S_{F,3}}{\left[\left(q_{3}^{2}+q_{s}^{2}-2\vec{q}.\vec{\ell}\right)^{2}+4 q_{3}^{2}\left(\ell_{s}^{2}+m_{F}^{2}\right)\right]}\\
& + \frac{ q_{3}\beta}{4} \frac{1}{\left[\left(q_{3}^{2}+q_{s}^{2}-2\vec{q}.\vec{\ell}\right)^{2}+4 q_{3}^{2}\left(\ell_{s}^{2}+m_{F}^{2}\right)\right]}\frac{\left(q_{3}^{2}+q_{s}^{2}-2\vec{q}.\vec{\ell}+2\ell_{s}^{2}+2m^{2}_{F}\right)}{\sqrt{(\vec{\ell}-\vec{q})^{2}+m^{2}_{F}}}S_{F,4}
\end{split}
\end{equation}

and, 
\begin{equation}
\label{matsumf5}
\begin{split}
  M_{F,3}& 
 = \frac{1}{4} \frac{\left(q_{3}^{2}+q_{s}^{2}-2\vec{q}.\vec{\ell}\right)}{\left[\left(q_{3}^{2}+q_{s}^{2}-2\vec{q}.\vec{\ell}\right)^{2}+4 q_{3}^{2}\left(\ell_{s}^{2}+m_{F}^{2}\right)\right]} \left(\sqrt{x^{2}+\mu_{F}^{2}}\hspace{0.04cm}S_{F,1}+\sqrt{z^{2}+\mu_{F}^{2}}\hspace{0.04cm}S_{F,2}\right)\\
& -\frac{i q_{3}\beta}{4} \frac{\left(q_{3}^{2}+q_{s}^{2}-2\vec{q}.\vec{\ell}+2\ell_{s}^{2}+2m^{2}_{F}\right)}{\left[\left(q_{3}^{2}+q_{s}^{2}-2\vec{q}.\vec{\ell}\right)^{2}+4 q_{3}^{2}\left(\ell_{s}^{2}+m_{F}^{2}\right)\right]} (S_{F,3}+S_{F,4})
\end{split}
\end{equation}

where $m_{F}=\beta^{-1}\mu_{F}$ and $S_{F,i}$ with $i\in (1,2,3,4)$ appearing in equations \ref{matsumf3}, \ref{matsumf4} and \ref{matsumf5} are given by 
\begin{equation}
\label{matsumf6}
\begin{split}
&  S_{F,1}= \tanh\left[\frac{1}{2}\sqrt{x^{2}+\mu_{F}^{2}}+i \pi |\lambda_{F}|u \right] + \text{c.c} ,\hspace{0.6cm} S_{F,2}= \tanh\left[\frac{1}{2}\sqrt{z^{2}+\mu_{F}^{2}}+i \pi |\lambda_{F}|u \right] + \text{c..c}\\ 
& S_{F,3}= \tanh\left[\frac{1}{2}\sqrt{x^{2}+\mu_{F}^{2}}+i \pi |\lambda_{F}|u \right] - \text{c.c}, \hspace{0.6cm} S_{F,4}=\tanh\left[\frac{1}{2}\sqrt{z^{2}+\mu_{F}^{2}}+i \pi |\lambda_{F}|u \right] - \text{c.c}       
\end{split}
\end{equation}

with $x= \beta \ell_{s}, \hspace{0.04cm} z=\beta\sqrt{(\vec{\ell}-\vec{q})^{2}}$. Now we turn to the holonomy integrals which in this case are as follows, 
\begin{equation}
\label{matsumf7}
\begin{split}
\mathcal{M}_{F,1}&=\int_{-1/2}^{1/2} du \hspace{0.1cm} M_{F,1}\\
& =  \frac{\beta}{4}\frac{1}{\left[\left(q_{3}^{2}+q_{s}^{2}-2\vec{q}.\vec{\ell}\right)^{2}+4 q_{3}^{2}\left(\ell_{s}^{2}+m_{F}^{2}\right)\right]} \left[\frac{q_{3}^{2}+q_{s}^{2}-2\vec{q}.\vec{\ell}}{\sqrt{\ell_{s}^{2}+m_{F}^{2}}}h(x,\mu_{F})+\frac{q_{3}^{2}-q_{s}^{2}+2\vec{q}.\vec{\ell}}{\sqrt{(\vec{\ell}-\vec{q})^{2}+m_{F}^{2}}}h(z,\mu_{F})\right] 
\end{split}
\end{equation}

\begin{equation}
\label{matsumf8}
\begin{split}
\mathcal{M}_{F,2}=\int_{-1/2}^{1/2} du \hspace{0.1cm} M_{F,2} = \frac{i \beta}{4}\frac{\left(q_{3}^{2}+q_{s}^{2}-2\vec{q}.\vec{\ell}\right)\left(h(x,\mu_{F})+h(z,\mu_{F})\right)}{\left[\left(q_{3}^{2}+q_{s}^{2}-2\vec{q}.\vec{\ell}\right)^{2}+4 q_{3}^{2}\left(\ell_{s}^{2}+m_{F}^{2}\right)\right]}
\end{split}
\end{equation}

\begin{equation}
\label{matsumf9}
\begin{split}
\mathcal{M}_{F,3}&=\int_{-1/2}^{1/2} du \hspace{0.1cm} M_{F,3} \\
&= \frac{1}{4}\frac{\left(q_{3}^{2}+q_{s}^{2}-2\vec{q}.\vec{\ell}\right)\left(\sqrt{x^{2}+\mu_{F}^{2}}\hspace{0.04cm}h(x,\mu_{F})+\sqrt{z^{2}+\mu_{F}^{2}}\hspace{0.04cm}h(z,\mu_{F})\right)}{\left[\left(q_{3}^{2}+q_{s}^{2}-2\vec{q}.\vec{\ell}\right)^{2}+4 q_{3}^{2}\left(\ell_{s}^{2}+m_{F}^{2}\right)\right]}
\end{split}
\end{equation} 

where $h(x,\mu_{F})$ is given by
\begin{equation}
\label{matsumf10}
\begin{split}
h(x,\mu_{F})& =\int_{-1/2}^{1/2} du \hspace{0.05cm}\left[\tanh\left(\frac{1}{2}\sqrt{x^{2}+\mu_{F}^{2}}-i \pi|\lambda_{F}|u\right)+\mathrm{c}.\mathrm{c}\right] \\
&=\frac{2 i}{|\lambda_{F}|\pi} \left[\log\left(\cosh\left(\frac{1}{2}\sqrt{x^{2}+\mu_{F}^{2}}-\frac{i \pi |\lambda_{F}|}{2}\right)\right)-\mathrm{c}.\mathrm{c}\right]
\end{split}
\end{equation}

\subsubsection*{Bosonic Theories}

In case of the Chern Simons theories coupled to fundamental bosons we have the same type of Matsubara sums as in equations \ref{matsumf1}, \ref{matsumf2}, \ref{matsumf3}. The only difference now is that the sum with respect to $n$ runs over integers due to periodic boundary conditions for the bosons around the thermal circle. Also, we have $\mu_{B}$, the bosonic theory's thermal mass instead of $\mu_{F}$. Then doing the Matsubara sums using Mathematica we get
\begin{equation}
\label{matsumb1}
\begin{split}
  M_{B,1} & =\sum_{n=-\infty}^{\infty} \frac{1}{\left((\tilde{\ell}-q)^{2}+\beta^{-2}\mu_{B}^{2}\right)\left(\tilde{\ell}^{2}+\beta^{-2}\mu_{B}^{2}\right)} \\
& =\frac{\beta}{4}\frac{1}{\left[\left(q_{3}^{2}+q_{s}^{2}-2\vec{q}.\vec{\ell}\right)^{2}+4 q_{3}^{2}\left(\ell_{s}^{2}+m_{B}^{2}\right)\right]} \left[\frac{q_{3}^{2}+q_{s}^{2}-2\vec{q}.\vec{\ell}}{\sqrt{\ell_{s}^{2}+m_{B}^{2}}}S_{B,1}+\frac{q_{3}^{2}-q_{s}^{2}+2\vec{q}.\vec{\ell}}{\sqrt{(\vec{\ell}-\vec{q})^{2}+m_{B}^{2}}}S_{B,2}\right] \\
& + \frac{1}{2}\frac{i q_{3}\beta}{\left[\left(q_{3}^{2}+q_{s}^{2}-2\vec{q}.\vec{\ell}\right)^{2}+4 q_{3}^{2}\left(\ell_{s}^{2}+m_{B}^{2}\right)\right]} \left[S_{B,3}+S_{B,4}\right]
\end{split}
\end{equation}

\begin{equation}
\label{matsumb2}
\begin{split}
M_{B,2}& =\sum_{n=-\infty}^{\infty} \frac{\tilde{\ell}_{3}}{\left((\tilde{\ell}-q)^{2}+\beta^{-2}\mu_{B}^{2}\right)\left(\tilde{\ell}^{2}+\beta^{-2}\mu_{B}^{2}\right)} \\
& = \frac{i \beta}{4}\frac{1}{\left[\left(q_{3}^{2}+q_{s}^{2}-2\vec{q}.\vec{\ell}\right)^{2}+4 q_{3}^{2}\left(\ell_{s}^{2}+m_{F}^{2}\right)\right]}\left(q_{3}^{2}+q_{s}^{2}-2\vec{q}.\vec{\ell}\right)\left(S_{B,1}+S_{B,2}\right)\\
& + \frac{ q_{3}\beta}{2}\frac{\sqrt{\ell_{s}^{2}+m^{2}_{B}}\hspace{0.05cm}S_{B,3}}{\left[\left(q_{3}^{2}+q_{s}^{2}-2\vec{q}.\vec{\ell}\right)^{2}+4 q_{3}^{2}\left(\ell_{s}^{2}+m_{B}^{2}\right)\right]}\\
& + \frac{ q_{3}\beta}{4} \frac{1}{\left[\left(q_{3}^{2}+q_{s}^{2}-2\vec{q}.\vec{\ell}\right)^{2}+4 q_{3}^{2}\left(\ell_{s}^{2}+m_{B}^{2}\right)\right]}\frac{\left(q_{3}^{2}+q_{s}^{2}-2\vec{q}.\vec{\ell}+2\ell_{s}^{2}+2m^{2}_{B}\right)}{\sqrt{(\vec{\ell}-\vec{q})^{2}+m^{2}_{B}}}S_{B,4}
\end{split}
\end{equation}

and
\begin{equation}
\label{matsumb3}
\begin{split}
 M_{B,3} & =\sum_{n=-\infty}^{\infty} \frac{\tilde{\ell}_{3}(\tilde{\ell}_{3}-q_{3})}{\left((\tilde{\ell}-q)^{2}+\beta^{-2}\mu_{B}^{2}\right)\left(\tilde{\ell}^{2}+\beta^{-2}\mu_{B}^{2}\right)} \\
&= \frac{1}{4} \frac{\left(q_{3}^{2}+q_{s}^{2}-2\vec{q}.\vec{\ell}\right)}{\left[\left(q_{3}^{2}+q_{s}^{2}-2\vec{q}.\vec{\ell}\right)^{2}+4 q_{3}^{2}\left(\ell_{s}^{2}+m_{B}^{2}\right)\right]} \left(\sqrt{x^{2}+\mu_{B}^{2}}\hspace{0.04cm}S_{B,1}+\sqrt{z^{2}+\mu_{B}^{2}}\hspace{0.04cm}S_{B,2}\right)\\
& -\frac{i q_{3}\beta}{4} \frac{\left(q_{3}^{2}+q_{s}^{2}-2\vec{q}.\vec{\ell}+2\ell_{s}^{2}+2m^{2}_{F}\right)}{\left[\left(q_{3}^{2}+q_{s}^{2}-2\vec{q}.\vec{\ell}\right)^{2}+4 q_{3}^{2}\left(\ell_{s}^{2}+m_{B}^{2}\right)\right]} (S_{B,3}+S_{B,4})
\end{split}
\end{equation}

where $m_{B}=\beta^{-1}\mu_{B}$ and $S_{B,i}$ with $i\in (1,2,3,4)$  in equations \ref{matsumb1}, \ref{matsumb2} and \ref{matsumb3} are given by 
\begin{equation}
\label{matsumb4}
\begin{split}
&  S_{B,1}= \coth\left[\frac{1}{2}\sqrt{x^{2}+\mu_{B}^{2}}+i \pi |\lambda_{B}|u \right] + \text{c.c} ,\hspace{0.6cm} S_{B,2}= \coth\left[\frac{1}{2}\sqrt{z^{2}+\mu_{B}^{2}}+i \pi |\lambda_{B}|u \right] + \text{c..c}\\ 
& S_{B,3}= \coth\left[\frac{1}{2}\sqrt{x^{2}+\mu_{B}^{2}}+i \pi |\lambda_{B}|u \right] - \text{c.c}, \hspace{0.6cm} S_{B,4}=\coth\left[\frac{1}{2}\sqrt{z^{2}+\mu_{B}^{2}}+i \pi |\lambda_{B}|u \right] - \text{c.c}       
\end{split}
\end{equation}

with $x= \beta \ell_{s}, \hspace{0.04cm} z=\beta\sqrt{(\vec{\ell}-\vec{q})^{2}}$. Next, the holonomy integrals can be straightforwadly done to give 
\begin{equation}
\label{matsumb5}
\begin{split}
\mathcal{M}_{B,1}&=\int_{-1/2}^{1/2} du \hspace{0.1cm} M_{B,1}\\
& =  \frac{\beta}{4}\frac{1}{\left[\left(q_{3}^{2}+q_{s}^{2}-2\vec{q}.\vec{\ell}\right)^{2}+4 q_{3}^{2}\left(\ell_{s}^{2}+m_{B}^{2}\right)\right]} \left[\frac{q_{3}^{2}+q_{s}^{2}-2\vec{q}.\vec{\ell}}{\sqrt{\ell_{s}^{2}+m_{B}^{2}}}h(x,\mu_{B})+\frac{q_{3}^{2}-q_{s}^{2}+2\vec{q}.\vec{\ell}}{\sqrt{(\vec{\ell}-\vec{q})^{2}+m_{B}^{2}}}h(z,\mu_{B})\right] 
\end{split}
\end{equation}

\begin{equation}
\label{matsumb6}
\begin{split}
\mathcal{M}_{B, 2}=\int_{-1/2}^{1/2} du \hspace{0.1cm} M_{B, 2} = \frac{i \beta}{4}\frac{\left(q_{3}^{2}+q_{s}^{2}-2\vec{q}.\vec{\ell}\right)\left(h(x,\mu_{B})+h(z,\mu_{B})\right)}{\left[\left(q_{3}^{2}+q_{s}^{2}-2\vec{q}.\vec{\ell}\right)^{2}+4 q_{3}^{2}\left(\ell_{s}^{2}+m_{B}^{2}\right)\right]}
\end{split}
\end{equation}

\begin{equation}
\label{matsumb7}
\begin{split}
\mathcal{M}_{B, 3}&=\int_{-1/2}^{1/2} du \hspace{0.1cm} M_{B, 3} \\
&= \frac{1}{4}\frac{\left(q_{3}^{2}+q_{s}^{2}-2\vec{q}.\vec{\ell}\right)\left(\sqrt{x^{2}+\mu_{B}^{2}}\hspace{0.04cm}h(x,\mu_{B})+\sqrt{z^{2}+\mu_{B}^{2}}\hspace{0.04cm}h(z,\mu_{B})\right)}{\left[\left(q_{3}^{2}+q_{s}^{2}-2\vec{q}.\vec{\ell}\right)^{2}+4 q_{3}^{2}\left(\ell_{s}^{2}+m_{B}^{2}\right)\right]}
\end{split}
\end{equation} 

In the above equations, $h(x,\mu_{B})$ is given by
\begin{equation}
\label{matsumb8}
\begin{split}
h(x,\mu_{B})& =\int_{-1/2}^{1/2} du \hspace{0.05cm}\left[\coth\left(\frac{1}{2}\sqrt{x^{2}+\mu_{B}^{2}}-i \pi|\lambda_{B}|u\right)+\mathrm{c}.\mathrm{c}\right] \\
&=\frac{2 i}{|\lambda_{B}|\pi} \left[\log\left(\sinh\left(\frac{1}{2}\sqrt{x^{2}+\mu_{B}^{2}}-\frac{i \pi |\lambda_{B}|}{2}\right)\right)-\mathrm{c}.\mathrm{c}\right]
\end{split}
\end{equation}


\subsection{Evaluating $\mathcal{F}_{B}(q_{3}\beta,\mu_{B})$ and $\mathcal{F}_{F}(q_{3}\beta,\mu_{F})$}
\label{fintegcalc}

The results for thermal correlation functions obtained in this paper involve the function $\mathcal{F}_{B}(q_{3}\beta,\mu_{B})$ and $\mathcal{F}_{F}(q_{3}\beta,\mu_{F}) $ in the bosonic and fermionic theories respectively. These admit integral expressions which are given in equations \ref{j0sdyson10} and \ref{j0fsdyson18}. In this section of the Appendix we will show that these integrals can indeed be explicitly done. We choose to focus on the bosonic case here. An entirely analogous result holds in the fermion case as well. \\

We have, 
\begin{equation}
\label{finteg}
\begin{split}
\mathcal{F}_{B}(q_{3}\beta,\mu_{B}) & = \int_{0}^{\infty} dx \hspace{0.1cm} x \hspace{0.05cm}\mathcal{H}_{B}(x, q_{3}\beta,\mu_{B}) \\
& =  \frac{ 4 i }{\pi} \hspace{0.03cm} \mathrm{sgn}(\lambda_{B}) \int_{0}^{\infty} dx \hspace{0.1cm} x \hspace{0.05cm} \frac{\left[  \log\left(\sinh\left(\frac{1}{2}\sqrt{x^{2}+\mu_{B}^{2}}-\frac{i\pi|\lambda_{B}|}{2}\right)\right)- \mathrm{c.c}  \right]}{\sqrt{x^{2}+\mu_{B}^{2}}\left(4 x^{2}+4 \mu_{B}^{2}+ q_{3}^{2}\beta^{2}\right)}
\end{split}
\end{equation}

Using Schwinger parametrisation we can write,
\begin{equation}
\label{finteg1}
\begin{split}
\frac{1}{4 x^{2}+4 \mu_{B}^{2}+q_{3}^{2}\beta^{2}}= \int_{0}^{\infty} du \hspace{0.2cm} e^{-\left(4 x^{2}+4 \mu_{B}^{2}+q_{3}^{2}\beta^{2}\right)u}
\end{split}
\end{equation}

Plugging this into \ref{finteg} we get, 
\begin{equation}
\label{finteg2}
\begin{split}
\mathcal{F}_{B}(q_{3}\beta,\mu_{B})  = \int_{0}^{\infty} du \hspace{0.2cm} e^{-\left(4\mu_{B}^{2}+q^{2}\beta^{2}\right)u} \hspace{0.1cm} \mathcal{I}\left(u, \mu_{B}\right)
\end{split}
\end{equation}

where we have defined
\begin{equation}
\label{finteg3}
\begin{split}
 \mathcal{I}\left(u, \mu_{B}\right) = \int_{0}^{\infty} dx \hspace{0.1cm} e^{-4 x^{2} u}\hspace{0.05cm} x \hspace{0.04cm} \frac{h(x,\mu_{B})}{\sqrt{x^{2}+\mu_{B}^{2}}}
\end{split}
\end{equation}

and
\begin{equation}
\label{finteg4}
\begin{split}
h(x,\mu_{B})= \frac{ 4 i }{\pi} \hspace{0.03cm} \mathrm{sgn}(\lambda_{B})  \left[  \log\left(\sinh\left(\frac{1}{2}\sqrt{x^{2}+\mu_{B}^{2}}-\frac{i\pi|\lambda_{B}|}{2}\right)\right)- \mathrm{c.c}  \right]
\end{split}
\end{equation}

Now we can do the $u$-integral in \ref{finteg2} by Taylor expanding $\mathcal{I}\left(u, \mu_{B}\right)$ around $u=0$. But this is a bit problematic because in this case the Taylor series coefficients will contain divergent integrals coming from expanding \ref{finteg3} around $u=0$. To avoid this issue let us proceed as follows.

Note that $h(x,\mu_{B})$ satisfies the following relation
\begin{equation}
\label{finteg5}
\begin{split}
h(x,\mu_{B})= \frac{ 4 i }{\pi} \hspace{0.03cm} \mathrm{sgn}(\lambda_{B})  \left[-i\pi |\lambda_{B}| +\frac{\sqrt{x^{2}+\mu_{B}^{2}}}{x}\hspace{0.05cm} \frac{dg(x,\mu_{B})}{d x} \right]
\end{split}
\end{equation}

where
\begin{equation}
\label{finteg6}
\begin{split}
g(x, \mu_{B})= \text{Li}_{2}\left(e^{-\sqrt{x^{2}+\mu_{B}^{2}}+i \pi |\lambda_{B}|}\right)- \text{c.c}
\end{split}
\end{equation}

Using the above in equation \ref{finteg3} we then obtain, 
\begin{equation}
\label{finteg7a}
\begin{split}
 \mathcal{I}\left(u, \mu_{B}\right) 
 & = \sum_{j=1}^{3} \mathcal{I}_{j}\left(u, \mu_{B}\right) 
\end{split}
\end{equation}

where,
\begin{equation}
\label{finteg7}
\begin{split}
\mathcal{I}_{1}\left(u, \mu_{B}\right) & = 4 \lambda_{B}  \int_{0}^{\infty} dx \hspace{0.2cm} \frac{x \hspace{0.04cm}e^{-4 x^{2} u} }{\sqrt{x^{2}+\mu_{B}^{2}}}=    \pi^{1/2} \lambda_{B} \hspace{0.05cm}\frac{e^{4 \hspace{0.04cm}u \mu_{B}^{2}}}{ u^{1/2}} \left(-1+ \text{Erfc}\left(2 \mu_{B} \sqrt{u} \right) \right)
\end{split}
\end{equation}

with $\text{Erfc}(x)$ denoting the error function. 

\begin{equation}
\label{finteg8}
\begin{split}
\mathcal{I}_{2}\left(u,\mu_{B}\right) & = \frac{4 i}{\pi}  \hspace{0.03cm} \mathrm{sgn}(\lambda_{B}) \left(g(x,\mu_{B}) e^{-4 x^{2}u}\right)_{0}^{\infty} = \frac{4 i}{\pi}  \hspace{0.03cm} \mathrm{sgn}(\lambda_{B})\left[\text{Li}_{2}\left(e^{-\mu_{B}-i \pi |\lambda_{B}|}\right)- \text{c.c}\right]
\end{split}
\end{equation}

\begin{equation}
\label{finteg9}
\begin{split}
\mathcal{I}_{3}\left(u,\mu_{B}\right) & =  \frac{16 \hspace{0.04cm}i}{\pi}  \hspace{0.03cm} \mathrm{sgn}(\lambda_{B}) u \int_{0}^{\infty} dx \hspace{0.2cm} x \hspace{0.04cm}g(x,m_{th}) \hspace{0.04cm}e^{-4 x^{2} u} \\
& =\frac{16 \hspace{0.04cm} i}{\pi}  \hspace{0.03cm} \mathrm{sgn}(\lambda_{B}) \sum_{n=0}^{\infty} \frac{(-4)^{n} u^{n+1}}{n!} \int_{0}^{\infty} dx \hspace{0.1cm} x^{2n+1} \hspace{0.04cm}g(x,\mu_{B}) \\
\end{split}
\end{equation}

where in the last line of the integral expression for $\mathcal{I}_{3}$ we have Taylor expanded around $u=0$. Since $g(x,\mu_{B})$ goes to zero exponentially fast for large $x$, the last integral in \ref{finteg9} is UV finite. Now we can do the $u$-integral in \ref{finteg2} to get,
\begin{equation}
\label{finteg10}
\begin{split}
\mathcal{F}_{B}(q_{3}\beta,\mu_{B}) = \sum_{i=1}^{3} \mathcal{A}_{i}(q_{3} \beta, \mu_{B})
\end{split}
\end{equation}

where,
\begin{equation}
\label{finteg11}
\begin{split}
\mathcal{A}_{1}(q_{3} \beta, \mu_{B})& = \int_{0}^{\infty} du \hspace{0.1cm} e^{-\left(4 \hspace{0.05cm}\mu_{B}^{2}+q_{3}^{2}\beta^{2}\right)u} \hspace{0.1cm} \mathcal{I}_{1}\left(u,\mu_{B}\right) = \frac{\lambda_{B}}{q_{3}\beta} \left(\pi- 2\text{arctan}\left(\frac{2 \mu_{B}}{q_{3}\beta}\right)\right)
\end{split}
\end{equation}

\begin{equation}
\label{finteg12}
\begin{split}
\mathcal{A}_{2}(q_{3} \beta, \mu_{B})& = \int_{0}^{\infty} du \hspace{0.1cm} e^{-\left(4 \hspace{0.05cm}\mu_{B}^{2}+q_{3}^{2}\beta^{2}\right)u} \hspace{0.1cm} \mathcal{I}_{2}\left(u,\mu_{B}\right) = \frac{4 i}{\pi}  \hspace{0.03cm} \mathrm{sgn}(\lambda_{B})\frac{\left[\text{Li}_{2}\left(e^{-\mu_{B}+i \pi |\lambda_{B}|}\right)- \text{c.c}\right]}{\left(q_{3}^{2}\beta^{2}+4 \hspace{0.05cm}\mu_{B}^{2}\right)}
\end{split}
\end{equation}

\begin{equation}
\label{finteg13}
\begin{split}
\mathcal{A}_{3}(q_{3} \beta, \mu_{B})& = \int_{0}^{\infty} du \hspace{0.1cm} e^{-\left(4 \hspace{0.05cm}\mu_{B}^{2}+q_{3}^{2}\beta^{2}\right)u} \hspace{0.1cm} \mathcal{I}_{3}\left(u,\mu_{B}\right) = \frac{16 \hspace{0.04cm} i}{\pi}  \hspace{0.03cm} \mathrm{sgn}(\lambda_{B})  \sum_{n=0}^{\infty} \frac{c_{n}}{\left(q_{3}^{2}\beta^{2}+4 \hspace{0.05cm}\mu_{B}^{2}\right)^{n+2}}
\end{split}
\end{equation}

and in the last expression, the coefficients $c_{n}$ are given by UV finite integrals of the following form,
\begin{equation}
\label{finteg14}
\begin{split}
c_{n} = (-4)^{n}(n+1) \int_{0}^{\infty} dx \hspace{0.1cm} x^{2n+1} \hspace{0.04cm}g(x,\mu_{B}) 
\end{split}
\end{equation}


\section{Thermal $2$-point Functions: Spin $s=2$} 
\label{s2}

In this appendix we compute the thermal two point functions of the stress tensor operator in the Regular snd Critical Boson theories as well as the Regular and Critical Fermion theories.

\subsection{Regular Bosons}
\label{s2rb}

The stress tensor operator in the Regular Boson theory is given by
\begin{equation}
\label{strb1}
\begin{split}
J^{(2)}_{B, \mu\nu} (x)= \phi^{\dagger}(x) \left[ \frac{3}{2} \overset{\leftarrow}{D_{(\mu}}\overset{\rightarrow}{D_{\nu)}} - \frac{1}{4}  \overset{\rightarrow}{D_{(\mu}}\overset{\rightarrow}{D_{\nu)}}- \frac{1}{4}\overset{\leftarrow}{D_{(\mu}}\overset{\leftarrow}{D_{\nu)}}\right] \phi(x) +g_{\mu\nu} \tilde{T}^{\sigma}_{\sigma}(x)
\end{split}
\end{equation}

where $A_{(\mu}B_{\nu)} =\frac{1}{2}\left(A_{\mu}B_{\nu}+A_{\nu}B_{\mu}\right)$ and 
\begin{equation}
\label{strb2}
\begin{split}
\tilde{T}^{\mu}_{\mu}(x) & = -\frac{1}{3}  \phi^{\dagger}(x) \left[ \frac{3}{2} \overset{\leftarrow}{D^{(\mu}}\overset{\rightarrow}{D_{\mu)}} - \frac{1}{4}  \overset{\rightarrow}{D^{(\mu}}\overset{\rightarrow}{D_{\mu)}}- \frac{1}{4}\overset{\leftarrow}{D^{(\mu}}\overset{\leftarrow}{D_{\mu)}}\right] \phi(x) \\
\end{split}
\end{equation}

The presence of this term ensures that $J^{(2)}_{B,\mu\nu}$ is traceless. In the case of the $U(1)$ current correlators considered in the previous sections, we had focussed only particular components of the currents. Here we will do the same and restrict attention to the $2$-point function where the two current insertions have components $(\mu,\nu)=(--)$ and $(\mu,\nu)=(+,+)$ respectively. Now for these specific components of the stress tensor the trace term $g_{\mu\nu} \tilde{T}^{\sigma}_{\sigma}$ does not contribute  since $g_{--}=0$. Then in lightcone gauge where $A_{-}=0$ we have

\begin{equation}
\label{strb3}
\begin{split}
J^{(2)}_{B,--}(x) &= \phi^{\dagger}(x) \left[ \frac{3}{2} \overset{\leftarrow}{\partial_{(-}}\overset{\rightarrow}{\partial_{-)}} - \frac{1}{4}  \overset{\rightarrow}{\partial_{(-}}\overset{\rightarrow}{\partial_{-)}}- \frac{1}{4}\overset{\leftarrow}{\partial_{(-}}\overset{\leftarrow}{\partial_{-)}}\right] \phi(x)  = \frac{3}{2} \partial_{-}\phi^{\dagger} \partial_{-}\phi - \frac{1}{4}\phi^{\dagger} \partial^{2}_{-}\phi - \frac{1}{4} \left(\partial^{2}_{-}\phi^{\dagger}\right)\phi  
\end{split}
\end{equation}
\begin{equation}
\label{strb4}
\begin{split}
J^{(2)}_{B,++}(x)& = \frac{3}{2} (D_{+}\phi)^{\dagger} (D_{+}\phi) -\frac{1}{4} \phi^{\dagger} D_{+} (D_{+}\phi)- \frac{1}{4} (D_{+}(D_{+}\phi))^{\dagger}\phi \\
& = \frac{3}{2} \partial_{+}\phi^{\dagger} \partial_{+}\phi - \frac{1}{4} \phi^{\dagger} \partial^{2}_{+}\phi -\frac{1}{4}( \partial ^{2}_{+} \phi)^{\dagger} \phi  + 2( \partial_{+} \phi)^{\dagger} A_{+}\phi -2 \phi^{\dagger} A_{+}\partial_{+}\phi  -2 \phi^{\dagger} A^{2}_{+} \phi
\end{split}
\end{equation}

In order to obtain Feynman rules for the current vertices, it is useful to have the momentum space expressions for the above currents. Further specialising to the kinematic regime where $q^{\pm}=0$ we get  
\begin{equation}
\label{strb5}
\begin{split}
J^{(2)}_{B,--}(q_{3}) &= \frac{1}{\beta} \sum_{n} \int \frac{d^{2}\vec{k}}{(2\pi)^{2}} \hspace{0.1cm} \phi^{\dagger}(k_{3}, \vec{k})\hspace{0.03cm}2 k^{2}_{-}\hspace{0.03cm}\phi(q_{3}-k_{3},-\vec{k}) 
\end{split}
\end{equation}

\begin{equation}
\label{strb6}
\begin{split}
J^{(2)}_{B,++} (q_{3}) &= \frac{1}{\beta} \sum_{n} \int \frac{d^{2}\vec{k}}{(2\pi)^{2}} \hspace{0.1cm} 2 \hspace{0.04cm}k^{2}_{+} \phi^{\dagger}(k_{3}, \vec{k})\phi(q_{3}-k_{3},-\vec{k}) \\
& +  \frac{1}{\beta^{2}} \sum_{n_{1},n_{2}} \int \frac{d^{2}\vec{k}\hspace{0.03cm} d^{2}\vec{p}}{(2\pi)^{4}} \hspace{0.1cm} (2 i )  (2k+p)_{+}\phi^{\dagger}(k_{3},\vec{k}) A_{+}(p_{3},\vec{p}) \phi(q_{3}-k_{3}-p_{3},-\vec{k}-\vec{p}) \\
& +  \frac{1}{\beta^{3}} \sum_{n_{1},n_{2},n_{3}} \int \frac{d^{2}\vec{k}\hspace{0.03cm} d^{2}\vec{p}\hspace{0.03cm} d^{2}\vec{\ell}}{(2\pi)^{6}} \hspace{0.1cm}  (-2)\phi^{\dagger}(k_{3}, \vec{k}) A_{+}(p_{3},\vec{p}) A_{+}(\ell_{3},\vec{\ell}) \phi(q_{3}-k_{3}-p_{3}-\ell_{3},-\vec{k}-\vec{p}-\vec{\ell}) \\
\end{split}
\end{equation} 

where $n_{1} =\frac{\beta \hspace{0.04cm} k_{3}}{2\pi},  n_{2} =\frac{\beta \hspace{0.04cm} p_{3}}{2\pi}, n_{3} =\frac{\beta \hspace{0.04cm} \ell_{3}}{2\pi}$. 


\subsubsection{Finite temperature vertex factor for $J^{(2)}_{B,--}$ } 

We now proceed towards calculating the exact vertex factor for $J^{(2)}_{B,--}$ which we will denote by $V^{(2)}_{B}(q,k)$ as follows, 
\begin{equation}
\label{tmm}
\begin{split}
\left\langle J^{(2)}_{B,--}(-q) \phi_{i}(p) (\phi^{\dagger})^{j}(-k)\right\rangle_{\beta} = V^{(2)}_{B}(q,k) \delta^{j}_{i} (2\pi)^{3} \delta_{n_{q}+n_{k},n_{p}}\delta^{(2)}(\vec{q}+\vec{k}-\vec{p}) \\
\end{split}
\end{equation}

We again restrict ourselves to the kinematic configuration $q^{\pm}=0$. Then noting from equation \ref{strb5} that  the bare vertex factor is given by $V^{(2)}_{B,free}(q,k) = 2k^{2}_{-}$, we can write down the Schwinger-Dyson equation for the vertex factor 
\begin{equation}
\label{tmm1}
\begin{split}
V^{(2)}_{B}(q,k) \delta_{i}^{j} &= 2 k_{-}^{2} \delta_{i}^{j}+ \int \mathcal{D}^{3}\ell \left[\mathcal{V}^{a,\mu} (k,\ell) G(\ell) V^{(2)}_{B}(q,\ell)  G(\ell+q) \mathcal{V}^{a,\nu}(\ell+q,k+q)\right]_{i}^{j}\mathcal{G}_{\nu\mu} (k-\ell) \\
& = 2(k^{+})^{2} \delta^{j}_{i}+ 4 \pi i \lambda_{B} q_{3} \hspace{0.05cm}\delta^{j}_{i}\int \frac{d^{2}\ell}{(2\pi)^{2}}\hspace{0.1cm} \frac{(\ell^{+}+k^{+})}{(\ell^{+}-k^{+})} \hspace{0.04cm}V^{(2)}_{B}(q,\ell) H_{B}(x,q_{3}\beta, \mu_{B})
\end{split}
\end{equation}

where $H_{B}(x,q_{3}\beta, \mu_{B})$ was defined in equation \ref{j0sdyson4}. Now let us consider the following ansatz for solving the above equation
\begin{equation}
\label{tmm2}
\begin{split}
V^{(2)}_{B}(q,k) = 2 (k^{+})^{2} \hspace{0.03cm}v(q, y)
\end{split}
\end{equation} 

Using this in equation \ref{tmm1} then gives us
\begin{equation}
\label{tmm3}
\begin{split}
2 (k^{+})^{2} v(q, y) &= 2 (k^{+})^{2}  + 8  \pi i \lambda_{B} q_{3}  \hspace{0.05cm}\int \frac{d^{2}\ell}{(2\pi)^{2}}\hspace{0.1cm} \frac{(\ell^{+}+k^{+})}{(\ell^{+}-k^{+})} \hspace{0.04cm} (\ell^{+})^{2} v(q, x) H_{B}(x,q_{3}\beta, \mu_{B})
\end{split}
\end{equation}

The angular part of the integral in \ref{tmm3} can be easily carried using the following result from section \ref{angintegs} of the Appendix
\begin{equation}
\label{tmm5}
\begin{split}
\int_{0}^{2\pi} d\theta \hspace{0.1cm} \frac{(\ell^{+})^{2}(\ell^{+}+k^{+})}{(\ell^{+}-k^{+})} = 4\pi (k^{+})^{2} \Theta(\ell_{s}-k_{s})
\end{split}
\end{equation}

Thus we get
\begin{equation}
\label{tmm6}
\begin{split}
v(q, y) &= 1 + \frac{4 i \lambda_{B} q_{3}}{\beta^{2}}  \hspace{0.05cm}\int_{y}^{\infty} dx \hspace{0.1cm} x  \hspace{0.04cm} v(q, x) H_{B}(x,q_{3}\beta, \mu_{B})
\end{split}
\end{equation}

The solution of the above integral equation is given by
\begin{equation}
\label{tmm7}
\begin{split}
v(q, y) &= \exp \left[ i q_{3}\beta  \int_{y}^{\infty} dx \hspace{0.1cm} x  \hspace{0.04cm} \mathcal{H}_{B}(x,q_{3}\beta, \mu_{B}) \right]
\end{split}
\end{equation}

where $\mathcal{H}_{B}(x,q_{3}\beta, \mu_{B}) =\frac{4 \lambda_{B}}{\beta^{3}} H_{B}(x,q_{3}\beta, \mu_{B}) $. Finally the exact vertex factor $V^{(2)}_{B}(q,k) $ in the large $N_{B}$ limit is given by
\begin{equation}
\label{tmm8}
\begin{split}
V^{(2)}_{B}(q,k) = 2 (k^{+})^{2} v(q,y) = 2 (k^{+})^{2} e^{ i \hspace{0.03cm} q_{3}\beta \hspace{0.04cm}\mathcal{F}_{B}(x,q_{3}\beta, \mu_{B}) }
\end{split}
\end{equation}

with $\mathcal{F}_{B}(x,q_{3}\beta, \mu_{B})$ being expressed by equation \ref{j0sdyson10}.

\subsubsection{Finite temperature vertex factor for $J^{(2)}_{B,++}$ } 

We now turn our attention towards computing the finite temperature vertex factor corresponding to $J^{(2)}_{B,++}$. This can be defined as 
\begin{equation}
\label{tbpp}
\begin{split}
\left\langle J^{(2)}_{B,++}(-q) \phi_{i}(p) (\phi^{\dagger})^{j}(-k)\right\rangle_{\beta} =U^{(2)}_{B}(q,k) \delta^{j}_{i} (2\pi)^{3} \delta_{n_{q}+n_{k},n_{p}}\delta^{(2)}(\vec{q}+\vec{k}-\vec{p})\\
\end{split}
\end{equation}

Let us also note the Feynman rules for that we will require shortly for calculating $U^{(2)}_{B}(q,k) $. These are given by the following tree-level correlators
\begin{equation}
\label{tbppu1}
\begin{split}
&\left\langle J^{(2)}_{B,++}(-q)  \phi_{i}(p) (\phi^{\dagger})^{j}(-k)\right\rangle_{\beta} = \mathcal{U}_{1}(q,k) \delta^{j}_{i} (2\pi)^{3} \delta_{n_{q}+n_{k},n_{p}}\delta^{(2)}(\vec{q}+\vec{k}-\vec{p}) \\
\end{split}
\end{equation}
\begin{equation}
\label{tbppu2}
\begin{split}
& \left\langle J^{(2)}_{B,++}(-q) \phi_{i}(p)A_{3}^{a}(r) (\phi^{\dagger})^{j}(-k)\right\rangle_{\beta} = \mathcal{U}_{2}^{a}(q,p,k) \delta^{j}_{i} (2\pi)^{3}\delta_{n_{q}+n_{k},n_{p}}\delta^{(2)}(\vec{q}+\vec{k}-\vec{p})\\
\end{split}
\end{equation}
\begin{equation}
\label{tbppu3}
\begin{split}
& \left\langle J^{(2)}_{B,++}(-q)  \phi_{i}(p)A_{3}^{a}(r)A_{3}^{b}(s) (\phi^{\dagger})^{j}(-k)\right\rangle_{\beta} = \mathcal{U}^{ab}_{3}(q,p,r,k) \delta^{j}_{i} (2\pi)^{3} \delta_{n_{q}+n_{k},n_{p}+n_{r}}\delta^{(2)}(\vec{q}+\vec{k}-\vec{p}-\vec{r})
\end{split}
\end{equation}

Now from \ref{strb6} we can easily determine the above tree-level vertex factors and we get
\begin{equation}
\label{tbppu1ans}
\begin{split}
\mathcal{U}_{1}(q,k)= 2 k_{+}^{2} 
\end{split}
\end{equation} 
\begin{equation}
\label{tbppu2ans}
\begin{split}
\mathcal{U}^{a}_{2}(q,p,k) =- 2i(p+ k)_{+}T^{a} 
\end{split}
\end{equation} 
\begin{equation}
\label{tbppu3ans}
\begin{split}
\mathcal{U}^{ab}_{3}(q,p,r,k)=-2 \{T^{a},T^{b}\}
\end{split}
\end{equation} 

Another object that we will require is the following exact $3$-point vertex
\begin{equation}
\label{tbpp2}
\begin{split}
\left\langle  \phi_{i}(p)  A^{a,\mu}(-r)(\phi^{\dagger})^{j}(-k)\right\rangle = ( \mathcal{K}^{a,\mu}(k,p))^{j}_{ i} (2\pi)^{3}\delta_{n_{k}+n_{r},n_{p}}\delta^{(2)}(\vec{r}+\vec{k}-\vec{p})
\end{split}
\end{equation}

We already computed this in section \ref{u1boson} and this is given by
\begin{equation}
\label{tbpp3a}
\begin{split}
( \mathcal{K}^{a,\mu}(k,p))^{j}_{ i}=( \mathcal{V}^{a,\mu}(k,p))_{i}^{j} &  + \mathcal{C}(k,p) T^{a}_{ji}\delta^{\mu 3}
\end{split}
\end{equation}

where $( \mathcal{V}^{b,\mu}(k,p_{2}))_{i}^{j}= i \left(T^{a}\tilde{ p}_{1}^{\mu} + \tilde{p}_{2}^{\mu}T^{a}\right)_{i}^{j} $ and $\mathcal{C}(k,p) $ is given by
\begin{equation}
\label{cqks2}
\begin{split}
 & \mathcal{C}(k,p) = \mathcal{X}(z) -\mathcal{X}(y)
\end{split}
\end{equation}

where $y=\beta k_{s}, z=\beta p_{s}$ and we have defined
\begin{equation}
\label{chidef}
\begin{split}
\mathcal{X}(y) = \frac{\lambda_{B}}{\beta} \left[\sqrt{y^{2}+\mu_{B}^{2}}+\frac{ i}{|\lambda_{B}|\pi} \left( \mathrm{Li}_{2}\left(e^{-\sqrt{y^{2}+\mu_{B}^{2}}+i \pi |\lambda|}\right)-\mathrm{c.c}\right)\right] 
\end{split}
\end{equation}



With the above ingredients in place we can now compute the vertex factor $U^{(2)}_{B}(q,k) $. As in the case of the $U(1)$ current, we will perform this computation by summing up a set of 1PI diagrams. The diagrams that we need to take into account have also been considered previously in the corresponding zero temperature calculations in \cite{Aharony:2012nh}. We simply need to consider the appropriate finite temperature generalisations thereof, i.e., replace all $T=0$ matter propagators with the corresponding exact thermal propagators.\\

First let us consider diagram labelled by $D_{1}(k,q,p) $ in Fig.\ref{s2rbD1}. Using the Feynman rules for the stress tensor vertex we get the following expression\footnote{We caution the reader that in section \ref{u1boson} we had also labelled the diagrams by symbols such as $D_{i}$, which are of course different from the ones being considered in this section.}  
\begin{equation}
\label{d1def}
\begin{split}
 D_{1} (k,q,p)  = & \int  \mathcal{D}^{3}\ell \left[  \mathcal{U}^{a}_{2}(q,k,\ell) \mathcal{G}_{3+}(p-\ell)G(\ell) \mathcal{K}^{a,3}(\ell,p)\right]_{i}^{j} + \text{reflection}
\end{split}
\end{equation} 

\begin{figure}[h!]
\begin{center}
\begin{tikzpicture}[scale=.5, transform shape]
 \draw[black,->] (-3.5,0) -- (-2,0);
 \draw[black,->] (-2,0) -- (0,0);
  \node at (0,0)[circle,fill,inner sep=5pt]{};
 \draw[black,->] (0,0) -- (1,0);
  \draw[black,->] (1,0) -- (3,0);
   \draw[black] (2,0) -- (3.5,0);
 \node at (1.5,0)[circle,fill,red,inner sep=4pt]{};
 \draw[draw=black, snake it] (1.5,0) arc (0:180:1.5cm);
 \draw [black,thick ,domain=120:60,->] plot ({2.5*cos(\x)}, {2.5*sin(\x)});
 \Huge{\node at (0,3.2) {$k+q-\ell$};}
 \node at (-2.5,-0.6) {$k$};
 \node at (0.7,-0.6) {$\ell $};
 \node at (2.6,-0.6) {$p$};
 \draw[black,->] (-1.5,-1.8) -- (-1.5,-0.6);
 \node at (-1.5,-2.4) {$q$};
  \Huge{\node at (-6.5,0) {$D_{1}(k,q,p)=$};}
  \Huge{\node at (6.5,0) {$+ ~~reflection$};}
  \draw (-1.5,0) node[cross] {};
\end{tikzpicture}
\caption{Feynman diagram contributing to $U^{(2)}_{B}$. The cross denotes the stress tensor vertex and the red dot is the exact vertex shown  in Fig.\ref{phiphiAvertex}. The internal propgator with the black dot is the exact thermal propagator for the fundamental bosons. The reflection diagram is the one where the stress tensor vertex lies to the right of the red vertex.  }
\label{s2rbD1}
  \end{center}
\end{figure} 

Then using equations \ref{tbppu1ans}, \ref{tbppu2ans} and \ref{tbppu3ans} the above takes the form
\begin{equation}
\label{d1simp}
\begin{split}
D_{1} (k,q,p) = & -\frac{4\pi }{\kappa_{B}} \int  \mathcal{D}^{3}\ell \hspace{0.1cm} \frac{1}{(\ell-p)^{+}}\left[  i (\tilde{\ell}_{3}+k_{3}+q_{3})+\mathcal{C}(\ell,p)\right] (\ell+k)_{+} \mathrm{Tr}(G(\ell)) \delta_{i}^{j} \\
&  + \frac{4\pi }{\kappa_{B}}  \int \mathcal{D}^{3}\ell \hspace{0.1cm} \frac{1}{(\ell-k)^{+}} \left[  i(\tilde{\ell}_{3}+k_{3})-\mathcal{C}(\ell,k)\right]  (\ell+p)_{+} \mathrm{Tr}(G(\ell)) \delta_{i}^{j} \\
& =-\frac{4\pi }{\kappa_{B}}  \int \mathcal{D}^{3}\ell \hspace{0.1cm} \frac{1}{(\ell-k)^{+}} \left[ i q_{3} +2\mathcal{C}(\ell,k)\right] (\ell+k)_{+} \mathrm{Tr}(G(\ell)) \delta_{i}^{j} 
\end{split}
\end{equation}

Performing the angular and holonomy integrals  then yields
\begin{equation}
\label{d1simp2}
\begin{split}
D_{1} (k,q,p)  & =  \lambda_{B} \frac{k_{s}^{2}}{(k^{+})^{2}}  \int_{0}^{k_{s}} d\ell_{s}  \hspace{0.05cm} \ell_{s} \hspace{0.1cm} \left[ i q_{3} +2\mathcal{C}(\ell,k)\right]  \left(1+\frac{\ell_{s}^{2}}{k_{s}^{2}}\right) \mathcal{M}_{B}(x,\mu_{B}) \delta_{i}^{j} \\
\end{split}
\end{equation} 

where $\mathcal{M}(x,\mu_{B})$ was also calculated in section \ref{u1pbose} and is given by
\begin{equation}
\label{d1simp3}
\begin{split}
\mathcal{M}(x,\mu_{B})&= \int_{-1/2}^{1/2}du \hspace{0.1cm} \frac{1}{\beta} \sum_{n=-\infty}^{\infty} \frac{1}{\ell_{s}^{2}+\frac{\mu_{B}^{2}}{\beta^{2}}+\frac{4\pi^{2}}{\beta^{2}}(n-|\lambda|u)^{2}}  \\
&= \frac{i \beta}{2 \pi |\lambda_{B}|} \frac{\left(\log\left[\sinh\left(\frac{1}{2}\sqrt{x^{2}+\mu_{B}^{2}}-\frac{i \pi |\lambda_{B}|}{2}\right)\right]-\mathrm{c.c}\right)}{ \sqrt{x^{2}+\mu_{B}^{2}}} 
\end{split}
\end{equation}

Then doing the radial integral we find
\begin{equation}
\label{d1final}
\begin{split}
D_{1} (k,q,p)  & = \frac{  k_{s}^{2}}{2(k^{+})^{2}}  \left[\left(i q_{3}+ \mathcal{C}(0,k)\right)\mathcal{C}(0,k) + \left(i q_{3}+ \mathcal{X}(y)\right)\mathcal{X}(y)\right] \\
& -\frac{1}{ \beta^{2}(k^{+})^{2}}\left[ \left(i q_{3}+2\mathcal{X}(y)\right)\int_{0}^{y} dx \hspace{0.1cm}x \hspace{0.05cm} \mathcal{X}(x) -\int_{0}^{y} dx \hspace{0.1cm}x \hspace{0.05cm} \mathcal{X}^{2}(x) \right]
\end{split}
\end{equation}

Next let us consider the diagram labelled by $D_{2}(k,q,p) $ in Fig.\ref{s2rbD2}.  This is given by
\begin{equation}
\label{d2def}
\begin{split}
D_{2} (k,q,p)  & = \int  \mathcal{D}^{3}\ell_{1}  \mathcal{D}^{3}\ell_{2}  \left[  \mathcal{K}^{a,3}(k,\ell_{1})  \mathcal{G}_{+3}(k-\ell_{1})G(\ell_{1}) \mathcal{U}^{a b}_{3} \mathcal{G}_{3+}(p-\ell_{2})G(\ell_{2}) \mathcal{K}^{b,3}(\ell_{2},p)\right]_{i}^{j}\\
& = \frac{1}{2}\left(\frac{4\pi i}{\kappa_{B}}\right)^{2} \int  \mathcal{D}^{3}\ell_{1}  \mathcal{D}^{3}\ell_{2} \hspace{0.1cm}  \frac{\left[i(k^{3}+(\tilde{\ell}_{1})^{3})+\mathcal{C}(k,\ell_{1})\right] \left[i(p^{3}+(\tilde{\ell}_{2})^{3})+\mathcal{C}(\ell_{2},p)\right]}{(\ell_{1}-k)^{+} (\ell_{2}-p)^{+}} \times  \\
& \hspace{4.5cm}\left[\delta_{ji} \mathrm{Tr}(G(\ell_{1})) \mathrm{Tr}(G(\ell_{2}))+ (G(\ell_{2})G(\ell_{1}))_{ji}\right] 
\end{split}
\end{equation}

\begin{figure}[h!]
\begin{center}
\begin{tikzpicture}[scale=.5, transform shape]
 \draw[black,->] (-3.5,0) -- (-2,0);
 \draw[black,->] (-2,0) -- (1,0);
 \draw[black] (1,0) -- (1.5,0);
  \draw[black,->] (1.5,0) -- (4,0);
   \draw[black] (4,0) -- (4.5,0);
   \draw[black,->] (4.5,0) -- (5.5,0);
   \draw[black] (5.5,0) -- (6.5,0);
 \node at (-1.5,0)[circle,fill,red,inner sep=4pt]{};
  \node at (4.5,0)[circle,fill,red,inner sep=4pt]{};
 \draw[draw=black, snake it] (1.5,0) arc (0:180:1.5cm);
 \draw [black,thick ,domain=120:60,->] plot ({2.5*cos(\x)}, {2.5*sin(\x)});
 \draw [black,thick ,domain=120:60,->] plot ({3+2.5*cos(\x)}, {2.5*sin(\x)});
 \Huge{\node at (-0.5,3) {$p-\ell_{1}$};}
 \Huge{\node at (3.5,3) {$p+q-\ell_{2}$};}
 \node at (-2.5,-0.6) {$k$};
 \node at (3.5,-0.8) {$\ell_{2}$};
 \node at (0.5,-0.8) {$\ell_{1}$};
 \node at (5.5,-0.6) {$p$};
 \draw[black,->] (1.5,-1.4) -- (1.5,-0.6);
 \node at (1.5,-1.8) {$q$};
  \Huge{\node at (-6.5,0) {$D_{2}(p,q,k)=$};}
  
  \draw (1.5,0) node[cross] {};
  \draw[draw=black, snake it] (4.5,0) arc (0:180:1.5cm);
  
  \node at (0,0)[circle,fill,black,inner sep=4pt]{};
   \node at (3,0)[circle,fill,black,inner sep=4pt]{};
\end{tikzpicture}
\caption{Feynman diagram contributing to $U^{(2)}_{B}$. The cross denotes the stress tensor vertex and the red dot is the exact vertex in Fig.\ref{phiphiAvertex}. The internal matter propagator is the exact thermal propagator.}
\label{s2rbD2}
  \end{center}
\end{figure}

Now in the above expression the connected term $ (G(\ell_{2})G(\ell_{1}))_{ji}$ is subleading compared to the disconnected term $\delta_{ji} \mathrm{Tr}(G(\ell_{1})) \mathrm{Tr}(G(\ell_{2}))$ in the large $N_{B}$ limit. We will henceforth neglect such contributions.  Thus we have
\begin{equation}
\label{d2simp}
\begin{split}
 D_{2} (k,q,p)  & = \frac{1}{2}\left(\frac{4\pi i}{\kappa_{B}}\right)^{2} \int  \mathcal{D}^{3}\ell_{1}  \mathcal{D}^{3}\ell_{2} \hspace{0.1cm}  \frac{\left[i(k^{3}+(\tilde{\ell}_{1})^{3})+\mathcal{C}(k,\ell_{1})\right] \left[i(p^{3}+(\tilde{\ell}_{2})^{3})+\mathcal{C}(\ell_{2},p)\right]}{(\ell_{1}-k)^{+} (\ell_{2}-p)^{+}} \times \\
 & \hspace{4.5cm} \mathrm{Tr}(G(\ell_{1})) \mathrm{Tr}(G(\ell_{2})) \delta_{ji} 
\end{split}
\end{equation}

Moving on to the diagram labelled by $D_{3}(k,q,p) $ in Fig. \ref{s2rbD3} we have
\begin{equation}
\label{d3def}
\begin{split}
D_{3} (k,q,p)  & = \int  \mathcal{D}^{3}\ell_{1}  \mathcal{D}^{3}\ell_{2} \left[\mathcal{U}^{ab}_{3}  \mathcal{G}_{3+}(p-\ell_{2})\mathcal{G}_{3+}(\ell_{2}-\ell_{1}) G(\ell_{1})\mathcal{K}^{b,3}(\ell_{1},\ell_{2})G(\ell_{2}) \mathcal{K}^{a,3}(\ell_{2},p)\right]_{i}^{j} +\text{reflection} \\
& = - \frac{1}{2}\left(\frac{4\pi i}{\kappa}\right)^{2} \int  \mathcal{D}^{3}\ell_{1}  \mathcal{D}^{3}\ell_{2} \hspace{0.1cm}  \frac{\left[i((\tilde{\ell}_{1})^{3}+(\tilde{\ell}_{2})^{3})+\mathcal{C}(\ell_{1},\ell_{2})\right] \left[i(p^{3}+(\tilde{\ell}_{2})^{3})+\mathcal{C}(\ell_{2},p)\right]}{(\ell_{2}-p)^{+} (\ell_{1}-\ell_{2})^{+}} \times  \\
& \left[\delta_{ji} \mathrm{Tr}(G(\ell_{1})) \mathrm{Tr}(G(\ell_{2}))+ (G(\ell_{2})G(\ell_{1}))_{ji}\right]  +\text{reflection}\\
& =- \frac{1}{2}\left(\frac{4\pi i}{\kappa_{B}}\right)^{2} \int  \mathcal{D}^{3}\ell_{1}  \mathcal{D}^{3}\ell_{2} \hspace{0.1cm} \frac{\left[ \mathcal{C}(\ell_{1},\ell_{2})(i q^{3}+ 2 \mathcal{C}(\ell_{2},p))- (\tilde{\ell}_{1}^{3}+\tilde{\ell}_{2}^{3})(2 k^{3}+q^{3}+2\tilde{\ell}_{2}^{3}) \right]}{(\ell_{2}-p)^{+} (\ell_{1}-\ell_{2})^{+}} \times \\
& \hspace{5.0cm} \mathrm{Tr}(G(\ell_{1})) \mathrm{Tr}(G(\ell_{2})) \delta_{ji} 
\end{split}
\end{equation}

where in the second line of the above we dropped the propagator terms which are subleading in the large $N_{B}$ limit. 
\begin{figure}[h!]
\begin{center}
\begin{tikzpicture}[scale=.6, transform shape]
 \draw[black,->] (-3.5,0) -- (-2,0);
 \draw[black,->] (-2,0) -- (1,0);
 \draw[black] (1,0) -- (1.5,0);
  \draw[black,->] (1.5,0) -- (4,0);
   \draw[black] (4,0) -- (4.5,0);
   \draw[black,->] (4.5,0) -- (6,0);
   \draw[black] (6,0) -- (6.5,0);
 \node at (1.5,0)[circle,fill,red,inner sep=4pt]{};
  \node at (4.5,0)[circle,fill,red,inner sep=4pt]{};
 \draw[draw=black, snake it] (1.5,0) arc (0:180:1.5cm);
 \draw [black,thick ,domain=120:60,->] plot ({1.5+3.5*cos(\x)}, {3.5*sin(\x)});
 \draw [black,thick ,domain=90:30,->] plot ({2.1*cos(\x)}, {2.1*sin(\x)});
 \draw[black,->] (-1.5,-1.4) -- (-1.5,-0.6);
  \Huge{\node at (-6.5,0) {$D_{3}(k,q,p)=$};}
  \node at (-2.5,-0.6) {$k$};
 \node at (3.5,-0.8) {$\ell_{2}$};
 \node at (0.5,-0.8) {$\ell_{1}$};
 \node at (5.5,-0.6) {$p$};
 \Huge{\node at (1.5,4.2) {$k+q-\ell_{2}$};}
 \Huge{\node at (2.1,2.2) {$\ell_{2}-\ell_{1}$};}
 \node at (-1.5,-1.8) {$q$};
  \draw (-1.5,0) node[cross] {};
  \draw[draw=black, snake it] (4.5,0) arc (0:180:3cm);
  \Huge{\node at (9.5,0) {$+ ~~reflection$};}

  \node at (0,0)[circle,fill,black,inner sep=4pt]{};
   \node at (3,0)[circle,fill,black,inner sep=4pt]{};
\end{tikzpicture}
\caption{Feynman diagram contributing to $U^{(2)}_{B}$. The cross stands for the stress tensor vertex and the red dot is the exact vertex in Fig.\ref{phiphiAvertex}. The internal matter propagator is the exact thermal propagator. The reflected diagram involves the stress tensor vertex being situated to the right of the vertices denoted in red.}
\label{s2rbD3}
  \end{center}
\end{figure} 

In order to evaluate $D_{2}(k,q,p)$ and $D_{3}(k,q,p)$ we find it convenient to add up their respective contributions. Then after doing some of the relevant integrals and Matsubara sums we end up with
\begin{equation}
\label{d2plusd3}
\begin{split}
D_{2}(k,q,p) +D_{3} (k,q,p)&= \frac{1}{6 (k^{+})^{2}}\left[3k_{3}\left(k_{3}+q_{3}\right) + 2 i q_{3} \mathcal{C}(0,k) +\mathcal{C}^{2}(0,k)\right]\mathcal{C}^{2}(0,k) \\
& + \left(\frac{4\pi i}{\kappa_{B}}\right)^{2}  \int  \mathcal{D}^{3}\ell_{1}  \mathcal{D}^{3}\ell_{2} \hspace{0.1cm}  \frac{(\tilde{\ell}_{2}^{3})^{2}}{(\ell_{2}-k)^{+} (\ell_{1}-\ell_{2})^{+}}  \mathrm{Tr}(G(\ell_{1})) \mathrm{Tr}(G(\ell_{2}))\delta_{ji} 
\end{split}
\end{equation}

Next let us consider the diagram labelled by $D_{4} (k,q,p)$ in Fig.\ref{s2rbD4} which is explicitly given by
\begin{equation}
\label{d4def}
\begin{split}
D_{4} (k,q,p)  & =  \int  \mathcal{D}^{3}\ell_{1}  \mathcal{D}^{3}\ell_{2}  \left[ \mathcal{V}^{a,+}(k,\ell_{1})\mathcal{G}_{3+}(k-\ell_{1})G(\ell_{1})\mathcal{U}_{2}^{b}(q,\ell_{1},\ell_{2})\mathcal{G}_{3+}(q+\ell_{1}-\ell_{2})G(\ell_{2})\{T^{a},T^{b}\} \right]_{i}^{j}\\
& +\text{reflection}\\
&=- \left(\frac{4\pi i}{\kappa_{B}}\right)^{2} \int  \mathcal{D}^{3}\ell_{1}  \mathcal{D}^{3}\ell_{2} \hspace{0.1cm}  \frac{(\ell_{1}^{+}+ k^{+}) (\ell_{1}+\ell_{2})_{+}}{(\ell_{1}-k)^{+} (\ell_{1}-\ell_{2})^{+}} \delta_{ji} \mathrm{Tr}(G(\ell_{1})) \mathrm{Tr}(G(\ell_{2}))
\end{split}
\end{equation}

\begin{figure}[h!]
\begin{center}
\begin{tikzpicture}[scale=.6, transform shape]
 \draw[black,->] (-3.5,0) -- (-2,0);
 \draw[black,->] (-2,0) -- (1,0);
 \draw[black] (1,0) -- (1.5,0);
  \draw[black,->] (1.5,0) -- (4,0);
   \draw[black] (4,0) -- (4.5,0);
   \draw[black,->] (4.5,0) -- (5.5,0);
   \draw[black] (5.5,0) -- (6.5,0);
 
 \draw[draw=black, snake it] (1.5,0) arc (180:0:1.5cm);
 \draw [black,thick ,domain=120:60,->] plot ({1.5+3.5*cos(\x)}, {3.5*sin(\x)});
 \Huge{\node at (9.5,0) {$+ ~~reflection$};}
 \draw [black,thick ,domain=170:120,->] plot ({3+2.1*cos(\x)}, {2.1*sin(\x)});
   \Huge{\node at (-6.5,0) {$D_{4}(k,q,p)=$};}
   \node at (-2.5,-0.6) {$k$};
   \node at (3,-0.8) {$\ell_{2}$};
   
 \node at (0,-0.8) {$\ell_{1}$};
 \node at (5.5,-0.6) {$p$};
 \node at (1.5,-1.8) {$q$};
 \Huge{\node at (1.5,2.2) {$q+\ell_{3}$};}
  \draw (1.5,0) node[cross] {};
  \Huge{\node at (1.5,4.2) {$k-\ell_{1}$};}
  \draw[black,->] (1.5,-1.4) -- (1.5,-0.6);
  \draw[draw=black, snake it] (-1.5,0) arc (180:0:3cm);
  
  \node at (0,0)[circle,fill,black,inner sep=4pt]{};
   \node at (3,0)[circle,fill,black,inner sep=4pt]{};
\end{tikzpicture}
\caption{Feynman diagram contributing to $U^{(2)}_{B}$. The cross again signifies the stress tensor vertex. The internal matter propagator is the exact thermal propagator. The reflected diagram involves the stress tensor vertex lying to the right of the seagull vertex.}
\label{s2rbD4}
  \end{center}
\end{figure} 

Simplifying the integral in equation \ref{d4def} by carrying out the holonomy, angular and radial integrals respectively we arrive at
\begin{equation}
\label{d4simp}
\begin{split}
D_{4} (k,q,p)  & =- \frac{1 }{4 \beta^{2} (k^{+})^{2}}\left[ 3 y^{2} \left(\mathcal{X}^{2}(y)-2\mathcal{X}(y)\mathcal{X}(0)\right) +2 \int_{0}^{y}dx \hspace{0.1cm}x \hspace{0.06cm}  \mathcal{X}^{2}(x) \right]\\
&  + \frac{1 }{\beta^{2} (k^{+})^{2}} \left(2  \mathcal{X}(y)-\mathcal{X}(0)\right) \int_{0}^{y}dx \hspace{0.1cm}x \hspace{0.06cm}  \mathcal{X}(x)  \\
& +  \frac{1 }{2} \left(\frac{4\pi i}{\kappa_{B}}\right)^{2}  \int  \mathcal{D}^{3}\ell_{1}  \mathcal{D}^{3}\ell_{2} \hspace{0.1cm}  \frac{\ell_{s,2}^{2}}{(\ell_{2}-k)^{+} (\ell_{1}-\ell_{2})^{+}}  \mathrm{Tr}(G(\ell_{1})) \mathrm{Tr}(G(\ell_{2})) \delta_{ji}
\end{split}
\end{equation}

\begin{figure}[h!]
\begin{center}
\begin{tikzpicture}[scale=.5, transform shape]
\draw[black,->] (-6,0) -- (-5,0);
\draw[black] (-5,0) -- (-4,0);
 \draw[black,->] (-4,0) -- (-2,0);
 \draw[black,->] (-2,0) -- (1,0);
 \draw[black] (1,0) -- (1.5,0);
  \draw[black,->] (1.5,0) -- (4,0);
   \draw[black] (4,0) -- (6,0);
   \draw[black,->] (4.5,0) -- (5.5,0);
   \draw[black] (5.5,0) -- (6.5,0);
 \node at (1.5,0)[circle,fill,red,inner sep=4pt]{};
 \draw[draw=black, snake it] (1.5,0) arc (0:180:1.5cm);
 \draw [black,thick ,domain=170:120,->] plot ({1.5+3.5*cos(\x)}, {3.5*sin(\x)});
  \draw [black,thick ,domain=90:30,->] plot ({2.1*cos(\x)}, {2.1*sin(\x)});
 \draw [black,thick ,domain=120:60,->] plot ({5.1*cos(\x)}, {5.1*sin(\x)});
   \Huge{\node at (-8.5,0) {$D_{5}(k,q,p)=$};}
   \node at (3.5,-0.8) {$\ell_{3}$};
   Huge{\node at (0,5.5) {$k-\ell_{1}$};}
     Huge{\node at (1.5,2.5) {$\ell_{3}-\ell_{2}$};}
   Huge{\node at (-0.3,3.8) {$q+\ell_{1}-\ell_{3}$};}
\node at (-5.5,-0.6) {$k$};
   \node at (0.5,-0.8) {$\ell_{2}$};
 \node at (-2.5,-0.8) {$\ell_{1}$};
 \node at (5.5,-0.6) {$p$};
 \node at (-1.5,-1.8) {$q$};   
   \draw[black,->] (-1.5,-1.4) -- (-1.5,-0.6);
  \draw (-1.5,0) node[cross] {};
  \draw[draw=black, snake it] (4.5,0) arc (0:180:3cm);
  \draw[draw=black, snake it] (4.5,0) arc (0:180:4.5cm);
  \Huge{\node at (9.5,0) {$+ ~~reflection$};}
  
  \node at (0,0)[circle,fill,black,inner sep=4pt]{};
   \node at (-3,0)[circle,fill,black,inner sep=4pt]{};
   \node at (3,0)[circle,fill,black,inner sep=4pt]{};
\end{tikzpicture}
\caption{Feynman diagram contributing to $U^{(2)}_{B}$. The cross stands for the stress tensor vertex and the red dots are the exact vertices in Fig.\ref{phiphiAvertex}. The internal matter propagator is the exact thermal propagator. The reflected diagram has the stress tensor vertex appearing to the right of the seagull vertex.}
\label{s2rbD5}
  \end{center}
\end{figure} 
There are two more diagrams that need to be considered. Denoting the penultimate one in Fig.\ref{s2rbD5} by $D_{5} (k,q,p)$ we have

\begin{eqnarray}
\label{d5def}\nonumber
D_{5} (k,q,p) &=&  \int  \mathcal{D}^{3}\ell_{1}  \mathcal{D}^{3}\ell_{2}   \mathcal{D}^{3}\ell_{3} \Big( \mathcal{V}^{a,+}(k,\ell_{1})  \mathcal{G}_{3+}(k-\ell_{1})G(\ell_{1}) \mathcal{U}_{3}^{bc}\times\\\nonumber
&&\mathcal{G}_{3+}(q+\ell_{1}-\ell_{3})
\mathcal{G}_{3+}(\ell_{3}-\ell_{2}) G(\ell_{2}) \mathcal{K}^{c,3}(\ell_{2},\ell_{3})G(\ell_{3})\{T^{a},T^{b}\}\Big)_{i}^{j}\\
 &+& \text{reflection}
\end{eqnarray}

Upon simplifying the above integrand we get
\begin{equation}
\label{d5simp}
\begin{split}
 & D_{5} (k,q,p)  = \frac{1}{4}  \left(\frac{4\pi i}{\kappa_{B}}\right)^{3}  \int  \mathcal{D}^{3}\ell_{1}   \mathcal{D}^{3}\ell_{2}  \mathcal{D}^{3}\ell_{3} \hspace{0.1cm}  \frac{i(k^{+}+ \ell_{1}^{+}) \left[i((\tilde{\ell}_{2})_{3}+(\tilde{\ell}_{3})_{3})+\mathcal{C}(\ell_{2},\ell_{3})\right]}{(\ell_{1}-k)^{+} (\ell_{3}-\ell_{1}-q)^{+}(\ell_{2}-\ell_{3})^{+}} \times  \\
& \hspace{-2.0cm} \left[\delta_{ji} \mathrm{Tr}(G(\ell_{1})) \mathrm{Tr}(G(\ell_{2}))\mathrm{Tr}(G(\ell_{3}))+\delta_{ji}\mathrm{Tr}(G(\ell_{1})G(\ell_{3})G(\ell_{2}))+ (G(\ell_{3})G(\ell_{1}))_{ji}\mathrm{Tr}(G(\ell_{2}))+\mathrm{Tr}(G(\ell_{1})G(\ell_{3})) G(\ell_{2})_{ji}\right] \\
& \hspace{1.5cm}+ \text{reflection} \\
& =\frac{1}{4} \left(\frac{4\pi i}{\kappa_{B}}\right)^{3}  \int  \mathcal{D}^{3}\ell_{1}   \mathcal{D}^{3}\ell_{2}  \mathcal{D}^{3}\ell_{3} \hspace{0.1cm}  \frac{2 i (k^{+}+ \ell_{1}^{+}) \mathcal{C}(\ell_{2},\ell_{3})}{(\ell_{1}-k)^{+} (\ell_{3}-\ell_{1})^{+}(\ell_{2}-\ell_{3})^{+}}  \mathrm{Tr}(G(\ell_{1})) \mathrm{Tr}(G(\ell_{2}))\mathrm{Tr}(G(\ell_{3}))  \delta_{ji}
\end{split}
\end{equation}

In the last line of the above we have taken $q^{\pm}=0$ and only kept the fully disconnected propagator piece which gives the dominant contribution in the large $N_{B}$ limit. Finally the last diagram that we have to evaluate is in Fig.\ref{s2rbD6} which is labelled by $D_{6} (k,q,p)$. For this we get
\begin{equation}
\label{d6def}
\begin{split}
&D_{6} (k,q,p)= \\
&  \hspace{-1.5 cm}  \int  \mathcal{D}^{3}\ell_{1}  \mathcal{D}^{3}\ell_{2}   \mathcal{D}^{3}\ell_{3} \left[ \mathcal{V}^{a,+}(k,\ell_{1})  \mathcal{G}_{3+}(k-\ell_{1})G(\ell_{1}) \mathcal{K}^{b,3}(\ell_{1},\ell_{2}) G(\ell_{2})\mathcal{G}_{+3}(\ell_{1}-\ell_{2})  \mathcal{U}_{3}^{bc} \mathcal{G}_{3+}(q+\ell_{1}-\ell_{3}) G(\ell_{3}) \{T^{a},T^{c}\}\right]_{i}^{j}\\
& + \text{reflection} \\
\end{split}
\end{equation}

\begin{figure}[h!]
\begin{center}
\begin{tikzpicture}[scale=.6, transform shape]
\draw[black,->] (-6,0) -- (-5,0);
\draw[black] (-5,0) -- (-3,0);
 \draw[black,->] (-3,0) -- (-2,0);
 \draw[black,->] (-2,0) -- (1,0);
 \draw[black] (1,0) -- (1.5,0);
  \draw[black,->] (1.5,0) -- (4,0);
   \draw[black] (4,0) -- (6,0);
   \draw[black,->] (4.5,0) -- (6,0);
   \draw[black] (6,0) -- (6.5,0);
 \node at (-1.5,0)[circle,fill,red,inner sep=4pt]{};
  
 \draw[draw=black, snake it] (1.5,0) arc (0:180:1.5cm);
 \draw [black,thick ,domain=170:120,->] plot ({2.2*cos(\x)}, {2.2*sin(\x)});
 \draw [black,thick ,domain=120:60,->] plot ({3+2*cos(\x)}, {2*sin(\x)});
 \draw [black,thick ,domain=120:60,->] plot ({5.1*cos(\x)}, {5.1*sin(\x)});
  \Huge{\node at (-8.5,0) {$D_{6}(k,q,p)=$};}
  
    Huge{\node at (0,5.5) {$k-\ell_{1}$};}
  \node at (-5.5,-0.6) {$k$};
   \node at (0.5,-0.8) {$\ell_{2}$};
 \node at (-2.5,-0.8) {$\ell_{1}$};
 \node at (5.5,-0.6) {$p$};
 \node at (1.5,-1.8) {$q$};   
  \node at (3.5,-0.8) {$\ell_{3}$};
   \draw[black,->] (1.5,-1.4) -- (1.5,-0.6);
   \Huge{\node at (9.5,0) {$+ ~~reflection$};}
    Huge{\node at (-2.0,2.5) {$\ell_{2}-\ell_{1}$};}
   Huge{\node at (1.7,2.5) {$q+\ell_{1}-\ell_{3}$};}
  \draw (1.5,0) node[cross] {};
  \draw[draw=black, snake it] (4.5,0) arc (0:180:1.5cm);
  \draw[draw=black, snake it] (4.5,0) arc (0:180:4.5cm);

   \node at (0,0)[circle,fill,black,inner sep=4pt]{};
   \node at (-3,0)[circle,fill,black,inner sep=4pt]{};
   \node at (3,0)[circle,fill,black,inner sep=4pt]{};
\end{tikzpicture}
\caption{Feynman diagram contributing to $U^{(2)}_{B}$. The cross denotes the stress tensor vertex and the red dots are the exact vertices in Fig.\ref{phiphiAvertex}. The internal matter propagator is the exact thermal propagator. The reflected diagram has the stress tensor vertex appearing to the right of the seagull vertex.}
\label{s2rbD6}
  \end{center}
\end{figure}

The integrand in \ref{d6def} can be simplified further to yield 
\begin{equation}
\label{d6simp}
\begin{split}
& D_{6} (k,q,p) =   \left(\frac{4\pi i}{\kappa_{B}}\right)^{3}  \int  \mathcal{D}^{3}\ell_{1}   \mathcal{D}^{3}\ell_{2}  \mathcal{D}^{3}\ell_{3} \hspace{0.1cm}  \frac{i(k^{+}+ \ell_{1}^{+}) \left[i((\tilde{\ell}_{1})_{3}+(\tilde{\ell}_{2})_{3})+\mathcal{C}(\ell_{1},\ell_{2})\right]}{(\ell_{1}-k)^{+} (\ell_{3}-\ell_{1}-q)^{+}(\ell_{1}-\ell_{2})^{+}} \times  \\
& \hspace{-2.0cm} \left[\delta_{ji} \mathrm{Tr}(G(\ell_{1})) \mathrm{Tr}(G(\ell_{2}))\mathrm{Tr}(G(\ell_{3}))+\delta_{ji}\mathrm{Tr}(G(\ell_{1})G(\ell_{3})G(\ell_{2}))+ (G(\ell_{3})G(\ell_{1}))_{ji}\mathrm{Tr}(G(\ell_{2}))+\mathrm{Tr}(G(\ell_{1})G(\ell_{3})) G(\ell_{2})_{ji}\right]\\
& \hspace{1.5cm}+ \text{reflection} \\
& = \frac{1}{4} \left(\frac{4\pi i}{\kappa_{B}}\right)^{3}  \int  \mathcal{D}^{3}\ell_{1}   \mathcal{D}^{3}\ell_{2}  \mathcal{D}^{3}\ell_{3} \hspace{0.1cm}  \frac{2 i (k^{+}+ \ell_{1}^{+}) \mathcal{C}(\ell_{1},\ell_{2})}{(\ell_{1}-k)^{+} (\ell_{3}-\ell_{1})^{+}(\ell_{1}-\ell_{2})^{+}}  \mathrm{Tr}(G(\ell_{1})) \mathrm{Tr}(G(\ell_{2}))\mathrm{Tr}(G(\ell_{3})) \delta_{ji}
\end{split}
\end{equation}

As before in the above equation we have once again only retained the contribution of the fully disconnected propagator term which is dominant at large $N_{B}$. Now it turns that it is simpler to evaluate the sum of  $D_{5}(k,q,p) $ and $D_{6}(k,q,p) $ rather than evaluating them individually. Then doing the necessary Matsubara sums and integrals using the results collected in the Appendix we obtain 
\begin{equation}
\label{d5plusd6}
\begin{split}
D_{5}(k,q,p) + D_{6}(k,q,p)  = - \frac{1 }{6  (k^{+})^{2}} \mathcal{C}^{4}(0,k)
\end{split}
\end{equation}

Finally let us add up all the above contributions in equations \ref{d1simp}, \ref{d2plusd3}, \ref{d4simp} and \ref{d5plusd6} to get
\begin{equation}
\label{d1to6final}
\begin{split}
\sum_{i=1}^{6} D_{i} (k,q,p) & =\frac{  i q_{3} }{6\beta^{2}(k^{+})^{2}}\left[3 y^{2} \left(\mathcal{C}(0,k)+\mathcal{X}(y)\right) -6\int_{0}^{y} dx \hspace{0.1cm}x \hspace{0.05cm} \mathcal{X}(x)+ 2 \beta^{2} \mathcal{C}^{3}(0,k)\right] \\
& +   \frac{1}{2(k^{+})^{2}}  \left[ k_{3}\left(k_{3}+q_{3}\right)  + \beta^{-2}  \left(y^{2}+\mu_{B}^{2}\right)\right] \mathcal{C}^{2}(0,k) - \frac{2\lambda }{\beta^{3} (k^{+})^{2}}  \left[\int_{0}^{y} dx \hspace{0.1cm} x \hspace{0.05cm} \mathcal{X}(x)\right](2\zeta(0) +1)\\ 
\end{split}
\end{equation} 

where $\zeta(s)$ is the Riemann Zeta function. The origin of this term is through a divergent sum over Matsubara modes which we encounter while adding $D_{3}(k,q,p)$ and $D_{4}(k,q,p)$. Now although $\zeta(s)$ is formally an asymptotic series, upon proper regularisation it admits a finite value  which is given by $\zeta(0)=-\frac{1}{2}$. As a result the last term in \ref{d1to6final} vanishes. 

Thus our final result in this section is that relevant finite temperature vertex factor for $J^{(2)}_{B,++}$ is given by the following
\begin{equation}
\label{tppvertexfinal}
\begin{split}
U^{(2)}_{B}(q,k) & = 2 k^{2}_{+} +\sum_{i=1}^{6} D_{i} (k,q,p) \\
& = 2 k^{2}_{+} +\frac{  i q_{3} }{6\beta^{2}(k^{+})^{2}}\left[3 y^{2} \left(\mathcal{C}(0,k)+\mathcal{X}(y)\right) -6\int_{0}^{y} dx \hspace{0.1cm}x \hspace{0.05cm} \mathcal{X}(x)+ 2 \beta^{2} \mathcal{C}^{3}(0,k)\right] \\
& +   \frac{1}{2(k^{+})^{2}}  \left[ k_{3}\left(k_{3}+q_{3}\right)  + \beta^{-2}  \left(y^{2}+\mu_{B}^{2}\right)\right] \mathcal{C}^{2}(0,k) 
\end{split}
\end{equation}

\subsubsection{Thermal $2$-point  function}

We are now equipped with the necessary ingredients to compute the thermal $2$-point function of the stress tensor operator in the Regular Boson theory. Sewing together the vertex factors $V^{(2)}_{B}(q,k)$ and $U^{(2)}_{B}(q,k)$ with the exact finite temperature propagators for the fundamental bosons yields
\begin{equation}
\label{tt}
\begin{split}
 \left\langle J^{(2)}_{B,--}(-q) J^{(2)}_{B,++}(q)\right\rangle_{\beta} & =\int \frac{d^{2}\ell}{(2\pi)^{2}}\hspace{0.04cm} \frac{1}{\beta} \sum_{n=-\infty}^{\infty} \mathrm{Tr}\left[V^{(2)}_{B}(q,\ell)G(\ell)U^{(2)}_{B}(-q,\ell+q)G (\ell+q)\right] \\
\end{split}
\end{equation}


The angular part of the integral is again trivial since the integrand is manifestly rotationally invariant in the spatial directions. Then executing the Matsubara sum and thermal holonomy integral we obtain the following integral expression
\begin{equation}
\label{tt1}
\begin{split}
& \left\langle J^{(2)}_{B,--}(-q) J^{(2)}_{B,++}(q) \right\rangle_{\beta} \\
 & =  \frac{N_{B}}{8 \pi \lambda_{B}\beta^{3}}\int_{0}^{\infty} dx \hspace{0.1cm} x \hspace{0.06cm} e^{ i q_{3}\beta \hspace{0.04cm}\mathcal{F}_{B}(x,q_{3}\beta,\mu_{B})} \hspace{0.06cm}  \mathcal{H}_{B}(x,q_{3}\beta,\mu_{B}) \left[ x^{4}+ 2 \left(x^{2}+\mu_{B}^{2}\right)\beta^{2}\mathcal{C}^{2}(0, \ell)  \right]\\
&  -  \frac{i N_{B}q_{3}}{24 \pi \lambda_{B}\beta } \int_{0}^{\infty} dx \hspace{0.1cm} x \hspace{0.06cm} e^{ i q_{3}\beta \hspace{0.04cm}\mathcal{F}_{B}(x,q_{3}\beta,\mu_{B})} \hspace{0.06cm}  \mathcal{H}_{B}(x,q_{3}\beta,\mu_{B})   \left[3 x^{2} \left(2\mathcal{X}(x)-\mathcal{X}(0)\right)  + 2 \beta^{2}  \mathcal{C}^{3}(0,\ell) - 6 \int_{0}^{x} dz \hspace{0.1cm}z \hspace{0.05cm} \mathcal{X}(z) \right]\\
\end{split}
\end{equation}

We now follow the exact identical steps outlined in the previous sections for evaluating the above radial integrals.  First of all the above integrals have UV divergences and we regulate them with the help of a cutoff $\Lambda$. Then employing the relations given in equation \ref{u1rbtherm2} and using integration by parts we can demonstrate that the sum of integrals in equation \ref{tt1} above reduces to just boundary terms. Therefore we have,
\begin{equation}
\label{tt2}
\begin{split}
  \left\langle  J^{(2)}_{B,--}(-q) J^{(2)}_{B,++}(q)\right\rangle_{\beta} & = \frac{i N_{B}}{128 \pi \lambda_{B} q_{3}\beta^{4}} \left(1-e^{ i q_{3}\beta \hspace{0.04cm}\mathcal{F}_{B}(q_{3}\beta,\mu_{B})}\right)\left(q_{3}^{2}\beta^{2}+4\mu_{B}^{2}\right)^{2}\\
& - \frac{N_{B} }{96\pi \lambda_{B} \beta^{2}}  \mathcal{C}(0, \Lambda) \left[ 2 \beta^{2} \left(3 i q_{3}-4\mathcal{C}(0, \Lambda)\right)\mathcal{C}(0, \Lambda)+ 3\left(q_{3}^{2}\beta^{2}+4 \mu_{B}^{2}\right) \right]\\
& + \frac{N_{B} \Lambda^{2}}{8 \pi \lambda_{B} } \mathcal{C}(0, \Lambda)  -\frac{N_{B}}{4\pi \lambda_{B} \beta^{2}}  \int_{0}^{\beta\Lambda} dx \hspace{0.1cm} x \hspace{0.06cm} \mathcal{X}(x) 
\end{split}
\end{equation}

In equation \ref{tt2}, the divergent contributions come from $\mathcal{C}(0, \Lambda)$ which according to \ref{cqks2} is given by $\mathcal{C}(0, \Lambda) \sim \lambda_{B}\Lambda-\mathcal{X}(0)$ for $\beta\Lambda \gg 1$. We need to remove these divergences by adding appropriate counterterms involving the background metric which couples to the stress tensor. We refer the reader to \citep{Aharony:2012nh} for a discussion concering the necessary counterterms. Thus, after removing the divergences and using \ref{chidef} we arrive at the following result for the renormalised thermal $2$-point function
\begin{equation}
\label{tt3}
\begin{split}
&  \left\langle  J^{(2)}_{B,--}(-q) J^{(2)}_{B,++}(q)\right\rangle_{\beta}\\
& = \frac{i N_{B}}{128 \pi \lambda_{B} q_{3}\beta^{4}} \left(1-e^{ i q_{3}\beta \hspace{0.04cm}\mathcal{F}_{B}(q_{3}\beta,\mu_{B})}\right)\left(q_{3}^{2}\beta^{2}+4\mu_{B}^{2}\right)^{2} - \frac{ N_{B}}{48 \pi \lambda_{B}} \mathcal{X}^{2}(0)\left( 3 i q_{3} + 4 \mathcal{X}(0)\right)\\
& + \frac{ N_{B}}{32 \pi \lambda_{B}\beta^{2}} \mathcal{X}(0)\left( q_{3}^{2}\beta^{2} + 4 \mu_{B}^{2}\right)  + \frac{N_{B} }{12 \pi \beta^{3} } \mu_{B}^{3} -\frac{i N_{B} }{4 \pi^{2} |\lambda_{B}| \beta^{3} } \int_{\mu_{B}}^{\infty}  dx \hspace{0.1cm} x \left[ \mathrm{Li}_{2}\left(e^{-x+i \pi |\lambda_{B}|}\right)- \mathrm{c.c} \right] 
\end{split}
\end{equation}

where
\begin{equation}
\label{tt4}
\begin{split}
 \mathcal{X}(0) & = \frac{\lambda_{B}}{\beta} \left[ \mu_{B} + \frac{i}{|\lambda_{B}|\pi} \left( \mathrm{Li}_{2}\left( e^{-\mu_{B}+ i \pi |\lambda_{B}|} \right)-\mathrm{c.c}\right)\right] \\
\end{split}
\end{equation}






\subsubsection{Zero temperature limit}

Let us now the check for the consistency of our result for the thermal $2$-point function above with the corresponding zero temperature correlation function that was obtained in \citep{Aharony:2012nh}. Now in the zero temperature limit as $\beta\rightarrow \infty$ we see from \ref{chidef} that $ \mathcal{X}(0)\rightarrow 0$ . Further taking the thermal mass to zero in this limit and using equations \ref{t0limrb1} and \ref{t0limrb2} we find, \begin{equation}
\label{st0temp2}
\begin{split}
& \lim_{\beta\to\infty}\left\langle  J^{(2)}_{B,--}(-q) J^{(2)}_{B,++}(q)\right\rangle_{\beta} = \frac{i N_{B} }{128 \pi \lambda_{B} } q_{3}^{3}\left(1-e^{ i\pi  \hat{\lambda}_{B} }\right)
\end{split}
\end{equation}

where $\hat{\lambda}_{B} = \lambda_{B} \hspace{0.04cm}\mathrm{sgn}(q_{3})$. This agrees precisely with the result for the zero temperature two function derived in \citep{Aharony:2012nh}.


\subsection{Critical Bosons}

It follows from our discussion in sections \ref{s0cb} and \ref{u1thermcb}, that in the Critical Boson theory the stress tensor thermal $2$-point function is simply given by the corresponding correlator in the Regular Boson theory with the replacement $\mu_{B}\rightarrow \mu_{B.c}$ where $\mu_{B.c}$ is the thermal mass in the critical theory. Then using \ref{tt3} and noting that $\mathcal{X}(0)=0$ for critical bosons, we get
\begin{equation}
\label{ttcb}
\begin{split}
 \left\langle  \tilde{J}^{(2)}_{B,--}(-q) \tilde{J}^{(2)}_{B,++}(q)\right\rangle_{\beta} & = \frac{i N_{B}}{128 \pi \lambda_{B} q_{3}\beta^{4}} \left(1-e^{ i q_{3}\beta \hspace{0.04cm}\mathcal{F}_{B}(q_{3}\beta,\mu_{B,c})}\right)\left(q_{3}^{2}\beta^{2}+4\mu_{B,c}^{2}\right)^{2}+ \frac{N_{B} }{12 \pi \beta^{3} } \mu_{B,c}^{3}\\
&    -\frac{i N_{B} }{4 \pi^{2} |\lambda_{B}| \beta^{3} } \int_{\mu_{B,c}}^{\infty}  dx \hspace{0.1cm} x \left[ \mathrm{Li}_{2}\left(e^{-x+i \pi |\lambda_{B}|}\right)- \mathrm{c.c} \right] 
\end{split}
\end{equation}


\subsection{Regular Fermions} 

The stress tensor in the Regular Fermion theory is given by
\begin{equation}
\label{strf1}
\begin{split}
J^{(2)}_{F,\mu\nu}(x)= \frac{1}{2}  \bar{\psi}(x) \gamma_{(\mu} \overset{\rightarrow}{D_{\nu)}}\psi(x)-\frac{1}{2}  \bar{\psi}(x) \gamma_{(\mu}\overset{\leftarrow}{D_{\nu)}}\psi(x) +g_{\mu\nu}T^{\sigma}_{F,\sigma}(x)
\end{split}
\end{equation}

where $T^{\sigma}_{F,\sigma}(x)$ is defined in a way that renders $J^{(2)}_{F,\mu\nu}(x)$ is traceless.  Following the discussion in preceding sections we will be interested in computing $2$-point thermal correlators where the components of the stress tensors involved are $(\mu,\nu)=(-,-)$ and $(\mu,\nu)=(+,+)$. In light cone gauge $A_{-}=0$ these components have the form
\begin{equation}
\label{strf2}
\begin{split}
J^{(2)}_{F,--}(x)=\frac{1}{2} \left(-\partial_{-} \bar{\psi}(x) \gamma_{-} \psi(x) + \bar{\psi}(x) \gamma_{-}\partial_{-}\psi(x) \right)   
\end{split}
\end{equation}

and
\begin{equation}
\label{strf3}
\begin{split}
J^{(2)}_{F,++}(x)= -\frac{1}{2}  \left( \partial_{+}\bar{\psi}(x) \gamma_{+}\psi(x)-\bar{\psi}(x)\gamma_{+}\partial_{+}\psi(x) \right)+\bar{\psi}(x) A_{+}(x) \gamma_{+}\psi(x)
\end{split}
\end{equation}

To facilitate the straightforward determination of the Feynman rules for vertices with insertions of $J^{(2)}_{F,\mu\nu}$, we note below the momentum space expressions corresponding to equations \ref{strf3} and \ref{strf4}. 
\begin{equation}
\label{strf4}
\begin{split}
J^{(2)}_{F,--}(q_{3})=-\frac{1}{2\beta} \sum_{n} \int \frac{d^{2}\vec{k}}{(2\pi)^{2}} \hspace{0.1cm} \bar{\psi}(k_{3},\vec{k}) \left(2 i  \gamma_{-} k_{-} \right) \psi(q_{3}-k_{3},\vec{q}-\vec{k})
\end{split}
\end{equation}

\begin{equation}
\label{strf5}
\begin{split}
J^{(2)}_{F,++}(q_{3}) = & -\frac{1}{2\beta} \sum_{n} \int \frac{d^{2}\vec{k}}{(2\pi)^{2}} \hspace{0.1cm} \bar{\psi}(k_{3},\vec{k}) \left(2 i \gamma_{+} k_{+}\right) \psi(q_{3}-k_{3},-\vec{k})\\
& + \frac{1}{\beta^{2}} \sum_{n_{1},n_{2}} \int \frac{d^{2}\vec{p} \hspace{0.04cm}d^{2}\vec{k}}{(2\pi)^{4}} \hspace{0.1cm} \bar{\psi}(p_{3},\vec{p}) \hspace{0.04cm}\gamma_{+}A^{a}_{+}(k_{3},\vec{k}) T^{a} \psi(q_{3}-p_{3}-k_{3},-\vec{p}-\vec{k})
\end{split}
\end{equation}

where $(n, n_{1}, n_{2})\in \mathbf{Z}$ are discrete Matsubara frequency modes. Also note that the expressions in equations \ref{strf4} and \ref{strf5} have been obtained by restricting to the special kinematic choice where the external spatial momentum satisfies $q^{\pm}=0$. 

\subsubsection{Finite temperature vertex factor for $J^{(2)}_{F,--}$ }

We define the exact finite temperature vertex factor corresponding to $J^{(2)}_{F,--}$ to be
\begin{equation}
\label{strschwingdyson}
\begin{split}
\left\langle J^{(2)}_{F,--}(-q) \bar{\psi}^{j}(-k) \psi_{i}(p) \right\rangle_{\beta} = V^{(2)}_{F}(q,k) \delta_{i}^{j}(2\pi)^{3} \delta_{n_{q}+n_{k},n_{p}} \delta^{(2)}(\vec{q}+\vec{k}-\vec{p})
\end{split}
\end{equation}

Then the Schwinger-Dyson equation for the vertex factor $V^{(2)}_{F}(q,k) $ is given by 
\begin{equation}
\label{strfsdyson}
\begin{split}
V^{(2)}_{F}(q,k) \delta_{i}^{j} = V^{(2)}_{F,free}(q,k)\delta^{j}_{i}+\int \mathcal{D}^{3}\ell \hspace{0.04cm} \left[\mathcal{V}^{a,\mu}(k,\ell)S(\ell)V^{(2)}_{F}(q,\ell)S(\ell+q)\mathcal{V}^{a,\nu}(\ell+q,k+q)\right]^{j}_{i} \mathcal{G}_{\nu\mu}(k-\ell)
\end{split}
\end{equation}

where the first term in the R.H.S. of the above is the vertex factor in the free theory. From equation \ref{strf4} it follows that this is simply given by $V^{(2)}_{F,free}(q,k) = i k_{-}\gamma_{-} $. Now the structure of the second term in the R.H.S. of the Schwinger-Dyson equation \ref{strfsdyson} is precisely of the same form that we have encountered before in the spin $s=0$ and $s=1$ cases. Therefore following our previous analysis we can decompose $V^{(2)}_{F}(q,k)$ as 
\begin{equation}
\label{strschwingdyson1}
\begin{split}
V^{(2)}_{F}(q,k) & = V^{(2)}_{F,+}(q,k) \gamma^{+}+ V^{(2)}_{F,I} (q,k)\mathbf{1} \\
\end{split}
\end{equation}

Then the Schwinger-Dyson equations satisfied by the components $V^{(2)}_{F,+}(q,k)$ and $V^{(2)}_{F,-}(q,k)$ are in turn given by 
\begin{equation}
\label{str6}
\begin{split}
& V^{(2)}_{F,+}(q,k) \\
 & =i k^{+} - \frac{4 i \pi  \lambda_{F} }{\beta^{2}}\int \frac{d^{2}\ell}{(2\pi)^{2}} \hspace{0.04cm}\frac{1}{(\ell-k)^{+}} \left[2 x^{2}(g(x)-1)V^{(2)}_{F,I}(q,\ell)-\beta\ell^{+}\left(q_{3}\beta -2\tilde{f}(x)\right) V^{(2)}_{F,+}(q,\ell)\right] \times\\
 & \hspace{12.0cm} H_{F}(x,q_{3}\beta,\mu_{F})
\end{split}
\end{equation}

\begin{equation}
\label{str7}
\begin{split}
 & V^{(2)}_{F,I}(q,k) \\
 & =-\frac{4 i \pi  \lambda_{F}}{\beta^{2}} \int \frac{d^{2}\ell}{(2\pi)^{2}} \hspace{0.04cm}\frac{1}{(\ell-k)^{+}}\hspace{0.04cm} \left[2 \beta^{2}(\ell^{+})^{2}V^{(2)}_{F,+}(q,\ell)-\beta \ell^{+} (q_{3}\beta+2\tilde{f}(x))V^{(2)}_{F,I}(q,\ell)\right] H_{F}(x,q_{3}\beta, \mu_{F})
\end{split}
\end{equation}

where in the above we have already carried out the integral over the gauge field holonomy and the sum over Matsubara modes. In order to solve the set of integral equations \ref{str6} and \ref{str7} we consider the following ansatz
\begin{equation}
\label{stvpvi}
\begin{split}
 V^{(2)}_{F,+}(q,k) = k^{+} v^{(2)}_{+}(q,y), \quad\quad   V^{(2)}_{F,I}(q,k) = \beta (k^{+})^{2} v^{(2)}_{I}(q, y)
 \end{split}
\end{equation}

where $y=\beta k_{s}$. Then using the results in section \ref{angintegs} of the Appendix we can do the angular part of the integrals which in this case are given by
\begin{equation}
\label{str8}
\begin{split}
 \int _{0}^{2\pi} d\theta \frac{(\ell^{+})^{2}}{(\ell-k)^{+}} = 2\pi k^{+} \Theta (\ell_{s}-k_{s}),  \quad  \int _{0}^{2\pi} d\theta \frac{(\ell^{+})^{3}}{(\ell-k)^{+}} = 2\pi (k^{+})^{2} \Theta (\ell_{s}-k_{s})
 \end{split}
\end{equation} 

As a result equations \ref{str6} and \ref{str7}  now take the form
\begin{equation}
\label{str10}
\begin{split}
v^{(2)}_{+}(q,k) & = i   -\frac{2 i \lambda_{F}}{\beta^{3}} \int_{y}^{\infty} d x \hspace{0.1cm} x \left(2 x^{2}(g(x)-1)v^{(2)}_{I}(q, x)- \left(q_{3}\beta-2\tilde{f}(x)\right)v^{(2)}_{+}(q, x)\right)H_{F}(x,q_{3}\beta,\mu_{F})
\end{split}
\end{equation}

\begin{equation}
\label{str11}
\begin{split}
v^{(2)}_{I}(q,k)  =-\frac{2 i \lambda_{F}}{\beta^{3}} \int_{y}^{\infty} dx \hspace{0.1cm} x \left[2 v^{(2)}_{+}(q, x)- (q_{3}\beta+ 2 \tilde{f}(x))v^{(2)}_{I}(q, x)\right]H_{F}(x,q_{3}\beta,\mu_{F})
\end{split}
\end{equation}

Let us now differentiate equations \ref{str10} and \ref{str11} with respect to $y$. This yields
\begin{equation}
\label{str12}
\begin{split}
 \partial_{y}v^{(2)}_{+}(q,y)  =\frac{2 i \lambda_{F}}{\beta^{3}} y \left[2 y^{2} (g(y)-1)v^{(2)}_{I}(q,y)-(q_{3}\beta-2\tilde{f}(y))v^{(2)}_{+}(q, y)\right]H_{F}(y,q_{3}\beta,\mu_{F})
\end{split}
\end{equation}

\begin{equation}
\label{str13}
\begin{split}
 \partial_{y}v^{(2)}_{I}(q,y)  =\frac{2 i \lambda_{F}}{\beta^{3}}y \left[2 v^{(2)}_{+}(q, y)- (q_{3}\beta+2\tilde{f}(y))v^{(2)}_{I}(q,y)\right]H_{F}(y,q_{3}\beta, \mu_{F})
\end{split}
\end{equation}

Then multiplying \ref{str13} with $(q_{3}\beta-2\tilde{f})/2$ and adding it to \ref{str12} we obtain
\begin{equation}
\label{str14}
\begin{split}
&  \partial_{y}\left[v^{(2)}_{+}(q,y)+\frac{(q_{3}\beta-2\tilde{f}(y))}{2}v^{(2)}_{I}(q,y) \right] =0\\
& \implies v^{(2)}_{+}(q,y)= \eta(q_{3})-\frac{(q_{3}\beta-2\tilde{f}(y))}{2}v^{(2)}_{I}(q,y)
\end{split}
\end{equation}

where $\eta(q_{3})$ is an integration constant. Note that from equations \ref{str10} and \ref{str11}, we have $v^{(2)}_{+}(q, \infty)=i$, and $v^{(2)}_{I}(q, \infty)=0$. Using this in equation \ref{str14} gives $\eta(q_{3})=i$. Now using the above results we obtain via integrating \ref{str13}
\begin{equation}
\label{str15}
\begin{split}
& v^{(2)}_{I}(q,y) = \frac{i}{q_{3}\beta} \left(1-e^{i q_{3}\beta \hspace{0.04cm}\mathcal{F}_{F}(y,q_{3}\beta,\mu_{F})}\right) 
\end{split}
\end{equation}

Substituting this in equation \ref{str14} we get 
\begin{equation}
\label{str16}
\begin{split}
& v^{(2)}_{+}(q,y) = i- \frac{i(q_{3}\beta-2\tilde{f}(y))}{2q_{3}\beta} \left(1-e^{i q_{3}\beta \hspace{0.04cm}\mathcal{F}_{F}(y,q_{3}\beta,\mu_{F})}\right) 
\end{split}
\end{equation}

where
\begin{equation}
\label{str17}
\begin{split}
\mathcal{F}_{F}(y,q_{3}\beta,\mu_{F})= \frac{4\lambda_{F}}{\beta^{3}}\int_{y}^{\infty}dx \hspace{0.1cm} x \hspace{0.04cm} H_{F}(x,q_{3}\beta,\mu_{F}) = \int_{y}^{\infty}dx \hspace{0.1cm} x \hspace{0.04cm} \mathcal{H}_{F}(x,q_{3}\beta,\mu_{F}) 
\end{split}
\end{equation}

Thus in the large $N_{F}$ limit, the finite temperature vertex factor $V^{(2)}_{F}(q,k)$ to all orders in $\lambda_{F}$  is given by
\begin{equation}
\label{j2fmvertexfinal}
\begin{split}
 V^{(2)}_{F}= k^{+} v^{(2)}_{+}(q,y) \gamma^{+}+ \beta (k^{+})^{2} v^{(2)}_{I}(q, y) \mathbf{1}
\end{split}
\end{equation}

with $v^{(2)}_{+}(q,y)$  and $v^{(2)}_{I}(q,y)$ being given by equations \ref{str16} and \ref{str15} respectively.

\subsubsection{Finite temperature vertex factor for $J^{(2)}_{F,++}$ } 

The finite temperature vertex factor for $J^{(2)}_{F,++}$ can be defined as
\begin{equation}
\label{s2rfpp}
\begin{split}
\left\langle J^{(2)}_{F,++}(-q) \bar{\psi}^{j}(-k)\psi_{i}(p) \right\rangle_{\beta} & = U^{(2)}_{F}(q,k) \delta_{i}^{j}(2\pi)^{3} \delta^{(3)}(q+k-p) \delta_{n_{q}+n_{k},n_{p}} \delta^{(2)}(\vec{q}+\vec{k}-\vec{p}) 
\end{split}
\end{equation}

From equation \ref{strf5} we see that the contribution to the vertex factor $U^{(2)}_{F}(q,k)$ from the free theory, i.e., where the coupling to Chern Simons gauge fields has been turned off is given by
\begin{equation}
\label{s2rfpp1}
\begin{split}
\mathcal{U}^{(2)}_{F, 1} (q,k) = i \gamma_{+} k_{+}
\end{split}
\end{equation}

\begin{figure}[h!]
\begin{center}
\begin{tikzpicture}[scale=.5, transform shape]
 \draw[black,->] (-3.5,0) -- (-2,0);
 \draw[black,->] (-2,0) -- (0,0);
  \node at (0,0)[circle,fill,inner sep=5pt]{};
 \draw[black,->] (0,0) -- (1,0);
  \draw[black,->] (1,0) -- (3,0);
   \draw[black] (2,0) -- (3.5,0);
 \draw[draw=black, snake it] (1.5,0) arc (0:180:1.5cm);
 \draw [black,thick ,domain=120:60,->] plot ({2.5*cos(\x)}, {2.5*sin(\x)});
 \Huge{\node at (0,3.2) {$k+q-\ell$};}
 \node at (-2.5,-0.6) {$k$};
 \node at (0.7,-0.6) {$\ell $};
 \node at (2.6,-0.6) {$p$};
 \draw[black,->] (-1.5,-1.8) -- (-1.5,-0.6);
 \node at (-1.5,-2.4) {$q$};
  \Huge{\node at (-6.5,0) {$\mathcal{U}^{(2)}_{F, 2} (q,k)=$};}
  \Huge{\node at (6.5,0) {$+ ~~reflection$};}
  \draw (-1.5,0) node[cross] {};
\end{tikzpicture}
\caption{Feynman diagram contributing to $U^{(2)}_{F}$. The cross denotes the stress tensor vertex The internal propagator with the black dot is the exact thermal propagator for the fundamental fermions. In additional reflected diagram the stress tensor vertex lies to the right of the $3$-point vertex due to the $\bar{\psi}\gamma^{\mu} A_{\mu}\psi$ interaction term.}
\label{s2rfD1}
  \end{center}
\end{figure} 

Now the non-trivial 1PI diagram that we need to consider is shown in Fig.\ref{s2rfD1}. Using the relevant Feynman rules for the interaction vertices we can evaluate this as follows
\begin{equation}
\label{s2rfpp2}
\begin{split}
\mathcal{U}^{(2)}_{F, 2} (q,k)\delta_{i}^{j} & = \int \mathcal{D}^{3} \ell \left[\gamma^{-} T^{a}S(\ell) \mathcal{G}_{3+}(k+q-\ell)T^{a}\gamma^{3}+ \gamma^{3}T^{a} S(\ell)\mathcal{G}_{+3}(k-\ell)T^{a}\gamma^{-}\right]_{i}^{j} \\ 
& =- \frac{2 i \pi}{\kappa_{F}} \int \mathcal{D}^{3} \ell \hspace{0.1cm} \left[\frac{\gamma^{-} S(\ell)\gamma^{3}}{(\ell-k-q)^{+}}+\frac{ \gamma^{3} S(\ell)\gamma^{-}}{(k-\ell)^{+}}\right]\delta_{i}^{j}\\
& = \frac{2 i \pi}{\kappa_{F}}  \int \mathcal{D}^{3} \ell \hspace{0.1cm}\frac{1}{(\ell-k)^{+}} \frac{2}{\tilde{\ell}^{2}+\beta^{-2}\mu_{F}^{2}} \left[ i \ell_{+} (g(\ell)-1)\mathbf{1}+f(\ell) \ell_{s} \gamma^{-} \right]\delta_{i}^{j}
\end{split}
\end{equation}

The angular integrals that we need to deal with here are given by
\begin{equation}
\label{s2rfpp3} 
\begin{split} 
& \int_{0}^{2\pi}  d\theta \hspace{0.1cm} \frac{\ell_{+}}{(\ell-k)^{+}} = -2\pi \frac{k^{-}}{k^{+}} \frac{\ell_{s}^{2}}{k_{s}^{2}} \Theta(k_{s}-\ell_{s})\\
&  \int_{0}^{2\pi}  d\theta \hspace{0.1cm}\frac{1}{(\ell-k)^{+}} =-\frac{2\pi}{k^{+}}\Theta(k_{s}-\ell_{s})
\end{split}
\end{equation}

Using the above result in equation \ref{s2rfpp2} gives rise to
\begin{equation}
\label{s2rfpp4}
\begin{split}
\left(\mathcal{U}^{(2)}_{F,2} (q,k)\right)_{i}^{j} &= -\frac{2 i N_{F}}{\kappa_{F} \beta } \delta_{i}^{j}\int_{0}^{k_{s}} d\ell_{s} \hspace{0.05cm}\ell_{s} \left[ i(g(x)-1) \frac{k^{-}}{k^{+}} \frac{\ell_{s}^{2}}{k_{s}^{2}} \mathbf{1} +f(x) \ell_{s} \frac{1}{k^{+}} \gamma^{-}\right] \int_{-1/2}^{1/2} du \sum_{n=-\infty}^{\infty}\frac{1}{\tilde{\ell}^{2}+\beta^{-2}\mu_{F}^{2}}\\
&  =- \frac{ 2 i \lambda_{F}  }{  k^{+} } \delta_{i}^{j}  \int_{0}^{k_{s}} d\ell_{s} \hspace{0.05cm}\ell_{s} \left[ i(g(x)-1)\frac{\ell_{s}^{2}}{k_{s}^{2}}  k^{-}  \mathbf{1} +f (x)\ell_{s}  \gamma^{-}\right] \mathcal{M}_{F}(x,\mu_{F})\\
\end{split}
\end{equation}

where we have defined
\begin{equation}
\label{s2rfpp5}
\begin{split}
\mathcal{M}_{F}(x,\mu_{F}) = \frac{i \beta}{2 \pi |\lambda_{F}|} \frac{\left(\log\left[\cosh\left(\frac{1}{2}\sqrt{x^{2}+\mu_{F}^{2}}-\frac{i \pi |\lambda_{F}|}{2}\right)\right]-\mathrm{c.c}\right)}{ \sqrt{x^{2}+\mu_{F}^{2}}} 
\end{split}
\end{equation}

Now using the following useful identity
\begin{equation}
\label{s2rfpp6}
\begin{split}
 \frac{\partial \tilde{f}(x)}{\partial x} = - \frac{2 i\lambda_{F}}{\beta} x \hspace{0.05cm} \mathcal{M}_{F}(x,\mu_{F})
\end{split}
\end{equation}

we find that equation \ref{s2rfpp4} becomes
\begin{equation}
\label{s2rfpp7}
\begin{split}
 \left(\mathcal{U}^{(2)}_{F,2}(q,k)\right)_{i}^{j} & =\frac{1 }{k^{+} } \delta_{i}^{j}  \int_{0}^{k_{s}} d\ell_{s} \hspace{0.05cm} \left[ i(g(x)-1) k^{-} \frac{\ell_{s}^{2}}{k_{s}^{2}} \mathbf{1} +f (x)\ell_{s}  \gamma^{-}\right] \frac{\partial \tilde{f}}{\partial x} 
\end{split}
\end{equation}

The above radial integral straightforward to carry out via integrating by parts and we find
\begin{equation}
\label{s2rfpp8}
\begin{split}
\mathcal{U}^{(2)}_{F,2}(q,k) & = -\frac{i}{2(k^{+})^{2}\beta^{3}}\left[ \frac{1}{3}(\tilde{f}^{3}(y)-\tilde{f}^{3}(0))+ (y^{2}+\mu_{F}^{2})\tilde{f}(y)-\mu_{F}^{2}\tilde{f}(0)-2 \int_{0}^{y} dx \hspace{0.05cm} x \tilde{f}\right] \mathbf{1} \\
&+\frac{i}{2 k^{+}\beta^{2}}\hspace{0.04cm} (-\tilde{ f}^{2}(y)+\tilde{f}^{2}(0) )\gamma^{-}
\end{split}
\end{equation}

Therefore, at large $N_{F}$ and to all orders in $\lambda_{F}$, the finite temperature vertex factor $U^{(2)}_{F}(q,k)$ is
\begin{equation}
\label{s2rfpp9}
\begin{split}
U^{(2)}_{F}(q,k) = \mathcal{U}^{(2)}_{F,1}(q,k) +\mathcal{U}^{(2)}_{F,2}(q,k) 
\end{split}
\end{equation}

where $\mathcal{U}^{(2)}_{F,1}(q,k) $ and $\mathcal{U}^{(2)}_{F,2}(q,k)$ are specified in equations \ref{s2rfpp1} and \ref{s2rfpp9} respectively.

\subsubsection{Thermal $2$-point function}

The stress tensor thermal $2$-point function in the Regular Fermion theory is given by connecting the vertices $V^{(2)}_{F}(q,k)$ and $U^{(2)}_{F}(q,k)$ by a pair of exact thermal propagators for the elementary fermions. 
\begin{equation}
\label{ttrf}
\begin{split}
 \left\langle J^{(2)}_{F,--}(-q) J^{(2)}_{F,++}(q)\right\rangle_{\beta} & =-\sum_{i=1}^{N_{F}}\int \frac{d^{2}\ell}{(2\pi)^{2}}\hspace{0.04cm} \frac{1}{\beta} \sum_{n=-\infty}^{\infty} \mathrm{Tr}\left[V^{(2)}_{F}(q,\ell)S_{i}(\ell)U^{(2)}_{F}(-q,\ell+q)S_{i}(\ell+q)\right] 
\end{split}
\end{equation}

where the trace in the above equation is over the gamma matrices coming from the vertex factors and fermion propagators. Then substituting the results for the vertex factors from equations \ref{str15},\ref{str16} and \ref{s2rfpp9} we get
\begin{equation}
\label{ttrf1}
\begin{split}
& \left\langle J^{(2)}_{F,--}(-q) J^{(2)}_{F,++}(q)\right\rangle_{\beta} \\
& = -\frac{i N_{F}}{8 \pi \lambda_{F} \beta^{3}}\int _{0}^{\infty} dx \hspace{0.1cm} x \hspace{0.05cm} t_{-} \left[ i x^{2} (g(x)-1)+\left(2x^{2}+2\mu_{F}^{2}+q_{3}\beta \tilde{f}\right)v^{(2)}_{+}(q, x)\right] \mathcal{H}_{F}(x,q_{3}\beta, \mu_{F})\\
& -\frac{i N_{F}}{8 \pi \lambda_{F} \beta^{3}}\int _{0}^{\infty} dx \hspace{0.1cm} x \hspace{0.05cm} t_{I} \left[2 x^{2} (g(x)-1)v^{(2)}_{I}(q,x)- (q_{3}\beta -2 \tilde{f}) v^{(2)}_{+}(q,x)\right] \mathcal{H}_{F}(x,q_{3}\beta, \mu_{F}) \\
\end{split}
\end{equation}

where $t_{-},\hspace{0.05cm} t_{I}$ are defined below
\begin{equation}
\label{ttrf2}
\begin{split}
t_{-}=  \left( x^{2} - \tilde{f}^{2}(x)+\tilde{f}^{2}(0)\right), \quad t_{I}=- \left[ \frac{1}{3}(\tilde{f}^{3}(y)-\tilde{f}^{3}(0))+ (y^{2}+\mu_{F}^{2})\tilde{f}(y)-\mu_{F}^{2}\tilde{f}(0)-2 \int_{0}^{y} dx \hspace{0.05cm} x \tilde{f}\right] 
\end{split}
\end{equation}

Now the integrals in \ref{ttrf1} involve UV divergences. We will regulate them employing the same cutoff regularization scheme that we have used so far. Then using the relations provided in equation \ref{j0thermalrf6} we can simplify the above integral  expressions to a significant degree via suitable application of integration by parts. Remarkably, in this case also the final contribution originates from boundary terms only and is given by
\begin{equation}
\label{ttrf3}
\begin{split}
 & \left\langle J^{(2)}_{F,--}(-q) J^{(2)}_{F,++}(q)\right\rangle_{\beta}\\
& = \frac{i N_{F} }{128 \pi\lambda_{F} q_{3}\beta^{4}} \left(1- e^{i q_{3}\beta \hspace{0.04cm}\mathcal{F}_{F}\left(q_{3}\beta,\mu_{F}\right)}\right)\left(q_{3}^{2}\beta^{2}+4 \mu_{F}^{2}\right)\left(q^{2}_{3}\beta^{2}-4 \mu_{F}^{2}-4 q_{3}\beta\tilde{f}(0)\right)\\
& + \frac{i N_{F} }{96 \pi\lambda_{F} \beta^{3}} \left[ 2 \mu_{F}^{2}\left(2 \tilde{f}(0)+3 q_{3}\beta \right)+3 q_{3}^{2}\beta^{2}\tilde{f}(0) \right] \\
& - \frac{i N_{F} }{96 \pi\lambda_{F} \beta^{3}} \tilde{f}(\beta\Lambda) \left[ 3 (q_{3}^{2}\beta^{2}+4\mu_{F}^{2})-2 \tilde{f}(\beta\Lambda)\left( 3 q_{3}\beta- 4 \tilde{f}(\beta\Lambda)\right)\right]\\
& + \frac{i N_{F} }{8\pi\lambda_{F} \beta} \Lambda^{2} \tilde{f}(\beta\Lambda)- \frac{i N_{F} }{4 \pi\lambda_{F} \beta^{3}} \int_{0}^{\beta\Lambda} dx \hspace{0.1cm} x \hspace{0.04cm} \tilde{f}(x)
\end{split}
\end{equation}

In order to obtain a UV finite answer we have to subtract off the divergent terms in the above which are contained in the terms that involve $\tilde{f}(\beta\Lambda)$, since for $\beta\Lambda \gg 1$ we have $\tilde{f}(\beta\Lambda) \sim - i \lambda_{F} \beta \Lambda $. These divergences should be cancelled by introducing appropriate counterterms for the background metric that couples to stress tensor. However we will not elaborate further on the systematic implementation of this procedure and hope to address it more thoroughly in future work.  Thus, after discarding the divergent pieces and using $\tilde{f}(0) = - i \hspace{0.04cm} \mathrm{sgn}(\lambda_{F}) \mu_{F}$  in equation \ref{ttrf3} we get the following renormalised thermal $2$-point function.
\begin{equation}
\label{strf1a}
\begin{split}
 & \left\langle J^{(2)}_{F,--}(-q) J^{(2)}_{F,++}(q)\right\rangle_{\beta}\\
& = \frac{i N_{F} }{128 \pi\lambda_{F} q_{3}\beta^{4}} \left(1- e^{i q_{3}\beta \hspace{0.04cm}\mathcal{F}_{F}\left(q_{3}\beta,\mu_{F}\right)}\right)\left(q_{3}^{2}\beta^{2}+4 \mu_{F}^{2}\right)\left(q^{2}_{3}\beta^{2}-4  \mu_{F}^{2}+4 i q_{3}\beta \hspace{0.04cm}\mathrm{sgn}(\lambda_{F}) \mu_{F}\right)\\
& + \frac{i N_{F} }{96 \pi\lambda_{F} \beta^{3}} \left[ 2 \mu_{F}^{2}\left(3 q_{3}\beta - 2i \hspace{0.04cm} \mathrm{sgn}(\lambda_{F}) \mu_{F} \right) - 3 i \hspace{0.04cm} q_{3}^{2}\beta^{2} \hspace{0.04cm} \mathrm{sgn}(\lambda_{F}) \mu_{F} \right]  \\
&+ \frac{N_{F} }{12 \pi  \beta^{3}} \mu_{F}^{3} - \frac{i N_{F} }{4 \pi^{2} \lambda_{F} \beta^{3}} \int_{\mu_{F}}^{\infty} dx \hspace{0.1cm} x \left[ \mathrm{Li}_{2}\left(-e^{-x+i \pi \lambda_{F}}\right) - \mathrm{c.c} \right] 
\end{split}
\end{equation}



\subsection{Critical Fermions}

The finite temperature $2$-point function of the stress tensor in the Critical Fermion theory can be immediately derived from the result in \ref{strf1a} corresponding to the Regular Fermion theory by replacing $\mu_{F}$ with $\mu_{F,c}$ which is the thermal mass of the fundamental fermions in the critical theory. The reasoning here is identical to what appeared in sections \ref{s0cf} and \ref{s1cf}. Then from equation \ref{strf1a} we obtain
\begin{equation}
\label{stcf}
\begin{split}
& \left\langle \tilde{J}^{(2)}_{F,--}(-q) \tilde{J}^{(2)}_{F,++}(q)\right\rangle_{\beta}\\
& = \frac{i N_{F} }{128 \pi\lambda_{F} q_{3}\beta^{4}} \left(1- e^{i q_{3}\beta \hspace{0.04cm}\mathcal{F}_{F}\left(q_{3}\beta,\mu_{F,c}\right)}\right)\left(q_{3}^{2}\beta^{2}+4 \mu_{F}^{2}\right)\left(q^{2}_{3}\beta^{2}-4  \mu_{F,c}^{2}+4 i q_{3}\beta \hspace{0.04cm}\mathrm{sgn}(\lambda_{F}) \mu_{F,c}\right)\\
& + \frac{i N_{F} }{96 \pi\lambda_{F} \beta^{3}} \left[ 2 \mu_{F,c}^{2}\left(3 q_{3}\beta - 2i \hspace{0.04cm} \mathrm{sgn}(\lambda_{F}) \mu_{F,c} \right) - 3 i \hspace{0.04cm} q_{3}^{2}\beta^{2} \hspace{0.04cm} \mathrm{sgn}(\lambda_{F}) \mu_{F,c} \right]  \\
&+ \frac{N_{F} }{12 \pi  \beta^{3}} \mu_{F,c}^{3} - \frac{i N_{F} }{4 \pi^{2} \lambda_{F} \beta^{3}} \int_{\mu_{F,c}}^{\infty} dx \hspace{0.1cm} x \left[ \mathrm{Li}_{2}\left(-e^{-x+i \pi \lambda_{F}}\right) - \mathrm{c.c} \right] 
\end{split}
\end{equation}

\subsection{Duality Check}

Let us now check if the above results are consistent with bosonization dualities. For this we will only consider here the Regular Fermion and Critical Boson theories. Applying the duality map from \ref{dualitymap1} and using \ref{dualtransf1} in \ref{strf1a} we find that the $2$-point function in the Regular Fermion theory transforms to the stress tensor $2$-point function in the Critical Boson theory upto the following extra terms
\begin{equation}
\label{dualextra}
\begin{split}
-  \frac{i N_{B} }{64 \pi\lambda_{B} \beta^{2}} q_{3}\left(q_{3}^{2}\beta^{2}+4 \mu_{B,c}^{2}\right)
\end{split}
\end{equation}

Now the first term in the above is a contact term since after Fourier transforming to position space it is the derivative of a delta function. This can be removed by adding a counterterm to the action of the Critical Boson theory which involves a gravitational Chern Simons term for the background metric that couples to the stress tensor. We again refer the reader to \citep{Closset2012}, for a discussion on contact terms in stress tensor correlators in $3$-$d$ QFTs. The second piece proportional to the thermal mass in \ref{dualextra} is however quite puzzling. We can not qualify it as a contact term in the standard sense even though in position space, it's contribution is a delta function. This is because the coefficient of this term is temperature dependent and so it is not clear if it can be removed via addition of local counterterms. At this moment we are unable to provide a satisfactory resolution of this problem. One possibility is that this piece is an artifact of the cutoff regularisation scheme that we have been using throughout this paper. It has to be checked whether this extra term survives in an alternative scheme such as dimensional regularisation. We leave a careful treatment of this problem to future work\footnote{We thank Shiraz Minwalla for discussions regarding this issue and for suggesting to us the implementation of dimensional regularisation scheme. }.

\bibliographystyle{JHEP}
\bibliography{tcsref}

\providecommand{\href}[2]{#2}\begingroup\raggedright\begin{thebibliography}{10}

\bibitem{Maldacena:1997re}
J.~M. Maldacena, \emph{{The Large N limit of superconformal field theories and
  supergravity}}, \href{http://dx.doi.org/10.1023/A:1026654312961,
  10.4310/ATMP.1998.v2.n2.a1}{\emph{Int. J. Theor. Phys.} {\bf 38} (1999)
  1113--1133}, [\href{http://arxiv.org/abs/hep-th/9711200}{{\tt
  hep-th/9711200}}].

\bibitem{Witten:1998zw}
E.~Witten, \emph{{Anti-de Sitter space, thermal phase transition, and
  confinement in gauge theories}},
  \href{http://dx.doi.org/10.4310/ATMP.1998.v2.n3.a3}{\emph{Adv. Theor. Math.
  Phys.} {\bf 2} (1998) 505--532},
  [\href{http://arxiv.org/abs/hep-th/9803131}{{\tt hep-th/9803131}}].

\bibitem{Giombi:2011kc}
S.~Giombi, S.~Minwalla, S.~Prakash, S.~P. Trivedi, S.~R. Wadia and X.~Yin,
  \emph{{Chern-Simons Theory with Vector Fermion Matter}},
  \href{http://dx.doi.org/10.1140/epjc/s10052-012-2112-0}{\emph{Eur. Phys. J.}
  {\bf C72} (2012) 2112}, [\href{http://arxiv.org/abs/1110.4386}{{\tt
  1110.4386}}].

\bibitem{Aharony:2011jz}
O.~Aharony, G.~Gur-Ari and R.~Yacoby, \emph{{d=3 Bosonic Vector Models Coupled
  to Chern-Simons Gauge Theories}},
  \href{http://dx.doi.org/10.1007/JHEP03(2012)037}{\emph{JHEP} {\bf 03} (2012)
  037}, [\href{http://arxiv.org/abs/1110.4382}{{\tt 1110.4382}}].

\bibitem{Aharony:2012nh}
O.~Aharony, G.~Gur-Ari and R.~Yacoby, \emph{{Correlation Functions of Large N
  Chern-Simons-Matter Theories and Bosonization in Three Dimensions}},
  \href{http://dx.doi.org/10.1007/JHEP12(2012)028}{\emph{JHEP} {\bf 12} (2012)
  028}, [\href{http://arxiv.org/abs/1207.4593}{{\tt 1207.4593}}].

\bibitem{GurAri:2012is}
G.~Gur-Ari and R.~Yacoby, \emph{{Correlators of Large N Fermionic Chern-Simons
  Vector Models}}, \href{http://dx.doi.org/10.1007/JHEP02(2013)150}{\emph{JHEP}
  {\bf 02} (2013) 150}, [\href{http://arxiv.org/abs/1211.1866}{{\tt
  1211.1866}}].

\bibitem{Geracie:2015drf}
M.~Geracie, M.~Goykhman and D.~T. Son, \emph{{Dense Chern-Simons Matter with
  Fermions at Large N}},
  \href{http://dx.doi.org/10.1007/JHEP04(2016)103}{\emph{JHEP} {\bf 04} (2016)
  103}, [\href{http://arxiv.org/abs/1511.04772}{{\tt 1511.04772}}].

\bibitem{Turiaci:2018dht}
G.~J. Turiaci and A.~Zhiboedov, \emph{{Veneziano Amplitude of Vasiliev
  Theory}}, \href{http://dx.doi.org/10.1007/JHEP10(2018)034}{\emph{JHEP} {\bf
  10} (2018) 034}, [\href{http://arxiv.org/abs/1802.04390}{{\tt 1802.04390}}].

\bibitem{Yacoby:2018yvy}
R.~Yacoby, \emph{{Scalar Correlators in Bosonic Chern-Simons Vector Models}},
  \href{http://arxiv.org/abs/1805.11627}{{\tt 1805.11627}}.

\bibitem{Kalloor:2019xjb}
R.~R. Kalloor, \emph{{Four-point functions in large $N$ Chern-Simons fermionic
  theories}},  \href{http://arxiv.org/abs/1910.14617}{{\tt 1910.14617}}.

\bibitem{Jain:2014nza}
S.~Jain, M.~Mandlik, S.~Minwalla, T.~Takimi, S.~R. Wadia and S.~Yokoyama,
  \emph{{Unitarity, Crossing Symmetry and Duality of the S-matrix in large N
  Chern-Simons theories with fundamental matter}},
  \href{http://dx.doi.org/10.1007/JHEP04(2015)129}{\emph{JHEP} {\bf 04} (2015)
  129}, [\href{http://arxiv.org/abs/1404.6373}{{\tt 1404.6373}}].

\bibitem{Dandekar:2014era}
Y.~Dandekar, M.~Mandlik and S.~Minwalla, \emph{{Poles in the $S$-Matrix of
  Relativistic Chern-Simons Matter theories from Quantum Mechanics}},
  \href{http://dx.doi.org/10.1007/JHEP04(2015)102}{\emph{JHEP} {\bf 04} (2015)
  102}, [\href{http://arxiv.org/abs/1407.1322}{{\tt 1407.1322}}].

\bibitem{Inbasekar:2015tsa}
K.~Inbasekar, S.~Jain, S.~Mazumdar, S.~Minwalla, V.~Umesh and S.~Yokoyama,
  \emph{{Unitarity, crossing symmetry and duality in the scattering of $
  \mathcal{N}=1 $ susy matter Chern-Simons theories}},
  \href{http://dx.doi.org/10.1007/JHEP10(2015)176}{\emph{JHEP} {\bf 10} (2015)
  176}, [\href{http://arxiv.org/abs/1505.06571}{{\tt 1505.06571}}].

\bibitem{Inbasekar:2017ieo}
K.~Inbasekar, S.~Jain, P.~Nayak and V.~Umesh, \emph{{All tree level scattering
  amplitudes in Chern-Simons theories with fundamental matter}},
  \href{http://dx.doi.org/10.1103/PhysRevLett.121.161601}{\emph{Phys. Rev.
  Lett.} {\bf 121} (2018) 161601}, [\href{http://arxiv.org/abs/1710.04227}{{\tt
  1710.04227}}].

\bibitem{Inbasekar:2019wdw}
K.~Inbasekar, S.~Jain, V.~Malvimat, A.~Mehta, P.~Nayak and T.~Sharma,
  \emph{{Correlation functions in ${\cal N}=2$ Supersymmetric vector matter
  Chern-Simons theory}},  \href{http://arxiv.org/abs/1907.11722}{{\tt
  1907.11722}}.

\bibitem{Jain:2012qi}
S.~Jain, S.~P. Trivedi, S.~R. Wadia and S.~Yokoyama, \emph{{Supersymmetric
  Chern-Simons Theories with Vector Matter}},
  \href{http://dx.doi.org/10.1007/JHEP10(2012)194}{\emph{JHEP} {\bf 10} (2012)
  194}, [\href{http://arxiv.org/abs/1207.4750}{{\tt 1207.4750}}].

\bibitem{Aharony:2012ns}
O.~Aharony, S.~Giombi, G.~Gur-Ari, J.~Maldacena and R.~Yacoby, \emph{{The
  Thermal Free Energy in Large N Chern-Simons-Matter Theories}},
  \href{http://dx.doi.org/10.1007/JHEP03(2013)121}{\emph{JHEP} {\bf 03} (2013)
  121}, [\href{http://arxiv.org/abs/1211.4843}{{\tt 1211.4843}}].

\bibitem{Jain:2013py}
S.~Jain, S.~Minwalla, T.~Sharma, T.~Takimi, S.~R. Wadia and S.~Yokoyama,
  \emph{{Phases of large $N$ vector Chern-Simons theories on $S^2 \times
  S^1$}}, \href{http://dx.doi.org/10.1007/JHEP09(2013)009}{\emph{JHEP} {\bf 09}
  (2013) 009}, [\href{http://arxiv.org/abs/1301.6169}{{\tt 1301.6169}}].

\bibitem{Yokoyama:2012fa}
S.~Yokoyama, \emph{{Chern-Simons-Fermion Vector Model with Chemical
  Potential}}, \href{http://dx.doi.org/10.1007/JHEP01(2013)052}{\emph{JHEP}
  {\bf 01} (2013) 052}, [\href{http://arxiv.org/abs/1210.4109}{{\tt
  1210.4109}}].

\bibitem{Jain:2013gza}
S.~Jain, S.~Minwalla and S.~Yokoyama, \emph{{Chern Simons duality with a
  fundamental boson and fermion}},
  \href{http://dx.doi.org/10.1007/JHEP11(2013)037}{\emph{JHEP} {\bf 11} (2013)
  037}, [\href{http://arxiv.org/abs/1305.7235}{{\tt 1305.7235}}].

\bibitem{Choudhury:2018iwf}
S.~Choudhury, A.~Dey, I.~Halder, S.~Jain, L.~Janagal, S.~Minwalla et~al.,
  \emph{{Bose-Fermi Chern-Simons Dualities in the Higgsed Phase}},
  \href{http://dx.doi.org/10.1007/JHEP11(2018)177}{\emph{JHEP} {\bf 11} (2018)
  177}, [\href{http://arxiv.org/abs/1804.08635}{{\tt 1804.08635}}].

\bibitem{Dey:2019ihe}
A.~Dey, I.~Halder, S.~Jain, S.~Minwalla and N.~Prabhakar, \emph{{The large N
  phase diagram of $ \mathcal{N} $ = 2 SU(N) Chern-Simons theory with one
  fundamental chiral multiplet}},
  \href{http://dx.doi.org/10.1007/JHEP11(2019)113}{\emph{JHEP} {\bf 11} (2019)
  113}, [\href{http://arxiv.org/abs/1904.07286}{{\tt 1904.07286}}].

\bibitem{Gross:1980he}
D.~J. Gross and E.~Witten, \emph{{Possible Third Order Phase Transition in the
  Large N Lattice Gauge Theory}},
  \href{http://dx.doi.org/10.1103/PhysRevD.21.446}{\emph{Phys. Rev.} {\bf D21}
  (1980) 446--453}.

\bibitem{Wadia:1980cp}
S.~R. Wadia, \emph{{$N$ = Infinity Phase Transition in a Class of Exactly
  Soluble Model Lattice Gauge Theories}},
  \href{http://dx.doi.org/10.1016/0370-2693(80)90353-6}{\emph{Phys. Lett.} {\bf
  93B} (1980) 403--410}.

\bibitem{Fradkin:1987ks}
E.~S. Fradkin and M.~A. Vasiliev, \emph{{On the Gravitational Interaction of
  Massless Higher Spin Fields}},
  \href{http://dx.doi.org/10.1016/0370-2693(87)91275-5}{\emph{Phys. Lett.} {\bf
  B189} (1987) 89--95}.

\bibitem{Vasiliev:1992av}
M.~A. Vasiliev, \emph{{More on equations of motion for interacting massless
  fields of all spins in (3+1)-dimensions}},
  \href{http://dx.doi.org/10.1016/0370-2693(92)91457-K}{\emph{Phys. Lett.} {\bf
  B285} (1992) 225--234}.

\bibitem{Vasiliev:1995dn}
M.~A. Vasiliev, \emph{{Higher spin gauge theories in four-dimensions,
  three-dimensions, and two-dimensions}},
  \href{http://dx.doi.org/10.1142/S0218271896000473}{\emph{Int. J. Mod. Phys.}
  {\bf D5} (1996) 763--797}, [\href{http://arxiv.org/abs/hep-th/9611024}{{\tt
  hep-th/9611024}}].

\bibitem{Vasiliev:1999ba}
M.~A. Vasiliev, \emph{{Higher spin gauge theories: Star product and AdS
  space}},  \href{http://arxiv.org/abs/hep-th/9910096}{{\tt hep-th/9910096}}.

\bibitem{GurAri:2016xff}
G.~Gur-Ari, S.~A. Hartnoll and R.~Mahajan, \emph{{Transport in
  Chern-Simons-Matter Theories}},
  \href{http://dx.doi.org/10.1007/JHEP07(2016)090}{\emph{JHEP} {\bf 07} (2016)
  090}, [\href{http://arxiv.org/abs/1605.01122}{{\tt 1605.01122}}].

\bibitem{Katz:2014rla}
E.~Katz, S.~Sachdev, E.~S. Sørensen and W.~Witczak-Krempa, \emph{{Conformal
  field theories at nonzero temperature: Operator product expansions, Monte
  Carlo, and holography}},
  \href{http://dx.doi.org/10.1103/PhysRevB.90.245109}{\emph{Phys. Rev.} {\bf
  B90} (2014) 245109}, [\href{http://arxiv.org/abs/1409.3841}{{\tt
  1409.3841}}].

\bibitem{Romatschke:2019ybu}
P.~Romatschke, \emph{{Finite-Temperature Conformal Field Theory Results for All
  Couplings: O(N) Model in 2+1 Dimensions}},
  \href{http://dx.doi.org/10.1103/PhysRevLett.122.231603}{\emph{Phys. Rev.
  Lett.} {\bf 122} (2019) 231603}, [\href{http://arxiv.org/abs/1904.09995}{{\tt
  1904.09995}}].

\bibitem{Mishra:2020wos}
A.~Mishra, \emph{{On thermal correlators and bosonization duality in
  Chern-Simons theories with massive fundamental matter}},
  \href{http://dx.doi.org/10.1007/JHEP01(2021)109}{\emph{JHEP} {\bf 01} (2021)
  109}, [\href{http://arxiv.org/abs/2010.03699}{{\tt 2010.03699}}].

\bibitem{Closset2012}
C.~Closset, T.~T. Dumitrescu, G.~Festuccia, Z.~Komargodski and N.~Seiberg,
  \emph{Comments on chern-simons contact terms in three dimensions},
  \href{http://dx.doi.org/10.1007/JHEP09(2012)091}{\emph{Journal of High Energy
  Physics} {\bf 2012} (Sep, 2012) 91}.

\bibitem{Iliesiu:2018fao}
L.~Iliesiu, M.~Koloğlu, R.~Mahajan, E.~Perlmutter and D.~Simmons-Duffin,
  \emph{{The Conformal Bootstrap at Finite Temperature}},
  \href{http://dx.doi.org/10.1007/JHEP10(2018)070}{\emph{JHEP} {\bf 10} (2018)
  070}, [\href{http://arxiv.org/abs/1802.10266}{{\tt 1802.10266}}].

\bibitem{Aharony:2018npf}
O.~Aharony, L.~F. Alday, A.~Bissi and R.~Yacoby, \emph{{The Analytic Bootstrap
  for Large $N$ Chern-Simons Vector Models}},
  \href{http://dx.doi.org/10.1007/JHEP08(2018)166}{\emph{JHEP} {\bf 08} (2018)
  166}, [\href{http://arxiv.org/abs/1805.04377}{{\tt 1805.04377}}].

\bibitem{Didenko:2009td}
V.~E. Didenko and M.~A. Vasiliev, \emph{{Static BPS black hole in 4d
  higher-spin gauge theory}},
  \href{http://dx.doi.org/10.1016/j.physletb.2013.04.021,
  10.1016/j.physletb.2009.11.023}{\emph{Phys. Lett.} {\bf B682} (2009)
  305--315}, [\href{http://arxiv.org/abs/0906.3898}{{\tt 0906.3898}}].

\bibitem{Maldacena:2011jn}
J.~Maldacena and A.~Zhiboedov, \emph{{Constraining Conformal Field Theories
  with A Higher Spin Symmetry}},
  \href{http://dx.doi.org/10.1088/1751-8113/46/21/214011}{\emph{J. Phys.} {\bf
  A46} (2013) 214011}, [\href{http://arxiv.org/abs/1112.1016}{{\tt
  1112.1016}}].

\bibitem{Maldacena:2012sf}
J.~Maldacena and A.~Zhiboedov, \emph{{Constraining conformal field theories
  with a slightly broken higher spin symmetry}},
  \href{http://dx.doi.org/10.1088/0264-9381/30/10/104003}{\emph{Class. Quant.
  Grav.} {\bf 30} (2013) 104003}, [\href{http://arxiv.org/abs/1204.3882}{{\tt
  1204.3882}}].

\end{thebibliography}\endgroup


\end{document}